\documentclass[letterpaper,journal]{IEEEtran}
\usepackage{amsmath,amsfonts}
\usepackage{algorithmic}
\usepackage{algorithm}
\usepackage{array}
\usepackage{textcomp}
\usepackage{stfloats}
\usepackage{url}
\usepackage{verbatim}
\usepackage{graphicx}
\usepackage{cite}
\newcommand {\ie} {{\em i.e., }}
\newcommand {\eg} {{\em e.g., }}
\newcommand {\beq} {\begin{equation}}
\newcommand {\eeq} {\end{equation}}
\newcommand {\bequn} {\begin{equation*}}
\newcommand {\eequn} {\end{equation*}}
\newcommand {\bear} {\begin{eqnarray}}
\newcommand {\eear} {\end{eqnarray}}
\newcommand {\bearun} {\begin{eqnarray*}}
\newcommand {\eearun} {\end{eqnarray*}}
\newcommand {\fig}[1]{Fig.~\ref{#1}}
\newcommand{\figs}[2]{Figs.~\ref{#1} and~\ref{#2}}
\newcommand {\Eqref}[1]{Eq.~(\ref{#1})}

\newcommand{\Greedy}{Greedy Reprofiling\xspace}

\newtheorem{theorem}{Theorem}
\newtheorem{lemma}[theorem]{Lemma}

\newtheorem{proposition}[theorem]{Proposition}

\usepackage{enumitem}
\usepackage{multirow}
\usepackage{supertabular}
\usepackage{subcaption}
\usepackage{xcolor}
\usepackage{hyperref}
\usepackage{makecell}
\usepackage{pifont}
\usepackage{tablefootnote}
\usepackage{makecell}
\usepackage{amssymb}
\usepackage{bm}
\usepackage{framed}
\usepackage[T1]{fontenc}
\usepackage{xspace}
\usepackage{xstring}
\usepackage{ifthen}
\usepackage{pgfkeys}

\begin{document}

\title{On the Benefits of Traffic ``Reprofiling'' \\ The Multiple Hops Case--Part II}

\author{Jiaming~Qiu,~\IEEEmembership{Member,~IEEE}
        and Roch~Gu\'{e}rin,~\IEEEmembership{Fellow,~ACM,~IEEE}
\thanks{J. Qiu and R. Gu\'{e}rin are with the Computer Science and Engineering department at Washington University in St.~Louis, Saint Louis, MO 63130, USA, e-mail: {\tt \{qiujiaming,guerin\}@wustl.edu}.}
\thanks{This work was supported by NSF grant CNS 2006530.}
\thanks{Any opinions, findings, and conclusions or recommendations expressed in this material are those of the author(s) and do not necessarily reflect the views of the National Science Foundation}
}

\maketitle
\thispagestyle{plain}
\pagestyle{plain}

\begin{abstract}
Delivering hard delay guarantees over packet networks is increasingly important to applications ranging from automotive systems, avionics, industrial control, etc.  Traffic control and schedulers play an essential role in enforcing such guarantees. In this paper, we focus on ``simple'' static priority and FIFO schedulers, and explore how \emph{reprofiling} flows entering the network, \ie proactively shaping them to a different traffic profile, can deliver delay guarantees with less bandwidth.  To that end, we formulate a joint optimization framework and develop efficient algorithms to solve it. Extensive evaluations across both realistic and synthetic topologies demonstrate that, as with more sophisticated schedulers, reprofiling flows is beneficial. They also highlight an intuitive coupling between a scheduler's capability and its ability to leverage more complex reprofiling solutions.
\end{abstract}


\section{Introduction}
\label{sec:intro}

Applications from domains as varied as automotive systems, avionics, industrial control, smart grids, etc., are increasingly deployed over packet networks where they demand predictable communication with bounded latency~\cite{automotive21,afdx,factory20,xu2023survey,satka2023comprehensive,smartgrid23,aws26,google-netw26,msft26}. This is reflected in recent standardization efforts such as Time-Sensitive Networking (TSN) and Deterministic Networking (DetNet)~\cite{tsn18,parsons22,seol21,zhang2024time,zanbouri2024comprehensive,detnet,rfc9023}, which both focus on enabling deterministic delay guarantees for regulated traffic under a range of scheduling mechanisms. This is the setting this paper targets.

Regulated flows indicate desired end-to-end delay bounds, with the \emph{token bucket}~\cite{le2018theory} often used as their traffic regulator, one that specifies both sustained transmission rates and burstiness constraints. The network task is then to provision sufficient resources (bandwidth) to guarantee those bounds.  Of interest in such a setting is to \emph{minimize} the required provisioning, \ie the amount of network bandwidth needed\footnote{A dual perspective maximizes the number of flows that can be accommodated for a given amount of bandwidth.}.

This paper studies the role of \emph{reprofiling}, \ie proactively modifying flows' profiles as they enter the network (and subsequently at each network hop), in reducing the bandwidth required to meet delay guarantees. The benefits of reprofiling with FIFO and static priority schedulers were demonstrated in~\cite{song2024benefits} in a single-hop setting. The extension to multi-hop networks was presented for service curve schedulers (SCED~\cite{sariowan2002sced}) in a precursor (Part I) to this paper~\cite{qiu2024benefits}. This paper extends the results to FIFO and static priority schedulers.

With such schedulers, reprofiling is realized through non-work-conserving traffic regulators, \ie \emph{shapers}\footnote{Hence, the paper uses the terms shaping and reprofiling interchangeably.}, that are used to reduce flows' burstiness.  Reprofiling, therefore, introduces a fundamental trade-off: it consumes part of the end-to-end delay budget upfront, leaving tighter delay constraints for in-network scheduling, while the smoother traffic it produces can reduce resource requirements at every hop. How to best leverage this trade-off is scheduler dependent.  An adaptive scheduler such as SCED was found~\cite{qiu2024benefits} to benefit most from ``middle-ground'' solutions, \ie using some of the delay budget to make flows smoother through reprofiling, while preserving enough of it for the scheduler to work with. Unlike SCED, FIFO and static-priority schedulers are static in their classification of packets.  How this affects their reprofiling solutions is unclear. Addressing this question is the focus of this paper.

The paper follows a similar methodology as~\cite{qiu2024benefits}.  It develops a unified framework to study bandwidth minimization under reprofiling for FIFO and static-priority schedulers, formulates a joint optimization for shaping, scheduling, and bandwidth provisioning, and designs efficient algorithms to solve it.

The investigation offers insights into how schedulers' ``expressiveness'' influences how reprofiling realizes bandwidth savings.  
For simple schedulers such as FIFO and static priority, the best or close to best reprofiling option is often ``full shaping,'' \ie allocating a flow's entire delay budget to making it smoother.
In contrast~\cite{qiu2024benefits}, SCED prefers less aggressive solutions because its greater scheduling flexibility is capable of exploiting some residual delay budget.

The remainder of the paper is structured as follows. Section~\ref{sec:background} offers relevant background on network calculus and traffic shaping. Our problem formulation is introduced in Section~\ref{sec:formulation}, with Section~\ref{sec:solution} presenting our bandwidth minimization algorithms for FIFO and static-priority schedulers. Section~\ref{sec:evaluation} evaluates their performance across a range of scenarios. Section~\ref{sec:related} discusses related work, while Section~\ref{sec:conclusion} concludes the paper. Proofs and supplementary material are relegated to appendices. For reproducibility, our solutions and the settings in which they are evaluated are available at~\url{https://github.com/qiujiaming315/traffic-reprofiling}.


\section{Background}
\label{sec:background}

\subsection{Network Calculus}
\label{sec:netcalc}

Network calculus~\cite{le2001network} provides a framework for computing worst-case delay and buffer bounds in packet networks under deterministic traffic models. In this section, we briefly review key concepts and results on which this paper relies. To the extent possible, we follow the notation of~\cite{qiu2024benefits}, and, as in~\cite{qiu2024benefits}, we adopt a fluid model to simplify the exposition.

\subsubsection{Arrival Curves}

Arrival curves constrain the amount of traffic a flow can generate over time~\cite[Definition 1.2.1]{le2001network}. Formally, given a wide-sense increasing function $\alpha(t)$ for $t \ge 0$, a flow with cumulative arrival function $A(t)$ is said to conform to the arrival curve $\alpha$ if
\begin{equation*}
\forall s \in [0,t], \quad A(t) - A(s) \le \alpha(t-s).
\end{equation*}
In other words, $\alpha(t-s)$ upper-bounds the amount of traffic that may arrive in any interval of duration $t-s$.

A commonly used arrival curve is the two-parameter token (or leaky) bucket~\cite{lb86}, denoted $(r,b)$. It corresponds to the affine arrival curve $\alpha(t)=rt+b$ for $t>0$, where $r$ represents the sustained rate of the flow and $b$ its maximum burst size. In this paper, we assume that flows entering the network are regulated by token buckets, which define their traffic profiles.

\subsubsection{Service Curves}

Service curves characterize the minimum service guaranteed to flows. If a flow with cumulative arrival function $A(t)$ is guaranteed a service curve $\beta$, then the cumulative service $S(t)$ it has received by time $t$ satisfies the condition: $\exists\, s \in [0,t]$ such that~\cite[Section 1.3]{le2001network}
\begin{equation*}
S(t) \ge A(s) + \beta(t-s).
\end{equation*}

\subsubsection{Delay Bounds}
\label{sec:delay_bounds}

Given a flow with arrival curve $\alpha$ and service curve $\beta$, its worst-case delay is upper-bounded by the maximum horizontal distance between $\alpha$ and $\beta$~\cite[Section 1.4]{le2001network}. This bound is expressed as
\begin{equation}
\label{eq:delay_bound}
\Delta(\alpha,\beta) =
\sup_{t \ge 0}
\left\{
\inf \{ s \ge 0 : \alpha(t) \le \beta(t+s) \}
\right\}.
\end{equation}

In the following sections, we use~\Eqref{eq:delay_bound} to derive the bandwidth required to guarantee per-hop scheduling delays.

\subsection{Traffic Shaping}
\label{sec:background_shaping}

Shapers regulate traffic to ensure compliance with a specified arrival curve. Following its arrival, a packet's \emph{eligibility time} is the earliest time at which it can depart the shaper without violating the arrival-curve constraint. Greedy shapers~\cite[Section 1.5.3]{le2001network} release packets as soon as they become eligible.

Traffic shaping plays an important role in multi-hop networks. Shapers can be placed before schedulers to (re)shape flows to their original traffic profiles without affecting worst-case delays\footnote{Based on the Pay Bursts Only Once (PBOO) property~\cite[Section 1.4.3]{le2001network}.}. This prevents burst accumulation across hops, which can significantly tighten end-to-end delay bounds~\cite{georgiadis96a}.

\begin{figure}[htbp]
    \centering
    \includegraphics[width=0.5\linewidth]{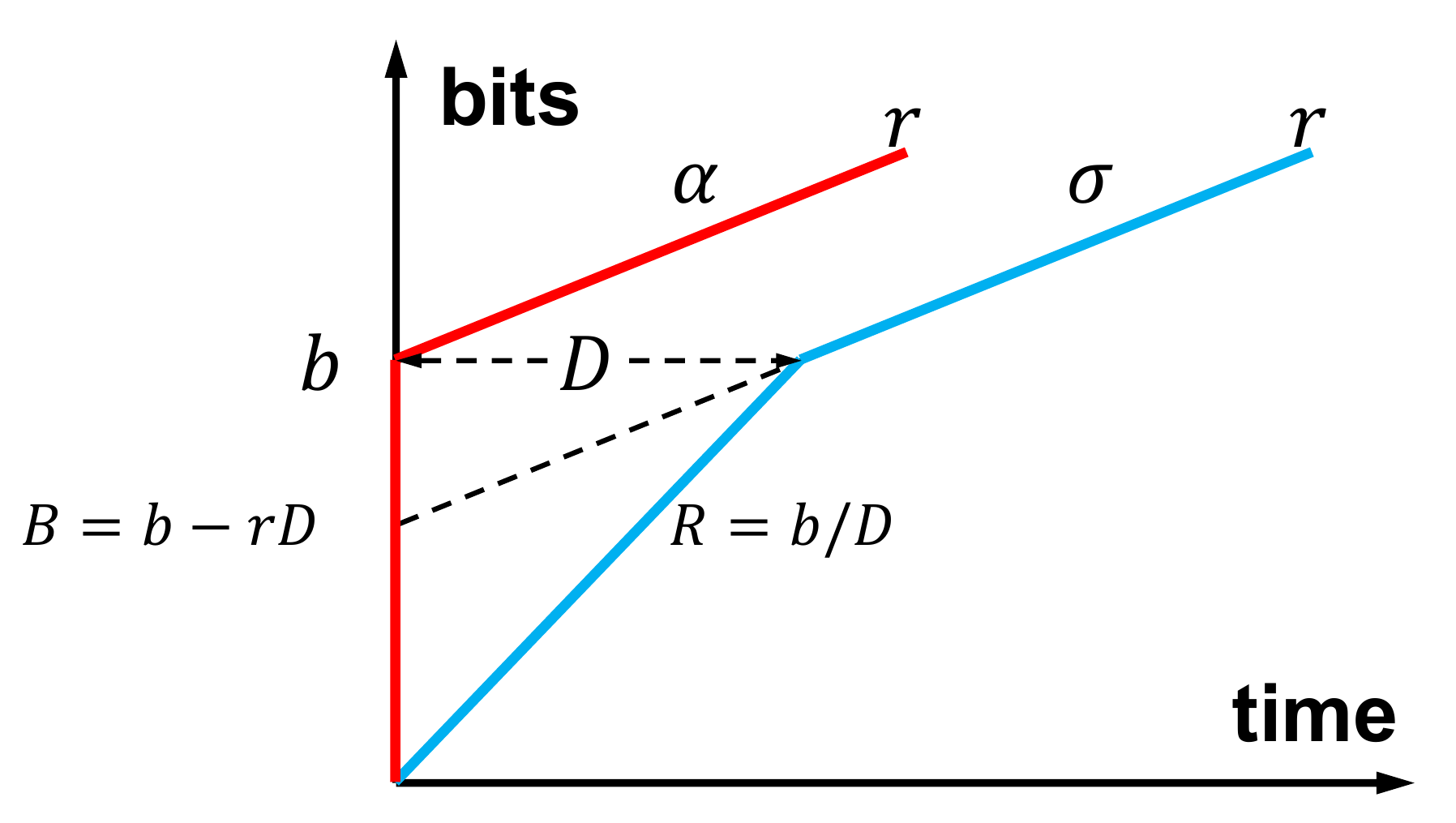}
    \caption{Two-Slope Reprofiling Curve (2SRC).}
    \label{fig:2src}
\end{figure}

In this paper, we consider the shaping strategy of~\cite{qiu2024benefits} based on a \emph{Two-Slope Reprofiling Curve} (2SRC), denoted by $\sigma$. The 2SRC profile introduces a peak-rate constraint to regulate the transmission of the burst of a flow with token-bucket arrival curve $\alpha=(r,b)$. This can be realized by concatenating\footnote{Concatenation maps to the min-plus convolution operator of~\cite[Section 3.1.6]{le2001network}.} two greedy shapers, each implemented as a two-parameter token bucket. Specifically, a 2SRC is realized by combining an $(r,B)$ token bucket with an $(R,0)$ token bucket, where $B=b-rb/R$ and $R \ge r$. The parameter $R$ represents the flow's peak rate, yielding the arrival curve $\sigma(t)=\min(Rt,\, B+rt)$ of \fig{fig:2src}.

As shown in~\cite[Section 1.4]{le2001network} and illustrated in \fig{fig:2src}, using $\sigma$ to reprofile a flow with arrival curve $\alpha$ introduces a \emph{shaping delay} $D$ of the form:
\begin{equation}
\label{eq:shap_delay}
D=\frac{b}{R}.
\end{equation}
Varying the peak rate $R$ yields a family of shaping profiles parameterized by their shaping delay $D$, where $D\in [0,\widehat{d}]$, and $\widehat{d} = \min(d,b/r)$ with $d$ the flow's delay bound.  Hence, $\widehat{d}$ is the maximum shaping delay a flow can afford\footnote{Our fluid-flow framework also assumes zero propagation delays.}. As per~\cite[Lemma~2]{qiu2024benefits}, given any shaping delay $D$, the 2SRC profile is the ``smallest'' (minimizes the flow's bandwidth requirements) among all concatenations of token buckets with that shaping delay. In the remainder of the paper, we use the shaping delay $D$ as the parameter that uniquely identifies a 2SRC profile.

\subsection{Shaping Enforcement}
\label{sec:background_shaping_enforcement}

As discussed earlier, traffic shaping plays an important role in the derivation of delay bounds. Achieving these bounds, however, requires \emph{in-network} shaping that reapplies a flow's traffic profile at every hop along its path.  This can be realized through \emph{Per-Flow} shaping, which reshapes individual flows at every hop. This is, however, complex and, therefore, at odds with the simplicity of static-priority and FIFO schedulers. 

\emph{Interleaved Shapers}~\cite{specht16,le2018theory} (ILS) offer an alternative. They enforce shaping per switch port rather than per flow. Specifically, an interleaved shaper processes packets from multiple flows in a single FIFO queue and only checks the packet at the \emph{head of the queue} against its shaping profile. This greatly reduces implementation overhead while preserving the same\footnote{As noted in~\cite{le2018theory}, in packetized settings, a small discrepancy may arise from heterogeneous packet sizes or processing delays.} worst-case delay guarantees as per-flow shaping\cite{le2018theory}. We, therefore, assume that in-network reprofiling is realized using per input port and per priority class interleaved shapers placed at each switch output port.


\section{Problem Formulation}
\label{sec:formulation}

\subsection{Problem Setting}
\label{sec:setting}

We consider a network with fixed routing on a topology consisting of $n$ links. The network carries $m$ token-bucket regulated flows $(r_i, b_i)$ with end-to-end delay bounds $d_i,\, 1 \le i \le m$. $\pmb{\mathcal{P}} = \{\mathcal{P}_1, \mathcal{P}_2, \ldots, \mathcal{P}_m\}$ denotes the set of flow paths, with $\mathcal{P}_i$ specifying the sequence of links traversed by flow~$i$. Given link $j, 1\leq j\leq n,$ $\mathcal{F}_j$ denotes the set of flows whose path includes link~$j$. Our objective is to satisfy the flows' delay bounds while minimizing the total network bandwidth, $\mathbf{C} = \sum_{j=1}^{n} C_j,$ where $C_j$ is the bandwidth of link~$j$.

To achieve this goal, we seek to proactively modify the traffic profiles of individual flows through traffic shaping. As mentioned in Section~\ref{sec:background_shaping}, we focus on 2SRC shaping profiles that for a token-bucket controlled $(r_i,b_i), 1 \le i \le m$, flow adds a peak-rate constraint $R_i$, and, as per \Eqref{eq:shap_delay}, a shaping delay $D_i=\frac{b_i}{R_i}$. The remaining delay budget, $d_i - D_i$, then becomes the flow's new allowable end-to-end \emph{network} delay.

Each network hop employs a static-priority scheduler with $k$ priority classes (for FIFO, $k=1$). At hop $j$, every flow $i \in \mathcal{F}_j$ is assigned a priority tag $p_{ij}$ indicating its priority class, where smaller values of $p_{ij}$ correspond to higher priorities. Flows within the same priority class $h$ are scheduled in FIFO order and share the same worst-case local scheduling deadline $T_{hj}$.

The priorities assigned to flows in $\mathcal{F}_j$ collectively define the \emph{priority assignment} $\Gamma_j$. Let $G_h(\Gamma_j)$ denote the subset of flows mapped to priority class $h$ under $\Gamma_j$. The assignment satisfies
\begin{align*}
    \bigcup_{1 \leq h \leq k} G_h(\Gamma_j) &= \mathcal{F}_j, \\
    G_h(\Gamma_j) \cap G_{h'}(\Gamma_j) &= \varnothing, \quad \forall\, 1 \le h < h' \le k.
\end{align*}
This ensures that each flow is assigned to exactly one priority class and that the sets $\{G_h(\Gamma_j)\}_{h=1}^k$ form a partition.

The inputs to our problem consist of the flows' parameter vectors $(\mathbf{r}, \mathbf{b}, \mathbf{d})$, representing the flows' rates, burst sizes, and delay bounds, and the flows' path set $\pmb{\mathcal{P}}$. The optimization variables are the vector of shaping delays $\mathbf{D}$, the global priority assignment $\mathbf{\Gamma}$, and the matrix $\mathbf{T}$ (per flow and link) of local scheduling deadlines. Our objective is to minimize the total network bandwidth $\mathbf{C}=\sum_{j=1}^{n} C_j$. The problem can then be formulated as a constrained optimization as follows:
\begin{align}
\label{eq:opt_sp}
\text{\bf{MIN}}_{SP}: \quad
& \min_{\mathbf{D}, \mathbf{\Gamma}, \mathbf{T}} \sum_{j=1}^{n} C_j,\\
& \text{s.t. } D_i + \sum_{j \in \mathcal{P}_i} T_{p_{ij}j} \le d_i,
\quad 1 \le i \le m.\nonumber
\end{align}

In the special FIFO case $k=1,$ the priority assignment becomes irrelevant and all flows traversing link $j$ share the same deadline $T_j$. The optimization then simplifies to
\begin{align}
\label{eq:opt_fifo}
\text{\bf{MIN}}_{FIFO}: \quad
& \min_{\mathbf{D}, \mathbf{T}} \sum_{j=1}^{n} C_j,\\
& \text{s.t. } D_i + \sum_{j \in \mathcal{P}_i} T_j \le d_i,
\quad 1 \le i \le m\,.\nonumber
\end{align}


\section{Solution}
\label{sec:solution}

With the optimization problem(s) defined, we now present the methods used to solve \textbf{MIN}.

\subsection{Link Bandwidth Provisioning}

Recall that in our formulation the required bandwidth $C_j$ of link $j$ depends on the local priority assignment $\Gamma_j$, the shaping delays $D_i$ of flows $i \in \mathcal{F}_j$, and the worst-case local scheduling deadlines $T_{hj}$ of each priority class $h$. We first show how to compute $C_j$ \emph{given} these variables.

To this end, we introduce the \emph{minimal service function} $S_{hj}$ for priority class $h$ at link $j$. We note that $S_{hj}$ is mainly used as an intermediate construct to facilitate the computation of $C_j$. The formal relationship between $C_j$ and $T_{hj}$ comes from Network Calculus via~\Eqref{eq:delay_bound}. Formally, $S_{hj}$ is defined as
\begin{equation}
\label{eq:minimal_service_function}
\hspace{-7pt} S_{hj}(t) =
\begin{cases}
0, & \hspace{-6pt} t \le T_{hj}, \\
\sum_{1 \le h' < h} H_{h'j}(t) + H_{hj}(t - T_{hj}), & \hspace{-6pt}  t > T_{hj},
\end{cases}
\end{equation}
where
\begin{equation*}
H_{hj}(t) = \sum_{i \in G_h(\Gamma_j)} \sigma_i(t),
\end{equation*}
denotes the aggregate arrival curve of flows assigned to class $h$, assuming that each flow $i$ is shaped according to its 2SRC arrival curve $\sigma_i$ prior to multiplexing by the scheduler.

Intuitively, the minimal service function specifies the minimum cumulative service that priority class $h$ must receive from the scheduler in order to meet its worst-case delay bound $T_{hj}$. The expression captures two competing sources\footnote{Under a fluid model, lower priority classes are transparent.} for bandwidth: (i) traffic from  higher-priority classes, \ie the aggregate traffic from classes $h' < h$, and (ii) traffic from flows within class $h$, which must be served within $T_{hj}$ after arrival. The minimal service function ensures that traffic from class $h$ arriving at time $t$ is guaranteed to depart by time $t + T_{hj}$.

Next, we provide an expression for the minimum bandwidth required to meet the deadlines of all priority classes at hop $j$.

\begin{proposition}
\label{prop:min_bw}
Consider a hop $j$ equipped with a static-priority scheduler with $k$ priority classes indexed in decreasing order of priority from $1$ to $k$ (priority~$1$ being the highest), serving the set of flows $\mathcal{F}_j$. Given a priority assignment $\Gamma_j$ and the corresponding minimal service function $S_{hj}$ for each priority class, the hop must provision a bandwidth of at least $C^*_j$ in order to satisfy the deadlines of all classes, where
\begin{equation}
\label{eq:min_bw}
C^*_j =
\max_{1 \le h \le k}
\left(
\sum_{i \in \mathcal{F}_j} r_i,\;
C^*_{hj}
\right),
\end{equation}
where
\begin{equation}
\label{eq:class_bw}
C^*_{hj} = \sup_{t > T_{hj}} \frac{S_{hj}(t)}{t},
\end{equation}
is the bandwidth needed to serve priority~$h$ traffic by $T_{hj}$. 
\end{proposition}
The proof is in Appendix~\ref{app:proof1}.

\begin{figure}[htbp]
    \centering
    \includegraphics[width=0.8\linewidth]{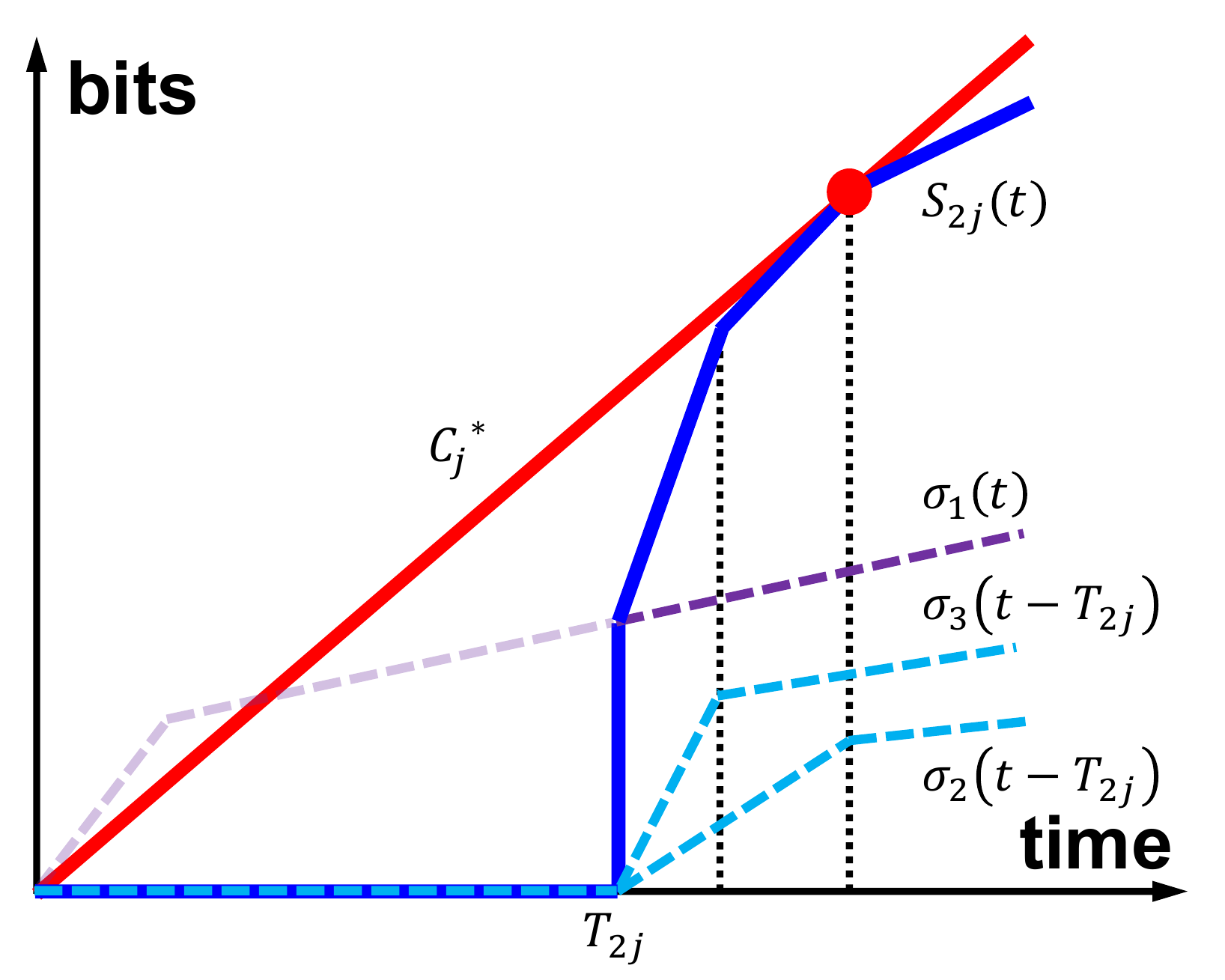}
    \caption{Minimal service function and required link bandwidth.}
    \label{fig:minimal_service_function}
\end{figure}

\fig{fig:minimal_service_function} illustrates the minimal service function of priority class $2$ and the corresponding bandwidth required to transmit its traffic within deadline $T_{2j}$ in an example with three flows and two priority classes. Flow~$1$ belongs to the highest priority class~$1$, and flows~$2$ and~$3$ to class~$2$. Consistent with~\Eqref{eq:minimal_service_function}, traffic from higher-priority classes (flow~$1$) arrives greedily according to its 2SRC starting at time $0$. Traffic from class~$2$ (flows 2 and 3) also arrives greedily according to their 2SRCs starting at time $0$, but their contributions to the minimal service function of priority class $2$ are shifted by its local deadline $T_{2j}$.

The intuition behind \fig{fig:minimal_service_function} is that we need to ensure that the service $S_{2j}(t)$ (blue line) that traffic from class~$2$ receives never ``falls behind'' by more than $T_{2j}$, where falling behind is captured by crossing the (red) line $C_j^*\cdot t$.  The initial shift of $T_{2j}$ in $S_{2j}(t)$ acknowledges that class~$2$ can tolerate a delay of $T_{2j}$, while the discontinuity it experiences at that point accounts for the amount of service that has been provided to class~$1$ by that time.  After $t=T_{2j}$, $S_{2j}(t)$ tracks the aggregate arrival curve of class~$2$ traffic $\sigma_2(t-T_{2j})+\sigma_3(t-T_{2j})$ with the added contribution of residual arrivals from class~$1$ ($r_1t$ in this case).

Repeating a similar reasoning across priority classes enables us to determine the minimum link bandwidth needed to ensure that the delay bounds of all priority classes are met.

We also note that determining the minimum link bandwidth for the highest priority class or under FIFO scheduling is considerably simpler. In such a setting, there is no higher-priority to account for, and the minimal service function depends solely on the aggregate traffic of the class itself.

\subsection{Bandwidth Minimization Algorithm}

With link bandwidths computable once given flows shaping delays (and, therefore, 2SRCs), local deadlines for priority classes on each link, and how flows map to priority classes, the remaining challenge to solve $\textbf{MIN}$ is to determine those quantities in a manner that minimizes the total network bandwidth. We next introduce the algorithms we developed for that purpose, beginning with the simpler FIFO case.

\subsubsection{The FIFO Case}

As in~\cite[Section IV.B]{qiu2024benefits}, solving $\textbf{MIN}_{FIFO}$ can be formulated as a collection of Non-Linear Programs (NLPs). This is achieved by expressing the worst-case link scheduling delay $T_j$ in terms of the flows' peak-rate shapers (determined by the shaping delays $D_i$) and the link bandwidth $C_j$ through~\Eqref{eq:min_bw}. Doing so requires a closed-form characterization of~\Eqref{eq:min_bw}, which can be obtained once the ordering (not the values) of the shaping delays $D_i$ is fixed.

Following the approach of~\cite{qiu2024benefits}, we exploit the fact that the aggregate arrival curve $H_j$ is piecewise-linear, concave, with segments of decreasing slopes, where segment boundaries are determined by the shaping delays $D_i$. We can enumerate all possible orderings of the $D_i$'s, and for each derive a closed-form expression of~\Eqref{eq:min_bw}.  The expressions can then be used to formulate a corresponding set of NLPs. As the derivation closely parallels that of~\cite{qiu2024benefits}, we relegate technical details to Appendix~\ref{app:fifo_nlp}.

Although the number of possible orderings grows combinatorially with the number of flows, the search space can be explored using standard randomized search heuristics. For each ordering encountered during the search, the corresponding NLP is solved and the resulting bandwidth requirement recorded. The best solution among the explored orderings is then retained. In principle, this procedure can recover the \emph{exact} optimal solution of $\textbf{MIN}_{FIFO}$ if all orderings are examined. In practice, however, computational complexity necessitates limiting the number of orderings explored.

This complexity motivates a simple heuristic introduced as Full Shaping (FS) in~\cite{qiu2024benefits}. Under FS, each flow $i$ is assigned the maximum feasible shaping delay,
\begin{equation}
\label{eqt:gs_shaping}
D_i = \widehat{d}_i = \min \left( d_i, \frac{b_i}{r_i} \right).
\end{equation}
The intuition for considering FS is that a FIFO scheduler treats all flows equally irrespective of their delay requirements. By allocating the maximum possible delay budget to shaping, FS pushes flows toward a common minimal in-network delay\footnote{Under a fluid model, FS commonly results in an in-network delay $T_j=0$.}, which reduces delay heterogeneity in addition to making flows smoother. Also of note is that the required bandwidth $C_j^*$ on FIFO link~$j$ under FS, is readily seen to be simply of the form
\begin{equation}
\label{eq:fs_fifo_bw}
C_j^*=\sum_{i\in{\cal F}_j}R_i\, ,
\end{equation}
${\cal F}_j$ is the set of link~$j$ flows and $R_i$ the shaping rate of flow~$i$.
\subsubsection{The Static Priority Case}
\label{sec:solution_sp}

As indicated by~\Eqref{eq:opt_sp}, solving $\textbf{MIN}_{SP}$ requires jointly determining the priority assignment $\mathbf{\Gamma}$ and allocating the delay budget $d_i$ of each flow $i$ between its shaping delay $D_i$ and per-hop scheduling delays $T_{hj}$. The inclusion of $\mathbf{\Gamma}$ significantly enlarges the combinatorial search space, rendering exact approaches based on solving Nonlinear Programs (NLPs) impractical.

Although an exact NLP-based solution is no longer feasible, \cite[Section V]{qiu2024benefits} suggested a progressive refinement heuristic that, in most cases, closely approximated the results of the NLP-based solution. Motivated by this observation, we adopt a similar heuristic, which we refer to as \emph{\Greedy}.

As in~\cite{qiu2024benefits}, \Greedy operates in two phases. For each flow~$i$, the \emph{exploration phase} generates candidate allocations of its delay budget $d_i$ across reprofiling delay and local link deadlines.   The \emph{adjustment phase} iteratively refines these allocations by updating local scheduling deadlines across links, and consequently the flow's reprofiling delay, towards reducing the total bandwidth. To facilitate this process, we allow independent initial \emph{per-flow local deadlines} $\widetilde{T}_{ij}$ (the delay permissible for flow $i$ at link $j$). The corresponding priority assignment $\Gamma_j$ and class-level deadlines $T_{hj}$ are then finalized in the adjustment phase.

The general structure of \Greedy and in particular its exploration phase mimics~\cite{qiu2024benefits}. As~\cite{qiu2024benefits}, exploration is based on a global reprofiling ratio $\gamma \in [0,1]$, with the shaping delay of flow~$i, 1 \le i \le m$ set to $D_i = \gamma \widehat{d}_i$, for all~$i$, where $\widehat{d}_i$ is defined in \Eqref{eqt:gs_shaping}. The remaining delay budget, $d_i - D_i$, is then evenly distributed across hops on the flow's path $\mathcal{P}_i$ to form an initial allocation that is subsequently refined in the adjustment phase. Similarly, \Greedy's exploration proceeds by searching over $\gamma \in [0,1]$ (with progressive refinement) and invokes the adjustment phase (see next) for each value of $\gamma$.  The process end with the $\gamma$ value that yields the minimum total bandwidth after adjustment. We refer interested readers to~\cite[Section V.B]{qiu2024benefits} for details.

\begin{algorithm}
\caption{Adjustment}
\begin{algorithmic}[1]
\label{algo:adjustment}
\renewcommand{\algorithmicrequire}{\textbf{Input:}}
\renewcommand{\algorithmicensure}{\textbf{Output:}}
\REQUIRE flow profiles $\mathbf{r} = (r_1, r_2, \ldots, r_m)$,\\
$\mathbf{b} = (b_1, b_2, \ldots, b_m)$, $\mathbf{d} = (d_1, d_2, \ldots, d_m)$\\
path matrix $\pmb{\mathcal{P}} = (\mathcal{P}_1, \mathcal{P}_2, \ldots, \mathcal{P}_m)$\\
initial reprofiling delays $\mathbf{D} = (D_1, D_2, \ldots, D_m)$\\
initial local deadlines $\mathbf{\widetilde{T}} = \{\widetilde{T}_{ij}: \forall 1 \leq i \leq m, j \in \mathcal{P}_i\}$\\
improvement threshold $\epsilon$
\ENSURE total bandwidth $\mathbf{C}^*$ after adjustment
\STATE initialize $\mathbf{C} = \infty, \mathbf{C}' = \infty, \mathbf{C}^* = \infty$
\STATE sort links in decreasing order of $|\bigcup_{i \in \mathcal{F}_j}\mathcal{P}_i|$
\WHILE {$(\mathbf{C}' - \mathbf{C}) / \mathbf{C}' > \epsilon$}
\FOR {$j = 1$ to $n$}
\STATE perform 1-d $k$-means clustering on $\widetilde{T}_{ij}, \forall j \in \mathcal{F}_j$ to determine the local priority assignment $\Gamma_j$
\STATE initialize $C^*_j = \sum_{i \in \mathcal{F}_j} r_i$
\FOR {$h = 1$ to $k$}
\STATE $T_{hj} = \min_{i \in G_h(\Gamma_j)}(\widetilde{T}_{ij})$
\STATE $D_i = \min(\widetilde{T}_{ij} + D_i - T_{hj}, b_i/r_i), \forall i \in G_h(\Gamma_j)$
\STATE compute $C^*_{hj}$ according to~\Eqref{eq:class_bw}
\STATE $C^*_j = \max(C^*_j, C^*_{hj})$
\STATE reduce $T_{hj}$ to $T^*_{hj}$
\STATE $D_i = \min(\widetilde{T}_{ij} + D_i - T^*_{hj}, b_i/r_i), \forall i \in G_h(\Gamma_j)$
\STATE $\widetilde{T}_{ij} = T^*_{hj}, \forall i \in G_h(\Gamma_j)$
\ENDFOR
\ENDFOR
\STATE update $C^*_j, \forall 1 \leq j \leq n$ according to~\Eqref{eq:min_bw}
\STATE $\widetilde{T}_{ij} = \widetilde{T}_{ij} + (d_i - D_i - \sum_{j\in\mathcal{P}_i}\widetilde{T}_{ij}) / \left|\mathcal{P}_i\right|$
\STATE $\mathbf{C}' = \mathbf{C}, \mathbf{C} = \sum_{1 \leq j \leq n}C^*_j$
\STATE $\mathbf{C}^*=\min(\mathbf{C}, \mathbf{C}^*)$
\ENDWHILE
\RETURN $\mathbf{C}^*$ 
\end{algorithmic}
\end{algorithm}

The adjustment phase involves key aspects specific to static-priority scheduling and is detailed next. Starting from the initial deadline allocations produced by the exploration phase, the algorithm iteratively processes each link $j$ in two steps as described in Algorithm~\ref{algo:adjustment}: (1) determining the local priority assignment $\Gamma_j$, and (2) \emph{adjusting} local deadlines by increasing shaping delays. We detail these two steps next. As the description is notation intensive, readers may wish to refer to Appendix~\ref{app:glossary} for a glossary of notation.

\begin{figure}[htbp!]
\centering
\begin{subfigure}{0.49\linewidth}
  \centering
  \includegraphics[width=\linewidth]{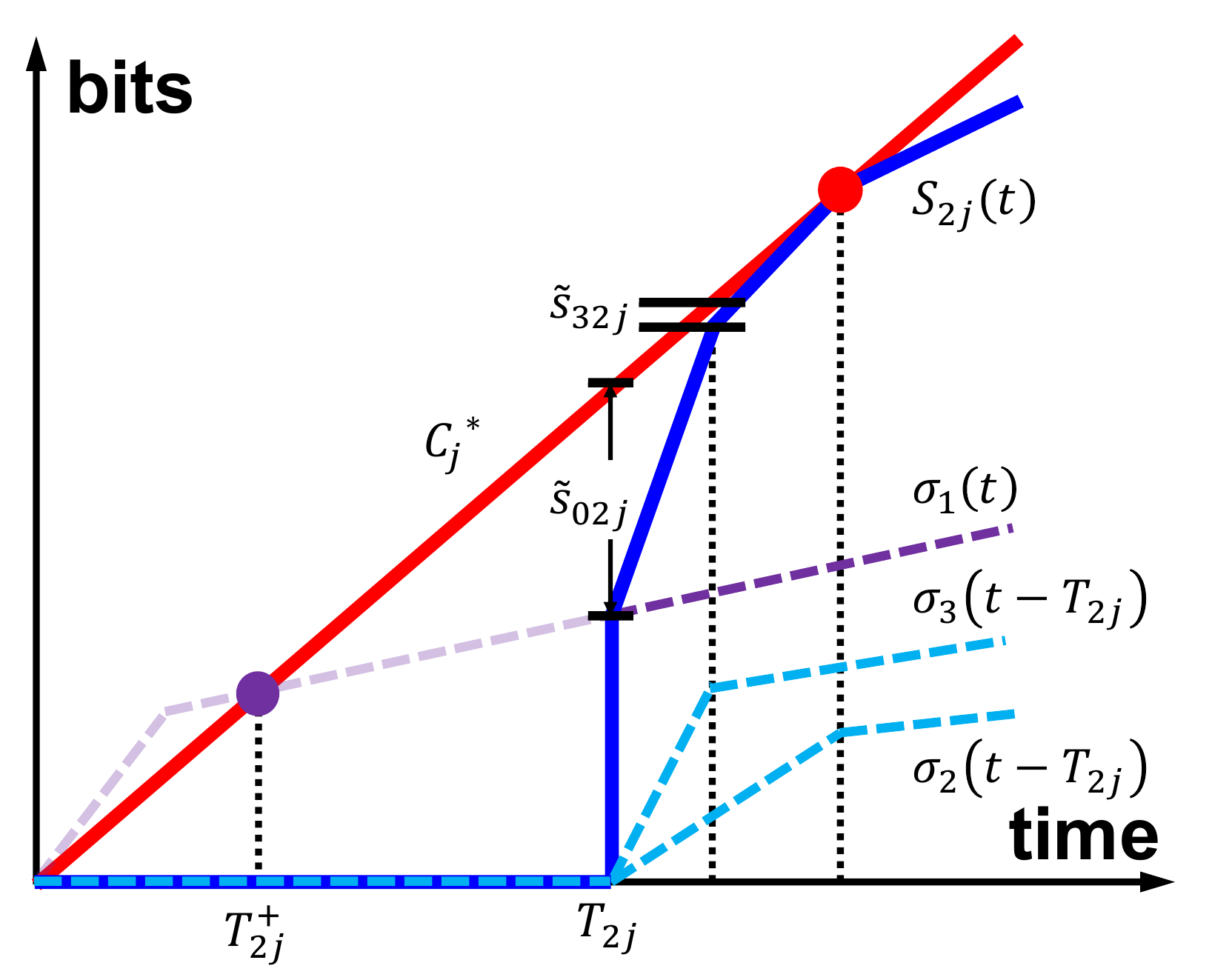}
  \caption{flow slack at inflection points}
  \label{fig:slack}
\end{subfigure}
\begin{subfigure}{0.49\linewidth}
  \centering
  \includegraphics[width=\linewidth]{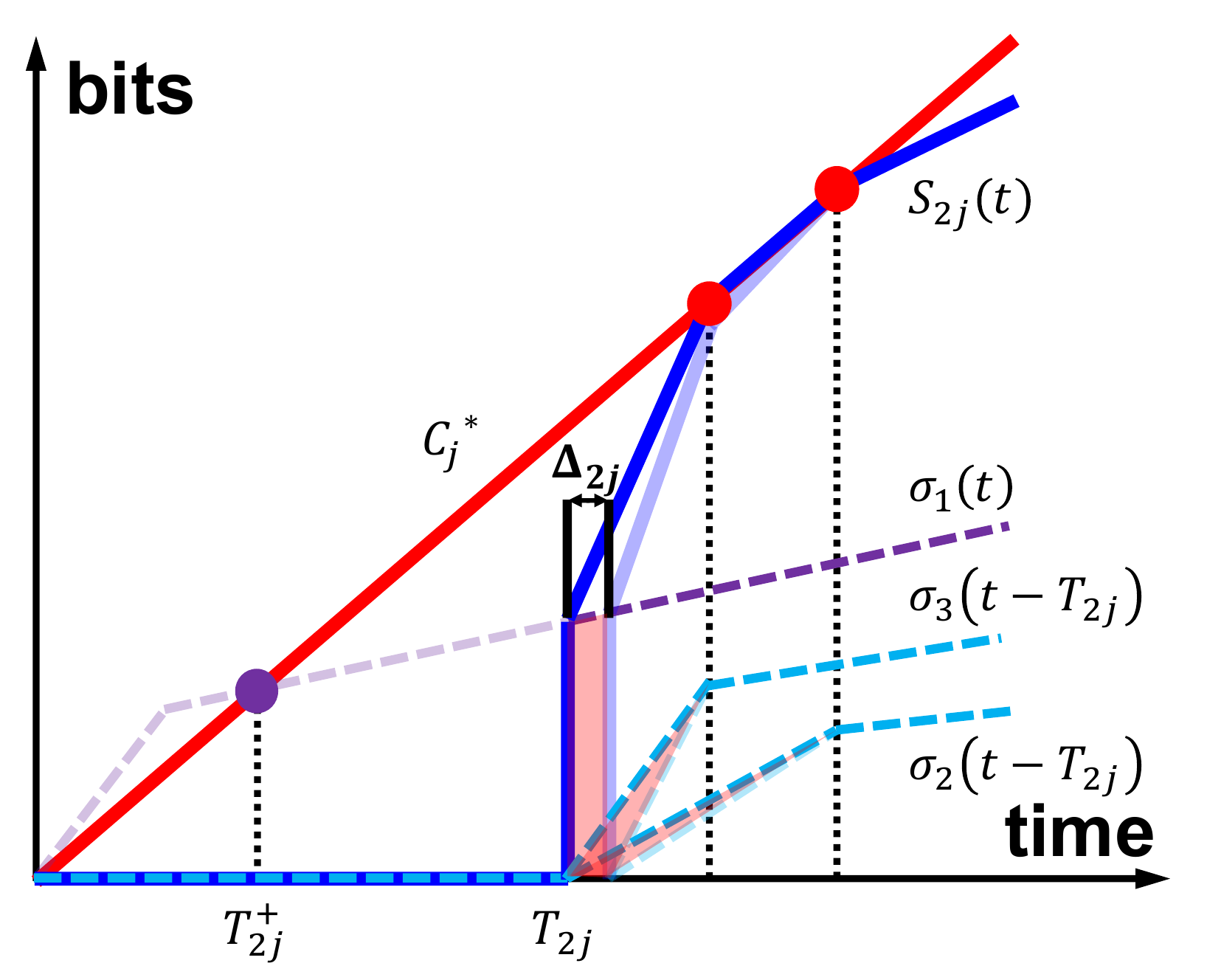}
  \caption{decrease $T_{2j}$ by $\Delta_{2j}$}
  \label{fig:adjustment}
\end{subfigure}
\caption{\Greedy adjustment.}
\label{fig:adjustment_phase}
\end{figure}

\paragraph{Determining the local priority assignment}

We start by establishing a key structural property satisfied by an optimal solution for $\Gamma_j$, the assignment of flows to priority classes.

\begin{proposition}
\label{prop:opt_assign}
Consider a hop $j$ serving a set of flows $\mathcal{F}_j$ and employing a static-priority scheduler with $k$ priority classes indexed in decreasing order of priority from $1$ to $k$ (with priority $1$ being the highest). Suppose the flows are 2SRC-shaped and indexed in non-decreasing order of their local deadlines $\widetilde{T}_{ij}$, \ie $\widetilde{T}_{ij} \le \widetilde{T}_{i'j}$ for $i \le i'$. Then there exists a priority assignment $\Gamma^*_j$ that minimizes the required link bandwidth $C_j$ while satisfying all local deadlines, such that a flow $i$ is assigned a strictly higher priority than flow $i'$ only if $\widetilde{T}_{ij} < \widetilde{T}_{i'j}$.
\end{proposition}

In other words, a flow with a larger deadline should never be assigned to a strictly higher priority class than a flow with a smaller deadline.  The result parallels~\cite[Proposition 4]{song2024benefits}, extending it from token-bucket arrival curves to 2SRCs. The proof is in Appendix~\ref{app:proof2}, with the next lemma a direct consequence of the proposition.
\begin{lemma}
\label{lemma:assign_ddl_boundary}
There exists an optimal priority assignment $\Gamma^*_j$ and a sequence of $k+1$ non-decreasing local boundaries $\{\check{T}_{hj}\}_{h=1}^{k+1}$, with $\check{T}_{hj} \le \check{T}_{h'j}$ for all $1 \le h < h' \le k+1$
\begin{equation}
\label{eq:cluster}
\text{such that}\quad
    \check{T}_{hj} \le \widetilde{T}_{ij} \le \check{T}_{h+1,j}, \quad \forall\, i \in G_h(\Gamma^*_j).
\end{equation}
\end{lemma}

The lemma implies that determining an optimal priority assignment $\Gamma^*_j$ is equivalent to selecting the $k+1$ boundary values $\{\check{T}_{hj}\}$ (the first and last boundaries can be readily dropped). In step~$1$ of the adjustment phase, we perform such a selection using a one-dimensional $k$-means clustering\footnote{We validate this choice in Appendix~\ref{app:k_means_performance}.} over the initial local deadlines at hop $j$, \ie $\{\widetilde{T}_{ij} : i \in \mathcal{F}_j\}$ (line~5 of Algorithm~\ref{algo:adjustment}). The resulting cluster boundaries are used as the $\check{T}_{hj}$'s to define an assignment $\Gamma^*_j$, where the set of flows assigned to priority class~$h$ are selected based on \Eqref{eq:cluster}. 

Intuitively, the approach seeks to group flows with similar local deadlines into the same priority class.  As shown in line~8 of Algorithm~\ref{algo:adjustment}, the deadline $T_{hj}$ assigned to priority class~$h$ on link~$j$ is then initially set to the smallest local deadline of the flows assigned to that class\footnote{Note that by construction, \ie \Eqref{eq:cluster}, we must have $T_{hj}\geq \check{T}_{hj}$.}.  The next step proceeds to adjust (reduce) $T_{hj}$ \emph{without increasing} the required link bandwidth.  The motivation is that reducing $T_{hj}$ frees-up delay that can then be allocated to shaping, \ie to make the flows smoother, which can benefit other links.

\paragraph{Adjusting the local deadline allocation}

The first step in determining if and by how much it is possible to reduce $T_{hj}$ involves computing the bandwidth\footnote{Recall the discussion of \fig{fig:minimal_service_function} regarding this computation.} $C^*_{hj}$ required by class $h$ using~\Eqref{eq:class_bw} (line~10), and updating the overall link bandwidth $C^*_j$ accordingly (line~11). As shown in~\cite{qiu2024benefits} and formalized in Proposition~1 of~\cite{song2024benefits}, $C^*_{hj}$ is attained at one of the inflection points $\{\widetilde{T}'_{ihj}\}$, where the slope of the minimal service function $S_{hj}$ decreases. Recalling that $G_h(\Gamma_j)$ denotes the subset of flows mapped to priority class $h$ under priority assignment $\Gamma_j$ on link~$j$, these inflection points arise from three sources:
\begin{enumerate}[label=(\alph*),wide=0pt]
\item The first inflection point $\widetilde{T}'_{0hj} = T_{hj}$ that coincides with the deadline $T_{hj}$ for class~$h$, and that accounts for the aggregate service that higher-priority flows must have received by time $T_{hj}$ when priority class~$h$ starts receiving service. 
\item A rate change for a flow $i$ in class $h$ ($i \in G_h(\Gamma_j)$), yielding $\widetilde{T}'_{ihj} = T_{hj} + D_i$.
\item A rate change from a flow $i$ in a higher-priority class $h' < h$ with shaping delay $D_i$ exceeding $T_{hj}$ ($i \in G_{h'}(\Gamma_j),\, D_i > T_{hj}$), yielding $\widetilde{T}'_{ihj} = D_i$.
\end{enumerate}
Following~\cite{qiu2024benefits}, the \emph{slack} at an inflection point $\widetilde{T}'_{ihj}$ is the excess service provided under the provisioned bandwidth $C^*_j$:
\begin{equation}
\label{eq:slack}
\widetilde{s}_{ihj} = C^*_j \widetilde{T}'_{ihj} - S_{hj}(\widetilde{T}'_{ihj}).
\end{equation}
The slack quantifies how much the current bandwidth exceeds the minimum service function at that point. Note that the use of $C^*_j$ instead of $C^*_{hj}$ allows accounting for the bandwidth provisioned across all priority classes, and not just class $h$.

Leveraging the possible presence of slack to better utilize $C_j^*$, the bandwidth provisioned at link $j$, is the focus of the second step of the adjustment phase illustrated in~\fig{fig:adjustment_phase}. Specifically, the second step of the adjustment phase decreases the local scheduling deadline $T_{hj}$ without increasing $C_j^*$. 
The resulting increase in the available delay budget is then redistributed for use in shaping, \ie by increasing the shaping delays of all flows in $G_h(\Gamma_j)$ without changing their inflection points $\widetilde{T}'_{ihj}$ This in turn is used to produce ``smoother'' 2SRCs, thereby reducing bandwidth requirements on other links. 

The adjustment process is illustrated in~\fig{fig:adjustment} for priority class~$2$ that includes flows~$2$ and~$3$, with \fig{fig:slack} showing non-zero slacks at the first inflection point $(s_{2j})$ and at the second inflection point ($\widetilde{s}_{32j}$ contributed by flow~$3$ from priority class $2$). The presence of slack means that the minimal service curve $S_{hj}$ can be increased without increasing $C_j^*$.  

The increases in $S_{hj}$ come from seeking to decrease $T_{hj}$ and come from two effects: (i) the 2SRC shaping profiles of flows in $G_h(\Gamma_j)$ increase with their shaping delays as transmissions start earlier, and (ii) more higher-priority traffic is accounted for within the service window. Since $S_{hj}$ increases monotonically as $T_{hj}$ decreases, this process continues until $T_{hj}$ reaches a critical value $T^*_{hj}$ beyond which the provisioned bandwidth $C^*_j$ is no longer sufficient to support $S_{hj}$ (line~12).

At this point, one of the following conditions must occur:
\begin{enumerate}[label=(\alph*),wide=0pt]
\item The slack $\widetilde{s}_{ihj}$ at the inflection point $\widetilde{T}'_{ihj}$ of some flow $i$ is depleted\footnote{This corresponds to flow~$3$ in \fig{fig:adjustment}.}.
\item The slack $\widetilde{s}_{0hj}$ at $\widetilde{T}'_{0hj}$ (\ie at $T_{hj}$) is depleted.
\end{enumerate}
In case~(a), the limiting condition can be identified by examining the slacks at the stationary inflection points $\widetilde{T}'_{ihj}$ (marked by $\textcolor{red}{\bullet}$). In case~(b), $T_{hj}$ reaches the intersection point between the service curve $C^*_j t$ and the aggregate higher-priority traffic, denoted by $T^+_{hj}$ (marked by $\textcolor{purple}{\bullet}$). 

This adjustment smooths the 2SRC arrival curves of flows in class~$h$. This can reduce both the bandwidth requirement $C^*_{h'j}$ of lower-priority classes $h' > h$ at link $j$, and the required bandwidth at other links traversed by the class~$h$ flows.

Since the minimal service function $S_{hj}$ depends on the 2SRC shaping profiles of higher-priority flows, the adjustment is performed in decreasing order of priority (starting from class~$1$, line~7). This ordering ensures that, when processing class $h$, the impact of higher-priority traffic has already been accounted for. However, it also implies that the link bandwidth $C_j$ available to class $h$ reflects only the updates from higher-priority classes processed thus far (line~11).

Consistent with~\cite{qiu2024benefits}, links are processed in decreasing order of $|\bigcup_{i \in \mathcal{F}_j} \mathcal{P}_i|$ (line~2), prioritizing those whose flows traverse the largest number of links and thus have the greatest impact on the network-wide bandwidth. After all links have been traversed, flows' shaping profiles may have changed (made smoother). Some flows may, therefore, now be fully shaped to their token rates $r_i$, possibly leaving a portion of their end-to-end delay budgets unused at some hops (cf. line~9 and~13). This creates opportunities for additional bandwidth reduction.

To exploit this, the unused delay budget is redistributed by evenly splitting it across the hops of each flow (line~18), and the adjustment phase is repeated. Since this redistribution modifies the local deadlines $\widetilde{T}_{ij}$, the priority assignments must be recomputed (line~5), and the total bandwidth $\mathbf{C}$ is not guaranteed to decrease monotonically across iterations. We therefore track the minimum bandwidth $\mathbf{C}^*$ observed so far (line~20), and terminate the algorithm when the improvement between successive iterations falls below a threshold $\epsilon$ (line~3), which is set to $0.1\%$ in all experiments.


\section{Evaluation}
\label{sec:evaluation}

\subsection{Evaluation Setup}

\begin{figure}[htbp]
    \centering
    \includegraphics[width=0.9\linewidth]{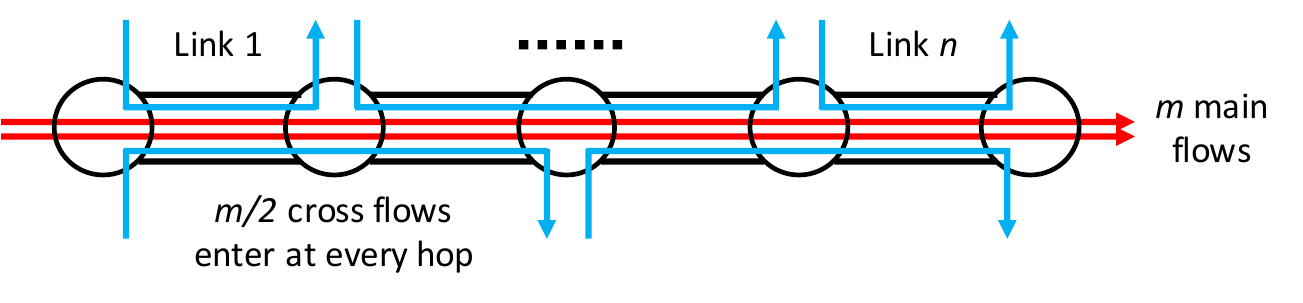}
    \caption{Parking lot topology.}
    \label{fig:parking_lot}
\end{figure}

Following the methodology of~\cite{qiu2024benefits}, we evaluate the performance of our proposed algorithms for solving \textbf{MIN} across three network topologies. First, \emph{Orion CEV}~\cite{paulitsch2018industrial} represents an in-vehicle network typical of automotive applications. Second, \emph{US-Topo} (Fig. 18 of~\cite{qiu2024benefits}) models a wide-area network interconnecting geographically distributed sites. Third, the \emph{parking lot} topology (shown in \fig{fig:parking_lot}) is a feed-forward synthetic network consisting of both main-path and cross traffic.

The Orion CEV topology captures a canonical Time-Sensitive Networking (TSN) scenario, while US-Topo represents Deterministic Networking use cases in cloud infrastructures.  These two topologies capture expected deployment settings where the work may be applicable. In contrast, the parking lot topology provides a controlled environment for systematic exploration by varying key structural parameters such as the number of flows and path lengths.

\subsubsection{Traffic Models}
In Orion CEV, we adopt flow profiles corresponding to standard TSN traffic classes~\cite{thangamuthu2015analysis}: Control-Data Traffic (CDT), class A, and class B, with delay bounds of $0.1$, $2$, and $50$\,ms, respectively. US-Topo uses inter-datacenter traffic characteristics from~\cite{roy2015inside} to model three application classes: Web, Cache read/replacement, and Hadoop, with delay bounds of $10$, $50$, and $200$\,ms, respectively.  The parking lot topology relies on synthetic traffic to facilitate exploring a broad parameter space. Flow rates and burst sizes are independently sampled from uniform distributions over $[1, 100]$\,Mb/s and $[1, 100]$\,Mb, respectively, with flows randomly assigned to one of four delay classes: $10$, $25$, $50$, $100$\,ms.

\subsubsection{Baselines}
We compare our algorithm against two baseline strategies: \emph{No Shaping (NS)} and \emph{Full Shaping (FS)}. Both assume that interleaved shaping is enforced at every hop, and differ only in the shaping profiles used.

Under NS, each flow retains its original token-bucket profile $(r_i, b_i)$, and, therefore, experiences zero shaping delay. In contrast, FS maximizes shaping by setting $D_i$ to the maximum feasible value $\widehat{d}_i$, which maximizes shaping delay and minimizes the remaining in-network delay budget.  

Realizing each baseline differs across schedulers.

Static-priority differentiates among flows based on their local deadline. Under both NS and FS, any remaining (after shaping) delay budget is evenly distributed across the hops a flow traverses\footnote{Even under FS, flows with large delay bounds can only be shaped down to their token rates $r_i$, leaving a residual in-network delay budget.}. Priority assignments are then determined via $k$-means clustering at each hop with flows with tighter local deadlines assigned to higher priority classes.

FIFO does not differentiate between flows that are all assigned to the same queue. Under FS, as all flows are fully shaped, there is little to no remaining (network) delay heterogeneity across flows, with most sharing a deadline of 0  (recall our fluid model assumption). As per \Eqref{eq:fs_fifo_bw}, the required link bandwidth is set to the sum of the flows' shaping rates.  Under NS, we instead apply an NLP-based approach to optimally distribute the flows' full delay budget across hops towards minimizing the required bandwidth\footnote{The NLPs are solely for deadline allocation absent any shaping.}.

\subsubsection{Evaluation Scope} It proceeds along three dimensions:
\begin{enumerate}[wide, labelwidth=0pt, labelindent=0pt, label=(\alph*)]

\item \emph{Bandwidth Minimization under FIFO.} 
We begin with FIFO networks, and first validate the statement that, because of FIFO's lack of flow differentiation, FS closely approximates the exact solution of $\textbf{MIN}_{FIFO}$. We then quantify the bandwidth reduction enabled by shaping relative to the NS baseline.

\item \emph{Bandwidth Minimization under Static Priority.} 
Next, we investigate static-priority networks with shaping realized through \Greedy. We first confirm the effectiveness of $k$-means clustering for priority assignment, before evaluating the bandwidth reduction achieved by \Greedy over NS and FS.  Finally, we explore when \Greedy outperforms FS and analyze the structural characteristics of its shaping decisions across priority classes.

\item \emph{Scheduler Comparison.} 
Finally, leveraging results from~\cite{qiu2024benefits}, we compare FIFO, static priority, and SCED schedulers. The comparison offers a quantitative assessment of the benefits of more sophisticated schedulers, and the extent to which shaping can mitigate them.  It also provides insight into how these benefits are affected by network topology and, for static priority, the number of priority classes.

\end{enumerate}

\subsection{Bandwidth Minimization under FIFO}
\label{sec:evaluation_fifo}

We begin by evaluating the effectiveness of shaping for bandwidth minimization in FIFO networks. As mentioned earlier, we anticipate that FIFO's inability to differentiate between flows will result in Full Shaping (FS) being close to the optimal solution of $\textbf{MIN}_{FIFO}$. To validate this observation, we compare the bandwidth achieved by FS with that obtained with an exact NLP-based solution.

The combinatorial nature of the NLP-based solution results in a high computational cost.  We, therefore, restrict our comparison to instances with $50$ flows on the Orion CEV and US-Topo networks. For each, we randomly sample flow profiles and source–destination pairs, and repeat the experiment $1000$ times for statistical significance. Given the expected ``optimality'' of the NLP-based approach, we use it as a baseline and report the bandwidth of FS relative to it.
\begin{table}[htb]
\renewcommand{\arraystretch}{1.3}
\begin{center}
\caption{Network bandwidth requirement of FS relative to the NLP-based solution}
\label{tab:bw_nlp_fs}
\begin{tabular}{|c|c|c|}
\hline
& \textbf{Average Bandwidth} & \textbf{95\% Confidence Interval}\\
\hline
Orion CEV & $99.97\%$ & $[99.93\%, 100.02\%]$\\
\hline
US-Topo & $100.03\%$ & $[100.02\%, 100.04\%]$\\
\hline
\end{tabular}
\end{center}
\end{table}

The results are in Table~\ref{tab:bw_nlp_fs}, It reports average bandwidth and $95\%$ confidence intervals. FS closely matches the NLP-based solution for both topologies, and even slightly outperforms it in Orion CEV\footnote{This is due to the use of randomized heuristics for exploring flow orderings and the possibility of the NLP solver converging to local optima.}. This is in contrast with observations from~\cite{qiu2024benefits} for SCED, where optimal reprofiling often differed significantly from FS.  This is because SCED can better leverage the residual scheduling flexibility these solutions preserve.  In contrast, FIFO has no such ability.  Allocating as much as possible of the delay budget to making flows more homogeneous (and smoother), as FS does, is then advantageous.

Since FS provides a near-optimal solution for $\textbf{MIN}_{FIFO}$, we use it in the remainder of FIFO's evaluation.

\begin{figure}[!h]
\centering
\begin{subfigure}{0.45\linewidth}
  \centering
  \includegraphics[width=\linewidth]{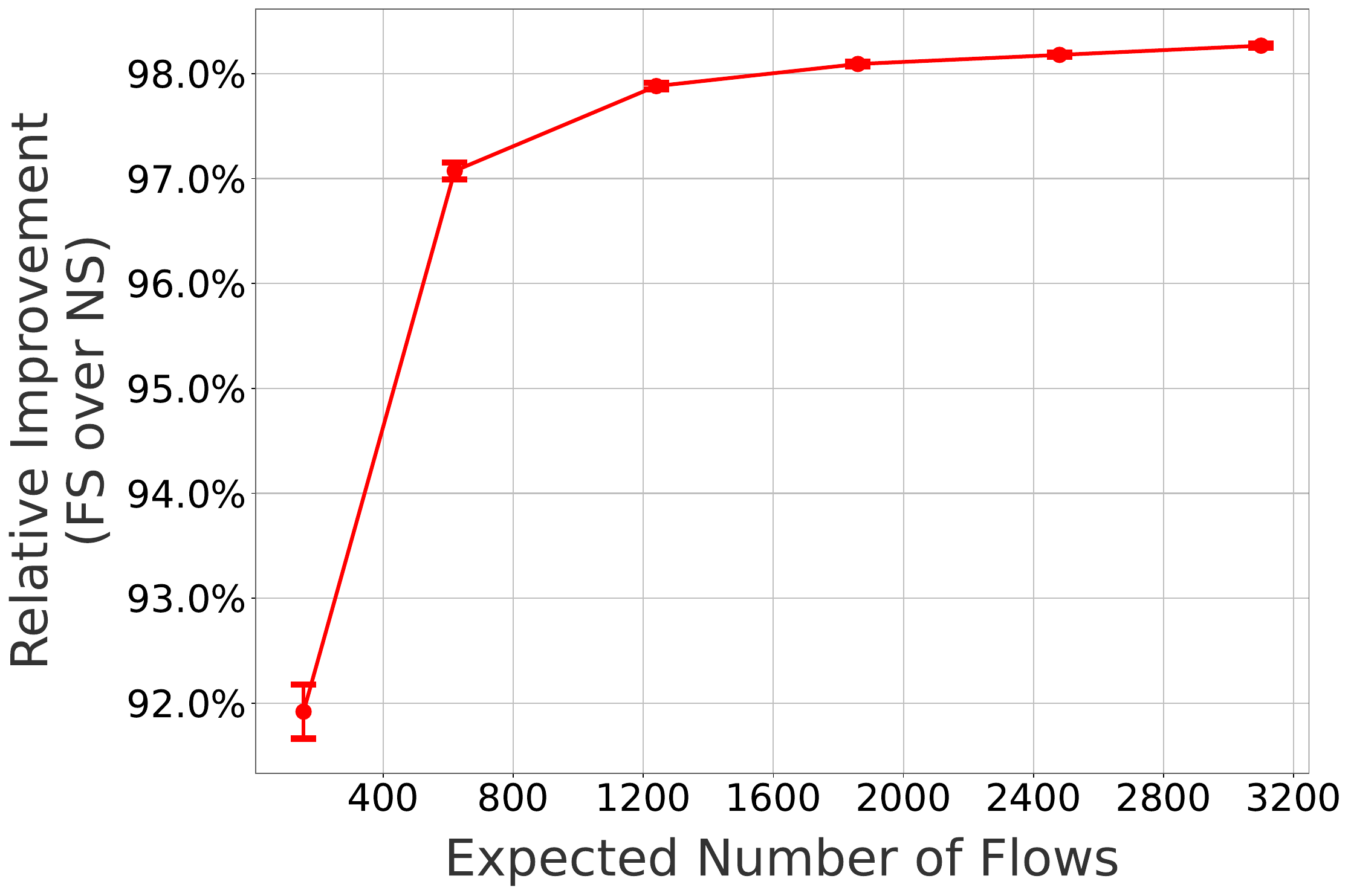}
  \caption{Orion CEV}
  \label{fig:cev_fifo_nr}
\end{subfigure}
\begin{subfigure}{0.45\linewidth}
  \centering
  \includegraphics[width=\linewidth]{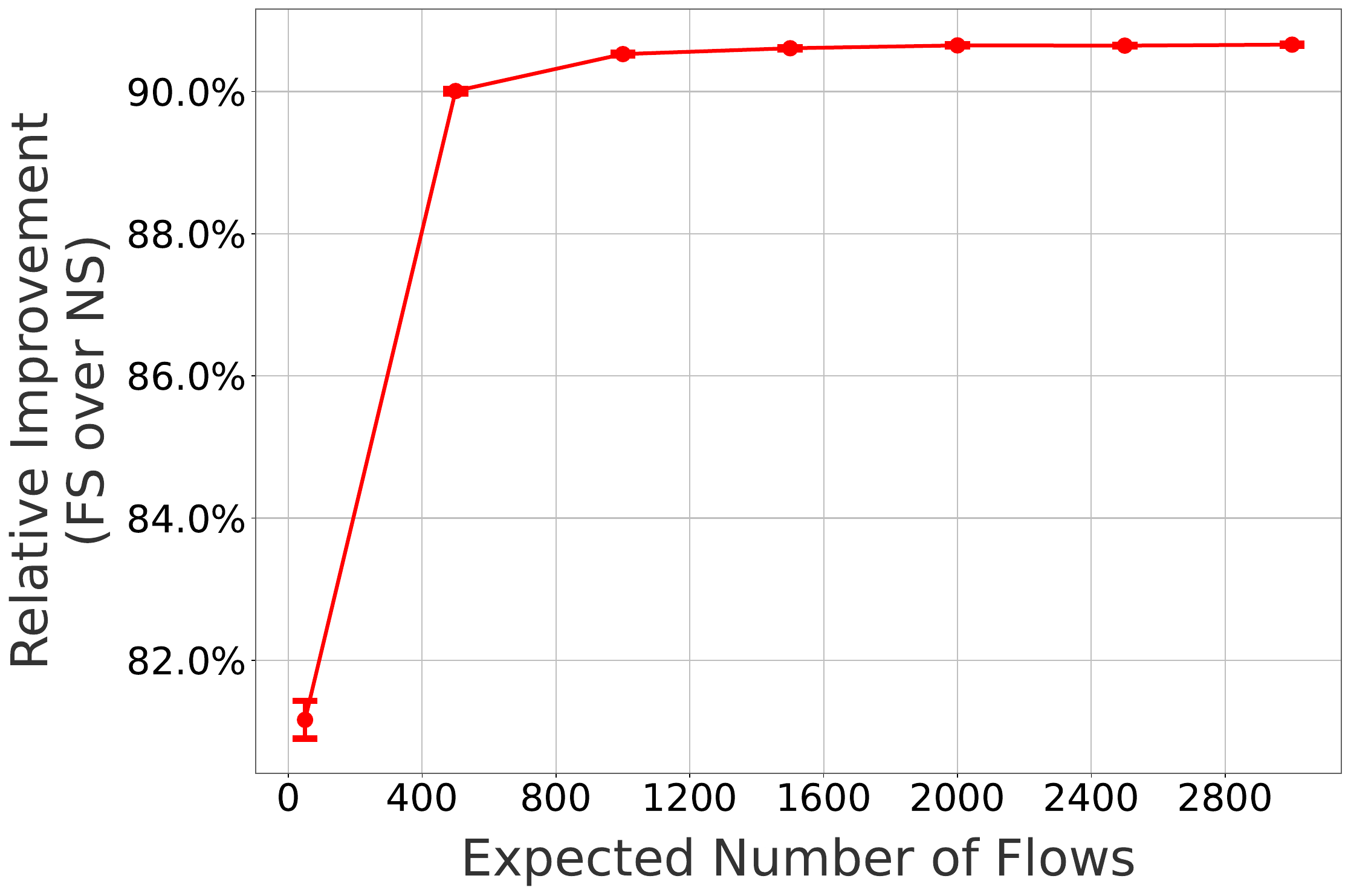}
  \caption{US-Topo}
  \label{fig:ustopo_fifo_nr}
\end{subfigure}
\caption{Bandwidth improvement of FS over NS under FIFO.}
\label{fig:fifo_nr}
\end{figure}

Next we compare FS against No Shaping (NS). \fig{fig:fifo_nr} reports the relative bandwidth reduction of FS over NS for both Orion CEV and US-Topo as a function of the number of flows.  In both networks, the benefits of FS are substantial (over $90\%$ in both) and eventually stabilize as the number of flows increase. This is because more flows means more homogeneous traffic mixes, which diminishes the relative impact of individual flow shaping decisions.  FS' improvements stem from the fact that it spreads bursts in time, reducing the amount of traffic simultaneously competing for bandwidth. In contrast, NS allows large bursts that may combine at any hop, therefore, requiring considerably more bandwidth on every link in spite of larger local deadlines.

\begin{figure}[!h]
\centering
\includegraphics[width=0.6\linewidth]{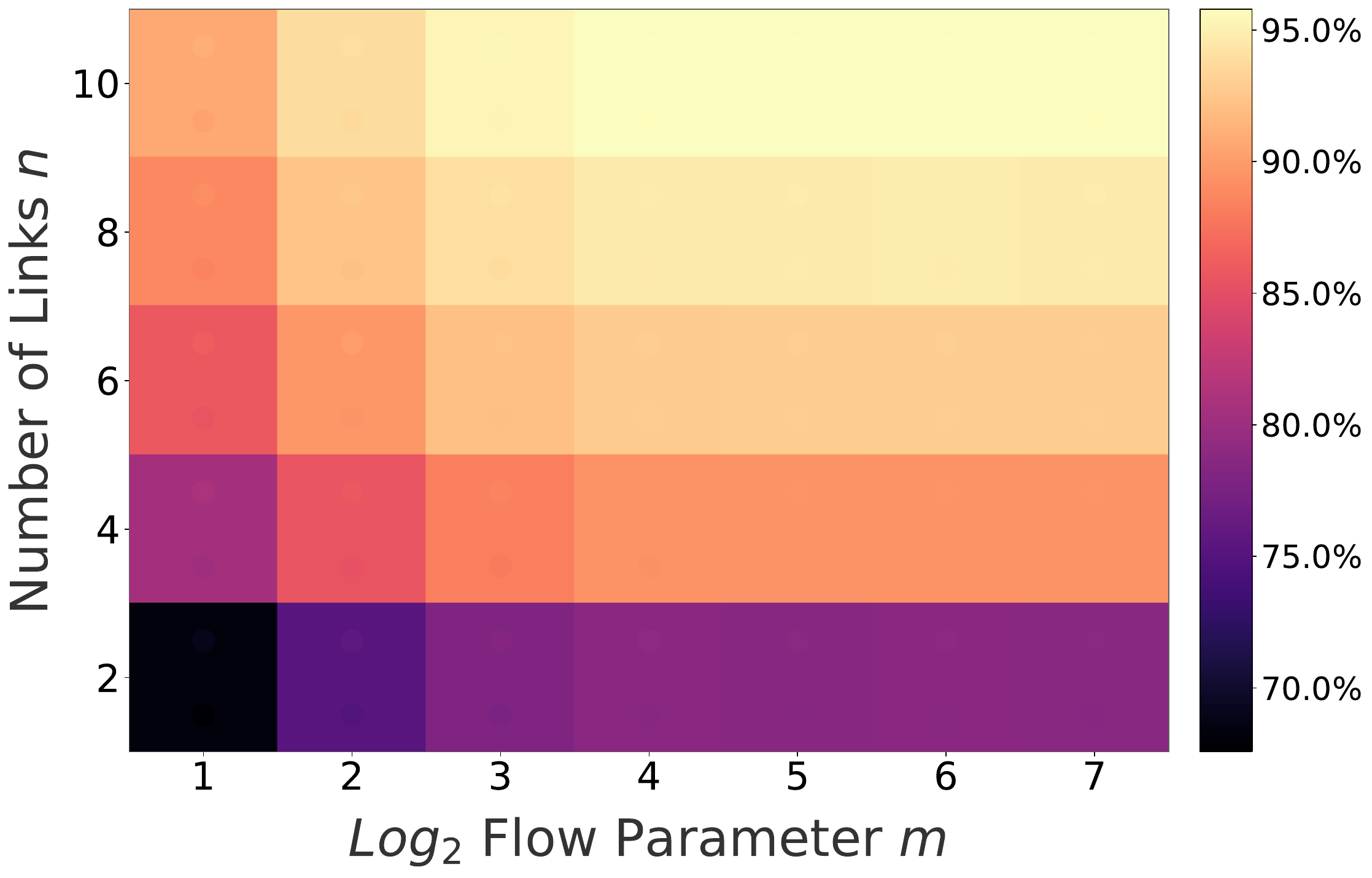}
\caption{Bandwidth improvement of FS over NS under FIFO schedulers on the parking lot topology.}
\label{fig:parking_lot_fifo_nr}
\end{figure}

Finally, we evaluate the impact of network scale using the parking lot topology, varying the number of links $n$ and main-path flows $m$. For each $(m, n)$ configuration, we generate $1000$ random instances. The results are in \fig{fig:parking_lot_fifo_nr} in the form of a heatmap that reports both average bandwidth reductions and $95\%$ confidence intervals (as vertically aligned markers within each cell\footnote{The confidence intervals are small and the markers hardly visible.}). FS again consistently outperforms NS across all configurations. Moreover, bandwidth reduction increases with path length, as the benefits of smoother traffic accrue over more hops. This aligns with similar observations in~\cite{qiu2024benefits}.

\subsection{Bandwidth Minimization under Static Priority}
\label{sec:evaluation_sp}

\begin{figure}[!h]
\centering
\begin{subfigure}{0.45\linewidth}
  \centering
  \includegraphics[width=\linewidth]{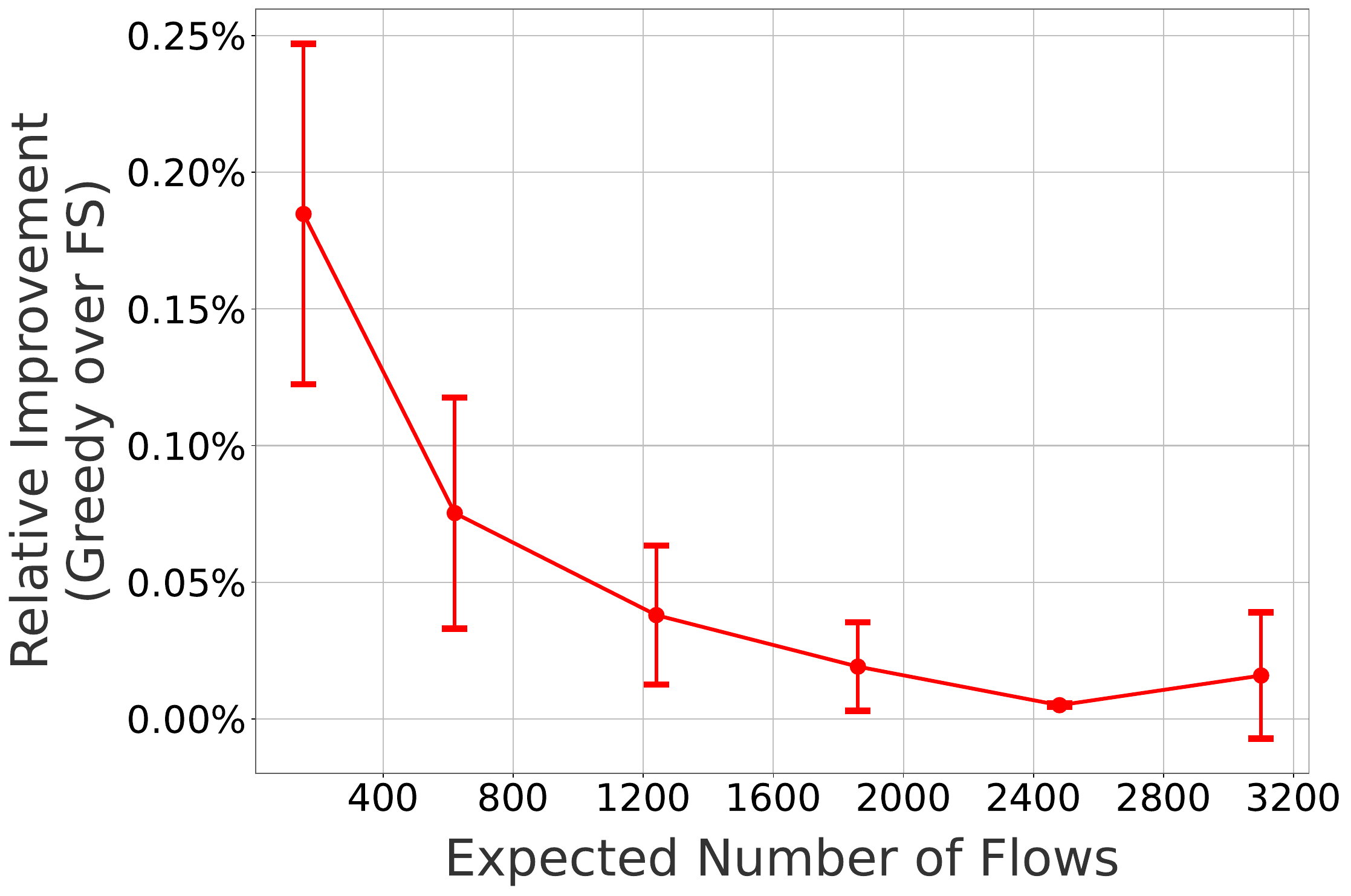}
  \caption{over FS}
  \label{fig:cev_fr}
\end{subfigure}
\begin{subfigure}{0.45\linewidth}
  \centering
  \includegraphics[width=\linewidth]{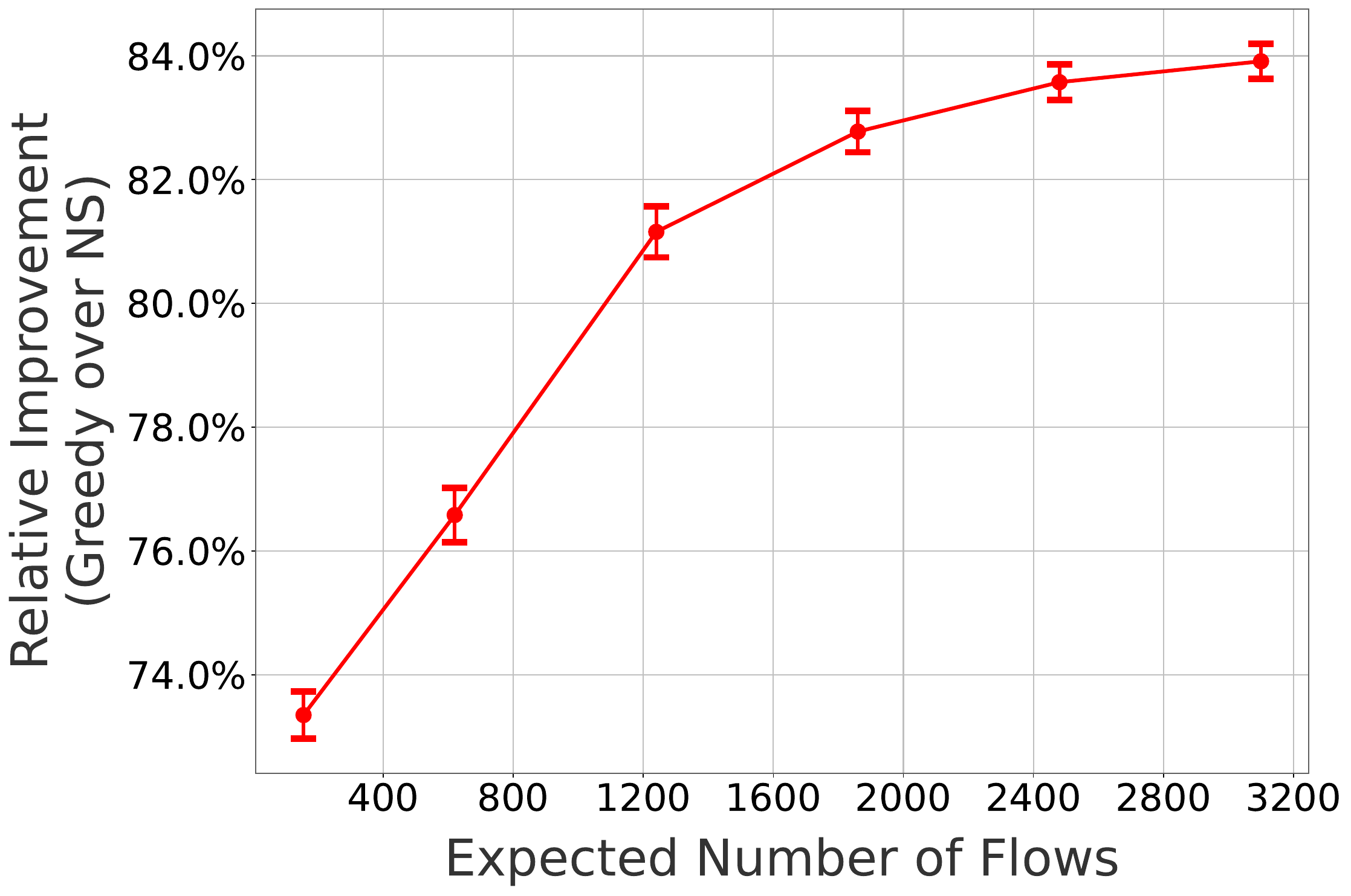}
  \caption{over NS}
  \label{fig:cev_nr}
\end{subfigure}
\caption{\Greedy's bandwidth improvement on Orion CEV (with 8 priority classes).}
\label{fig:cev_baseline}
\end{figure}

We first evaluate the performance of \Greedy (rather than FS) for static-priority scheduling. Recall that \Greedy parallels the structure of a similar solution from~\cite{qiu2024benefits}.  In particular, it shares its exploration phase and deadline adjustment mechanism, whose effectiveness were already established in~\cite[Appendix~B.G]{qiu2024benefits}.  As a result, we begin our evaluation by assessing the efficacy of our proposed $k$-means clustering algorithm for mapping flows (deadlines) to priority classes; a step specific to static priority schedulers. Appendix~\ref{app:k_means_performance} compares $k$-means to several alternative assignment strategies for different numbers of priority classes, with $k$-means consistently performing the best.

Having established the soundness of the priority assignment step used in  \Greedy, we evaluate next its bandwidth reduction capability relative to the two baselines, FS and NS. We begin with the Orion CEV topology, with a number of priority classes $k=8$ commonly available in practice\footnote{IEEE 802.1Q and TSN standards~\cite{xue2023real} define a 3-bit Priority Code Point (PCP) field, supporting $2^3=8$ priority classes.}.

\fig{fig:cev_baseline} shows bandwidth reductions from \Greedy compared to FS and NS as a function of the number of flows. 

As with FIFO, FS closely approximates the solution found by \Greedy, with less than a $0.2\%$ difference across all configurations, and diminishing with the number of flows. This gap is significantly smaller than the $\sim 16\%$ improvements observed under SCED (Fig.~12a of~\cite{qiu2024benefits}). 

Similarly, \Greedy (and FS) achieves substantial gains over NS, reducing bandwidth by up to $84\%$. This exceeds the $\sim73\%$ gain observed under SCED (Fig.~12b of~\cite{qiu2024benefits}), but remains below the $\sim98\%$ improvement under FIFO (cf. \fig{fig:cev_fifo_nr}). This parallels the schedulers' progressively increasing flow differentiation capabilities. More powerful schedulers are less dependent on the proactive actions of shaping.

\begin{figure}[!h]
\centering
\begin{subfigure}{0.45\linewidth}
  \centering
  \includegraphics[width=\linewidth]{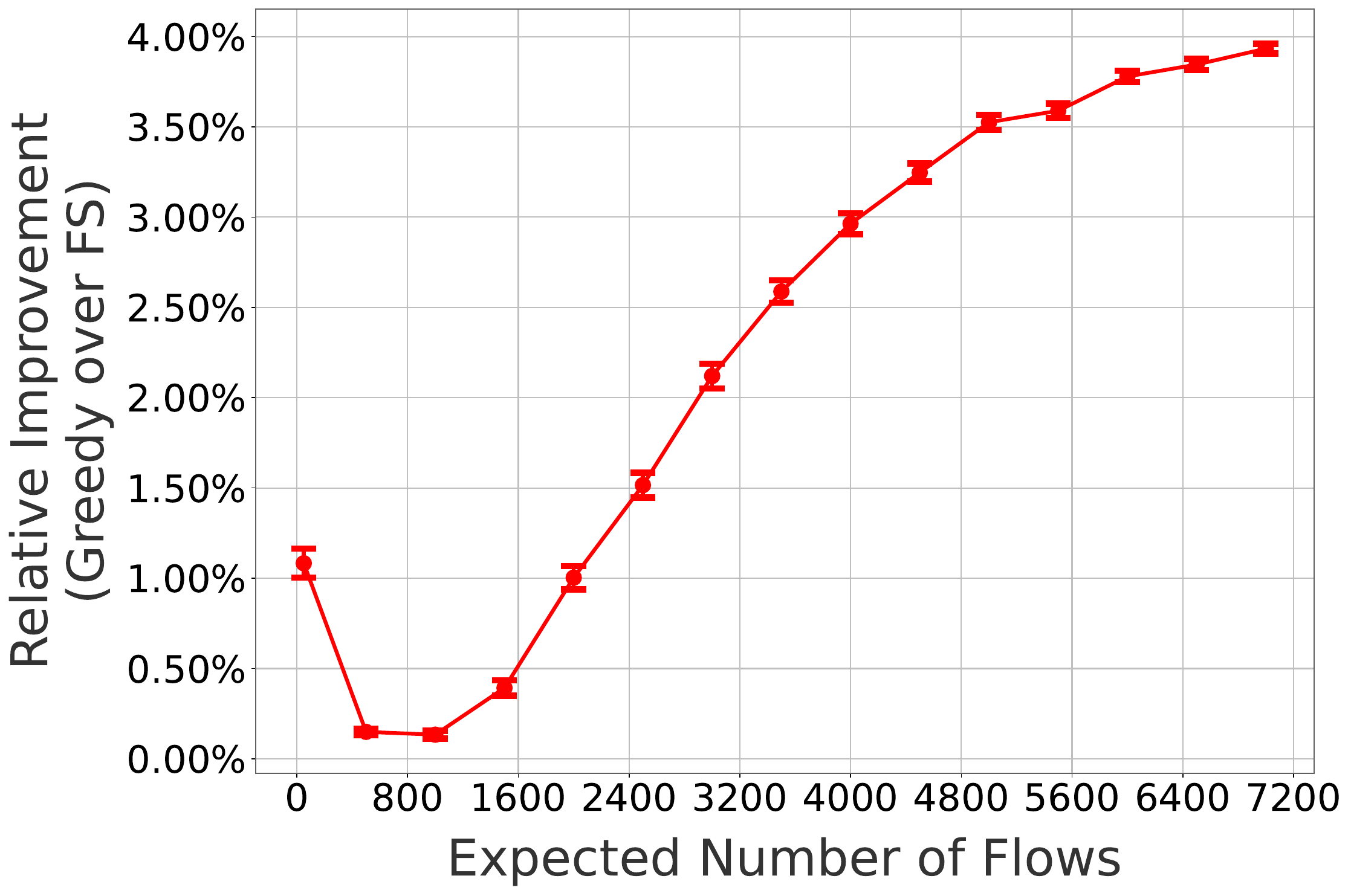}
  \caption{over FS}
  \label{fig:ustopo_fr}
\end{subfigure}
\begin{subfigure}{0.45\linewidth}
  \centering
  \includegraphics[width=\linewidth]{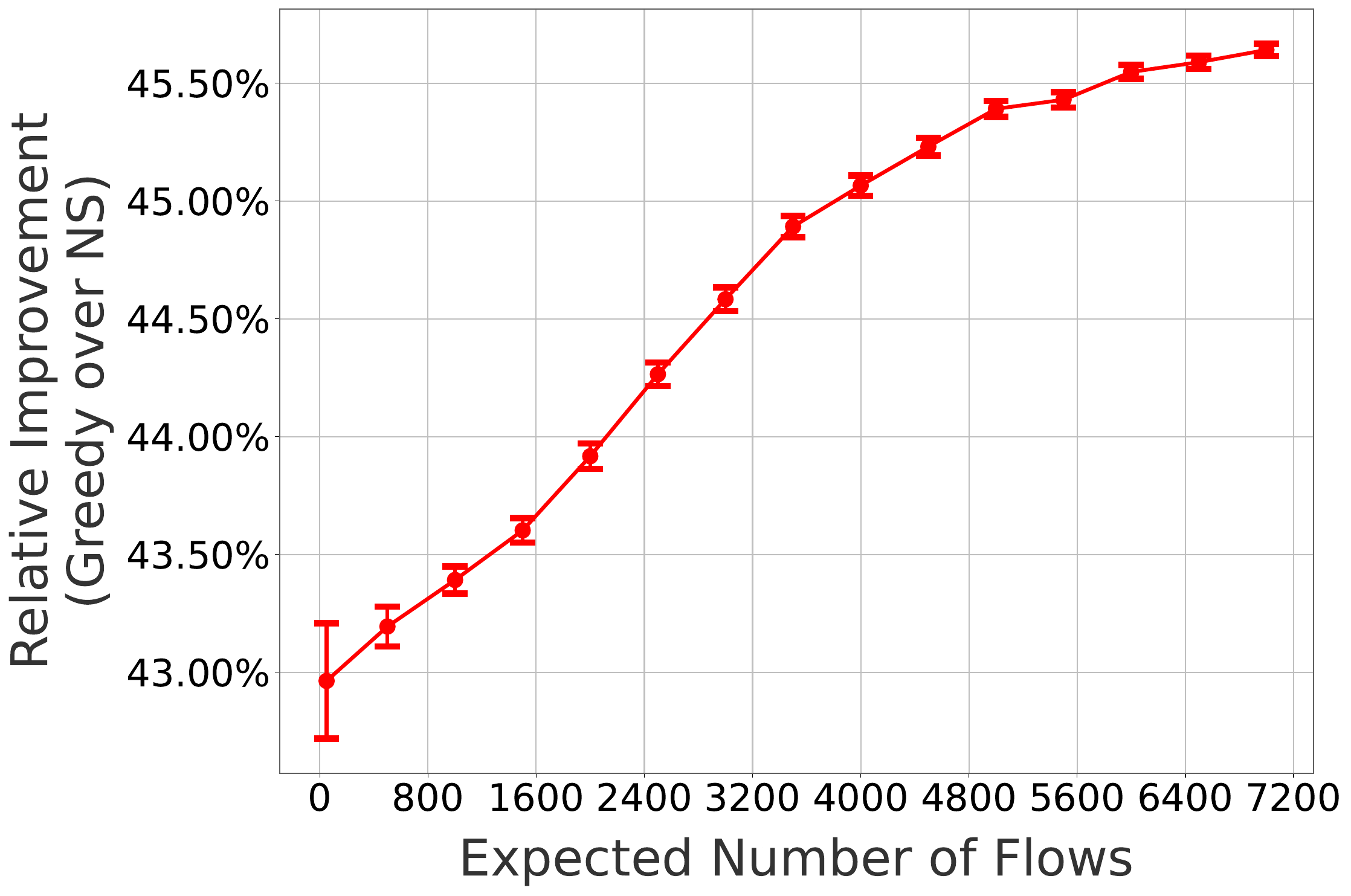}
  \caption{over NS}
  \label{fig:ustopo_nr}
\end{subfigure}
\caption{\Greedy's bandwidth improvement on US-Topo (with 8 priority classes).}
\label{fig:ustopo_baseline}
\end{figure}

We next evaluate \Greedy on US-Topo, with results in \fig{fig:ustopo_baseline} that are largely consistent with those of Orion CEV. \Greedy achieves an improvement of about $2\%$ over FS, again smaller than the $\sim8\%$ gain under SCED (Fig.~20a of~\cite{qiu2024benefits}), and an improvement of $\sim 46\%$ over NS, similar to that of SCED (also $\sim46\%$ from Fig.~20b of~\cite{qiu2024benefits}), but much less than the $\sim90\%$ under FIFO (cf. \fig{fig:ustopo_fifo_nr}).

While the delay targets of flows used with Orion CEV and US-Topo are derived from realistic application traces, it is instructive to examine how the benefit of shaping scales with delay requirements. To this end, we introduce a scaling factor $\omega$ and uniformly scale the delay bounds of all flows.
\begin{figure}[!h]
\centering
\begin{subfigure}{0.45\linewidth}
  \centering
  \includegraphics[width=\linewidth]{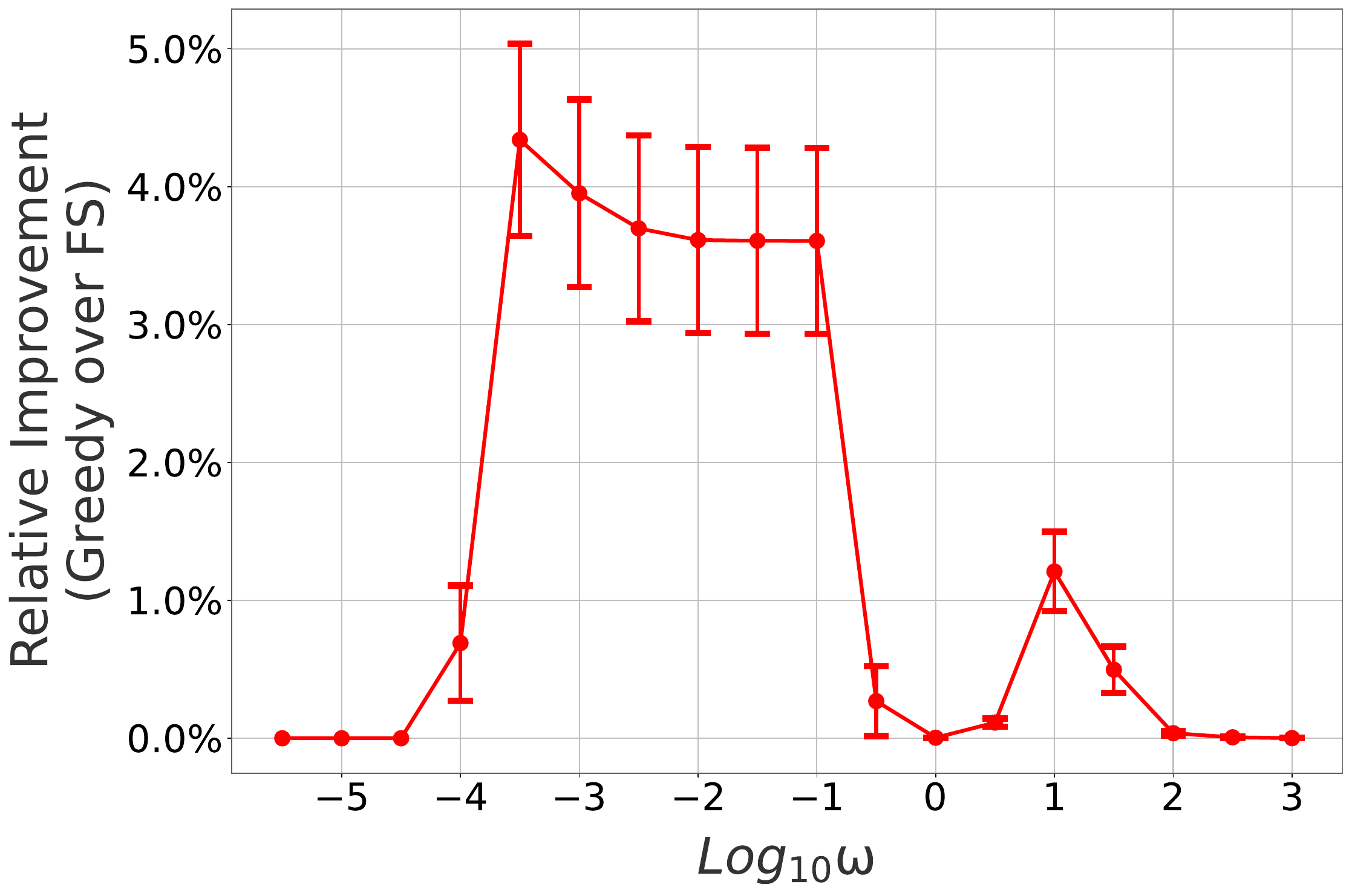}
  \caption{Orion CEV (3100 flows)}
  \label{fig:cev_ddl_scale}
\end{subfigure}
\begin{subfigure}{0.45\linewidth}
  \centering
  \includegraphics[width=\linewidth]{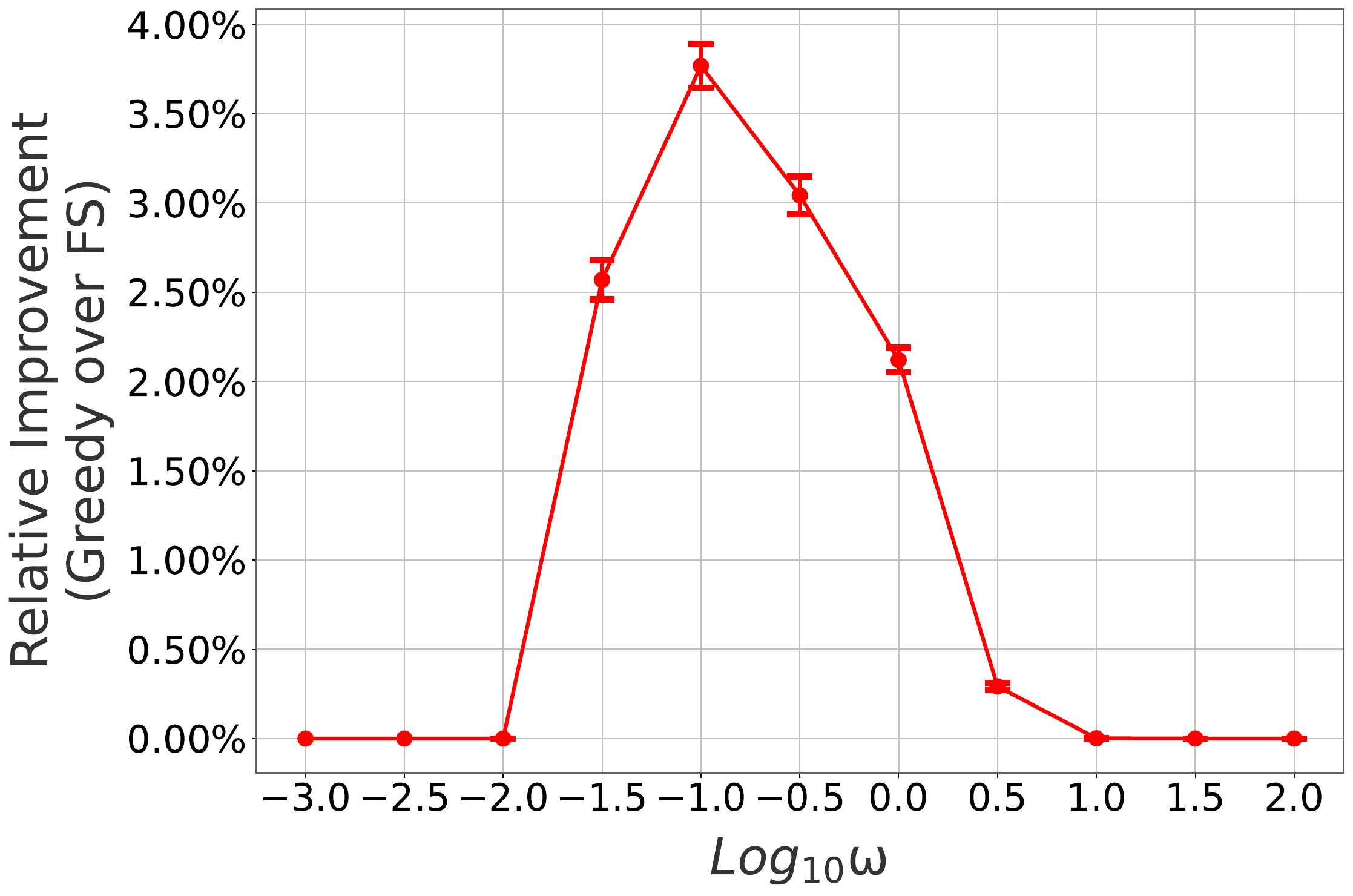}
  \caption{US-Topo (3000 flows)}
  \label{fig:ustopo_ddl_scale}
\end{subfigure}
\caption{Bandwidth improvement of \Greedy over FS as a function of deadline scaling $\omega$ (with 8 priority classes).}
\label{fig:ddl_scale}
\end{figure}

\fig{fig:ddl_scale} reports the bandwidth improvement of \Greedy over FS as a function of $\omega$ for both topologies. We observe that the improvement peaks at around $4\%$ for both Orion CEV and US-Topo, although for different values of $\omega$.  Consistent with intuition, all improvements disappear for sufficiently large or small deadlines. When deadlines are large, all flows can be fully shaped to their token rates, making FS optimal. Conversely, when deadlines are very tight, per-hop deadlines all approach $0$, leaving no room for differentiation, with \Greedy again converging to FS.

\begin{figure}[!h]
\centering
\begin{subfigure}{0.45\linewidth}
  \centering
  \includegraphics[width=\linewidth]{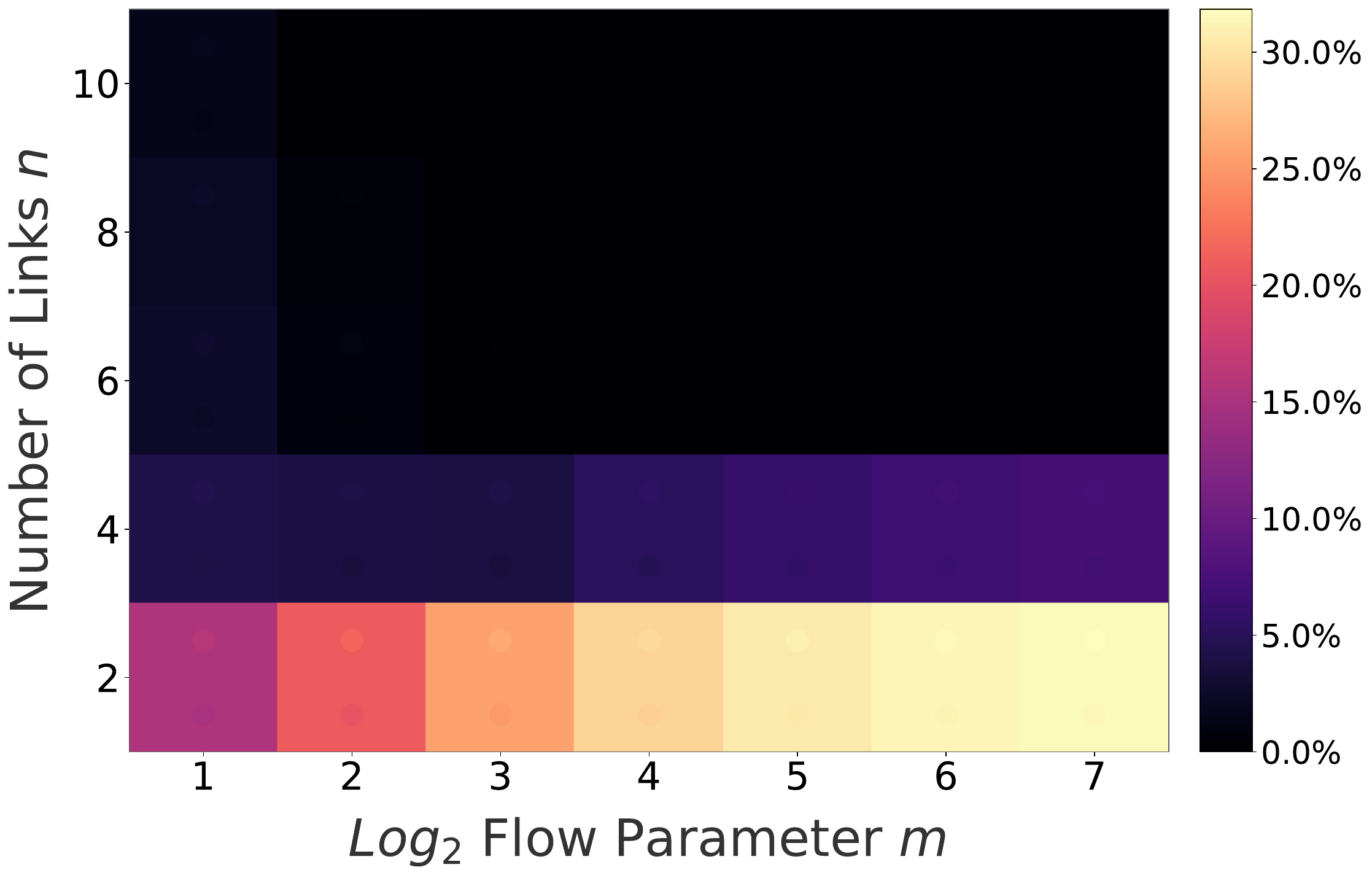}
  \caption{over FS}
  \label{fig:parking_lot_fr}
\end{subfigure}
\begin{subfigure}{0.45\linewidth}
  \centering
  \includegraphics[width=\linewidth]{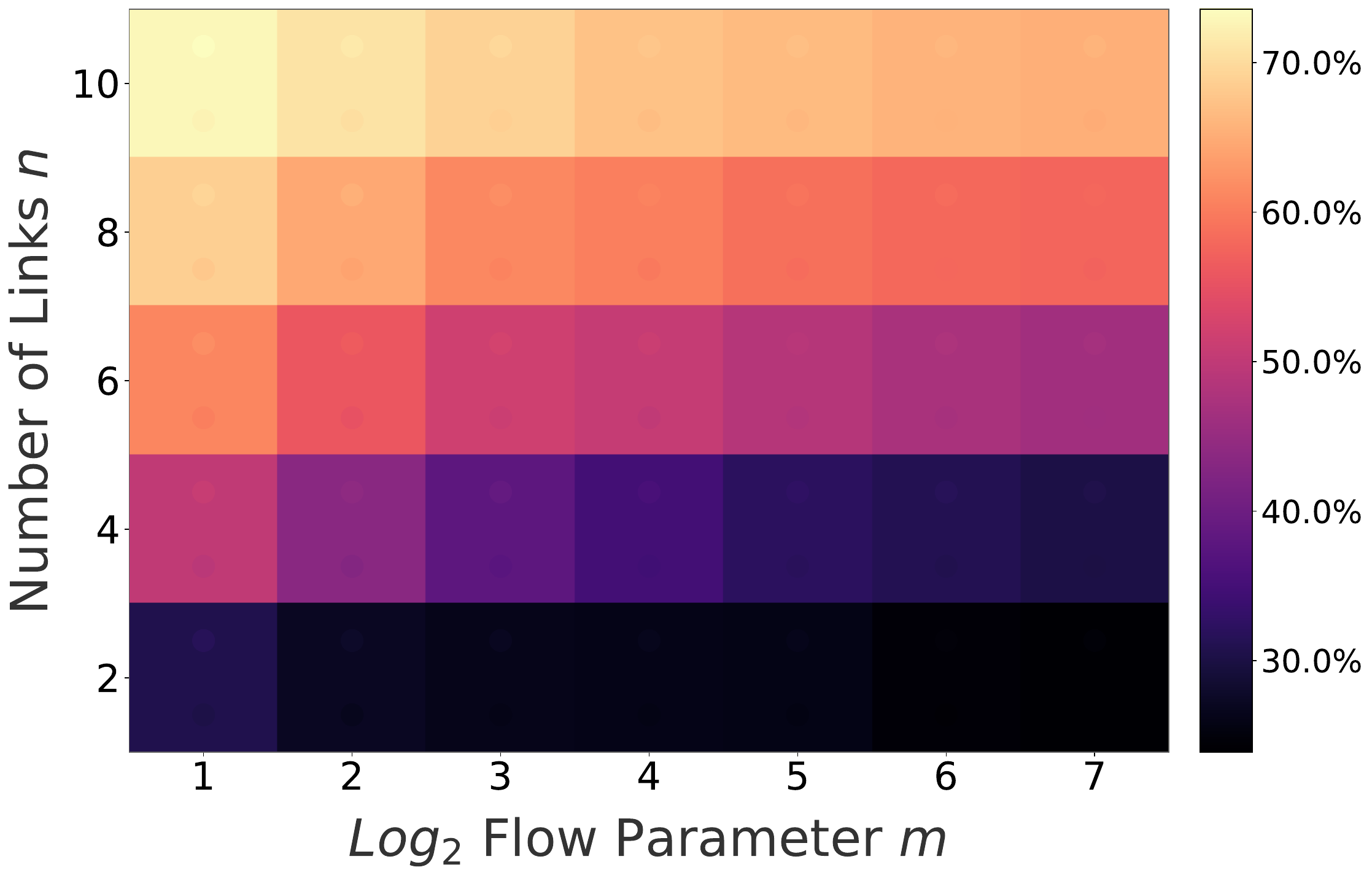}
  \caption{over NS}
  \label{fig:parking_lot_nr}
\end{subfigure}
\caption{\Greedy's bandwidth improvement on parking-lot topology (with 8 priority classes).}
\label{fig:parking_lot_baseline}
\end{figure}

Finally, we examine how the bandwidth improvement of \Greedy scales with network size by evaluating it against FS and NS on the parking lot topology over a range of $(m,n)$ configurations (\fig{fig:parking_lot_baseline}). As expected (\fig{fig:parking_lot_fr}), as path length $(n)$ increases, \Greedy converges to FS.  This partially explains why, in multi-hop topologies such as Orion CEV and US-Topo, \Greedy yields only marginal gains over FS.  When flow paths are short (\eg $n=2$), the improvement over FS can, however, reach up to $30\%$ when the number of flows is large (the greater diversity in flows' deadlines enables a more effective use of static-priority's scheduling flexibility).  Appendix~\ref{app:shaping_ratio} explores this aspect in greater details.  The comparison to NS (\fig{fig:parking_lot_nr}) is also intuitive, the benefits of \Greedy, as those of FS, increase with path length.

\subsection{Scheduler Comparison}

Since the role of shaping (reprofiling) has been explored for SCED~\cite{qiu2024benefits}, static-priority, and FIFO schedulers, it is natural to examine how that role varies across schedulers.

We begin by comparing the three schedulers under NS and FS, with FIFO serving as the baseline.  The comparison under NS (flow profiles remain unchanged) helps gauge the benefits afforded by schedulers of increasing complexity.  For static priority, results are reported while varying the number of priority classes (from 2 to 20).  Conversely, the comparison under FS offers insight into how a common shaping strategy can narrow the performance gap between the three schedulers. The comparisons are carried out on Orion CEV and US-Topo and, for each, involve a combination of $3000$ flows.
\begin{figure}[!h]
\centering
\begin{subfigure}{0.45\linewidth}
  \centering
  \includegraphics[width=\linewidth]{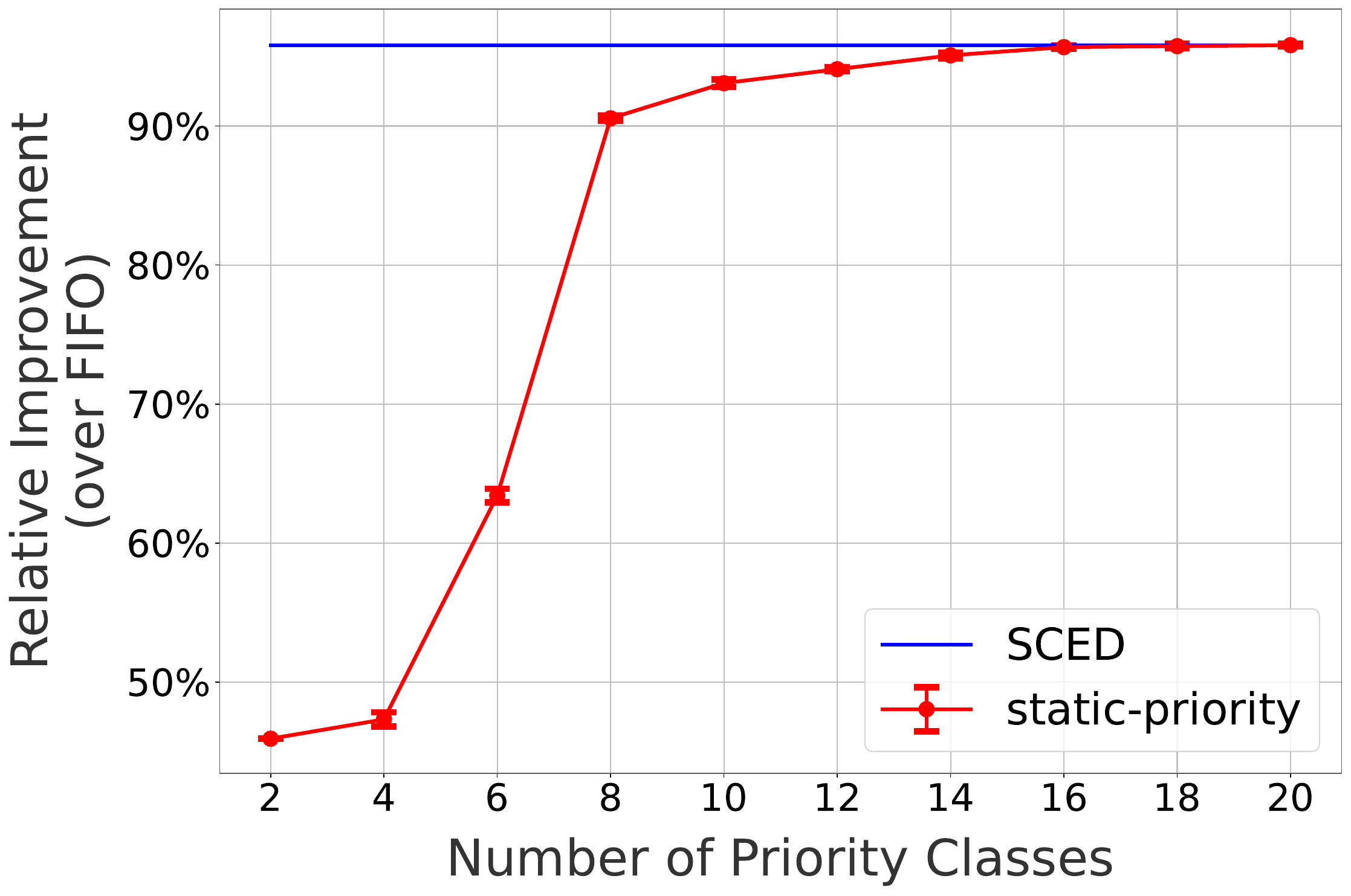}
  \caption{Orion CEV (3100 flows)}
  \label{fig:cev_num_class_nr}
\end{subfigure}
\begin{subfigure}{0.45\linewidth}
  \centering
  \includegraphics[width=\linewidth]{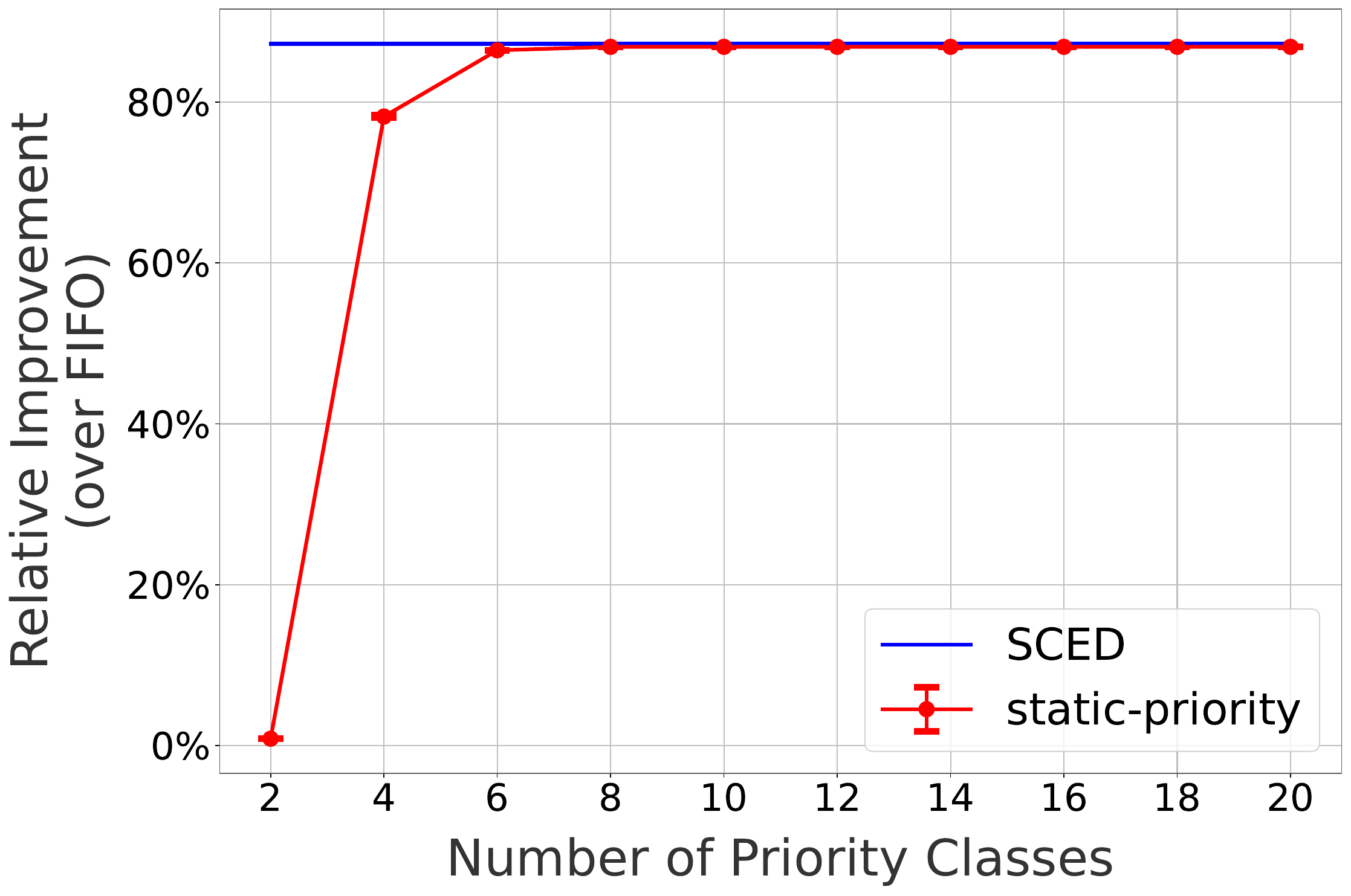}
  \caption{US-Topo (3000 flows)}
  \label{fig:ustopo_num_class_nr}
\end{subfigure}
\caption{SCED and static priority schedulers' bandwidth improvement (with NS) compared to FIFO (with NS) as a function of the number of priority classes.}
\label{fig:scheduling_gap_num_class_nr}
\end{figure}

Results of the comparison under NS are in \fig{fig:scheduling_gap_num_class_nr} that illustrates the advantages of greater scheduling flexibility.  SCED and static priority both significantly outperform FIFO, by up to $90\%$ or $80\%$ depending on the topology.  Of interest is the fact that, as the number of priority classes increases, the performance of static priority becomes indistinguishable from that of SCED.  This is an artifact of SCED being limited to use the flows' original token bucket profiles as their service curves. This results in SCED behaving like EDF, which, at least for the worst-case scenarios we consider, can be well approximated using a (sufficient) number of fixed priority classes.  
\begin{figure}[!h]
\centering
\begin{subfigure}{0.45\linewidth}
  \centering
  \includegraphics[width=\linewidth]{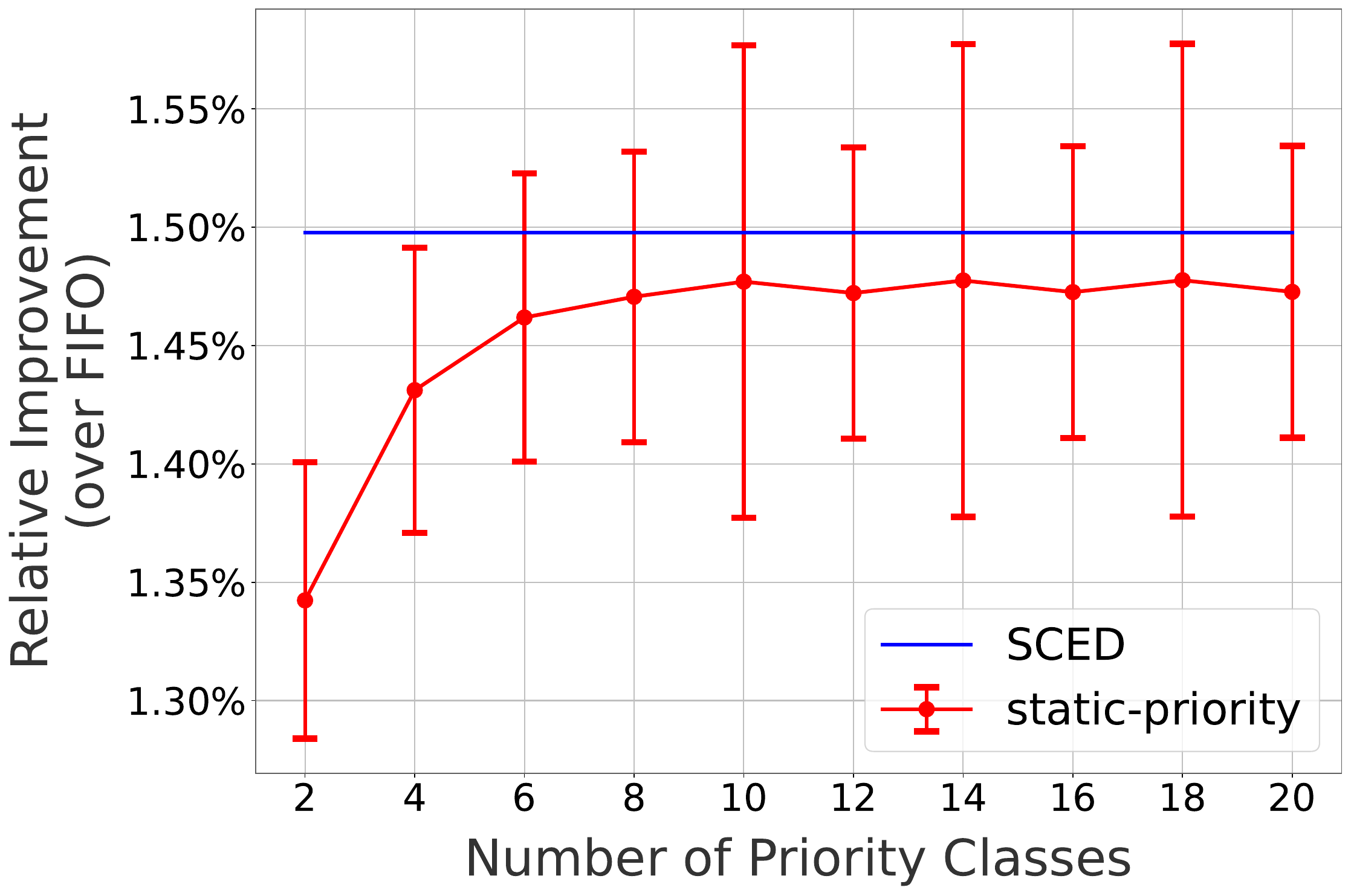}
  \caption{Orion CEV (3100 flows)}
  \label{fig:cev_num_class_fr}
\end{subfigure}
\begin{subfigure}{0.45\linewidth}
  \centering
  \includegraphics[width=\linewidth]{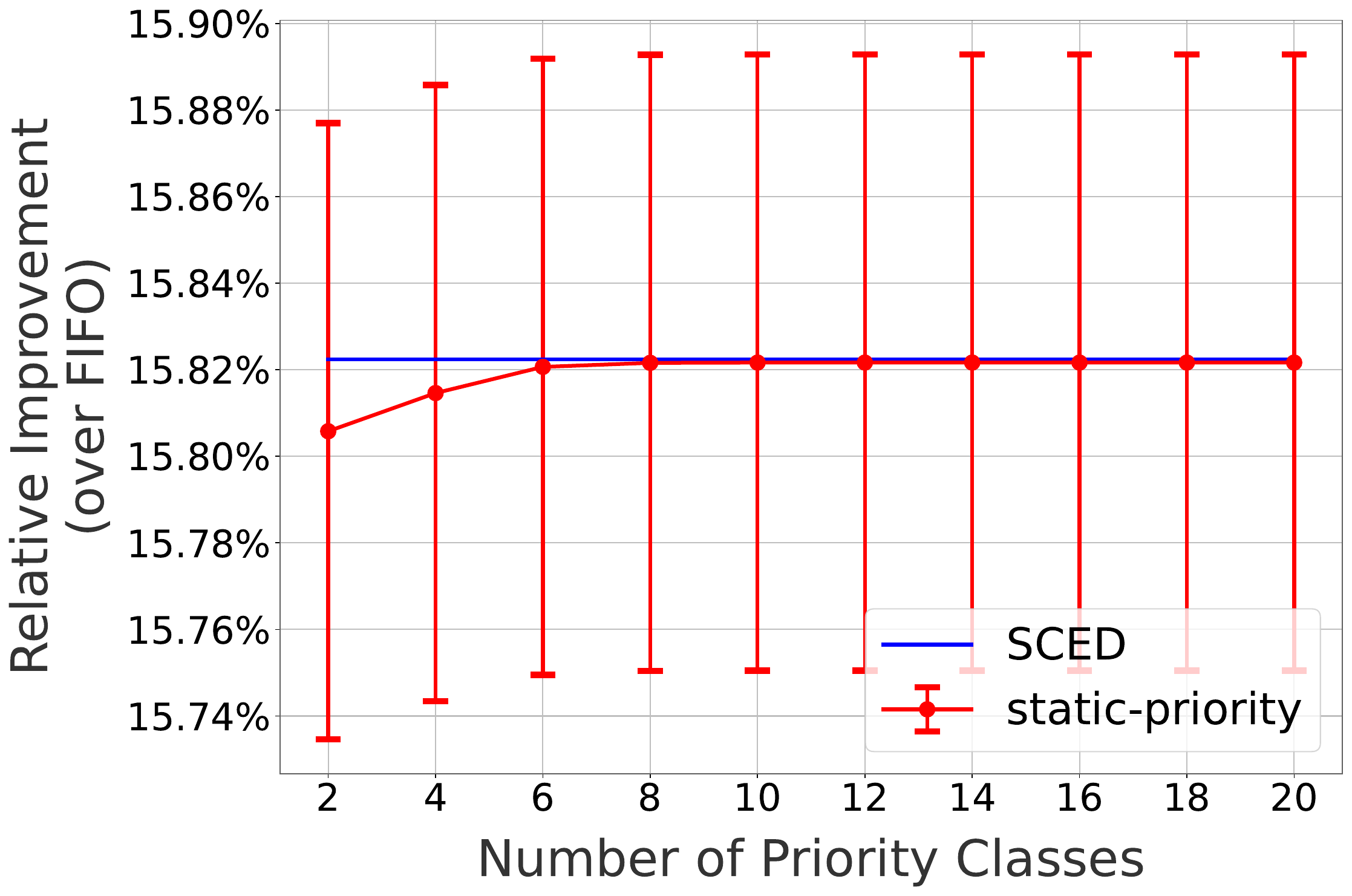}
  \caption{US-Topo (3000 flows)}
  \label{fig:ustopo_num_class_fr}
\end{subfigure}
\caption{SCED and static priority schedulers' bandwidth improvement (with FS) compared to FIFO (with FS) as a function of the number of priority classes.}
\label{fig:scheduling_gap_num_class_fr}
\end{figure}

\fig{fig:scheduling_gap_num_class_fr} reports a similar comparison as \fig{fig:scheduling_gap_num_class_nr}, but now under FS.  By construction, FS devotes as much of a flow's delay budget to making it smoother, and while this often results in flows with no residual network deadline\footnote{Recall our assumption of a fluid model.}, flows with large initial deadlines may retain an unused delay budget (they cannot be reshaped below their token rate).  This offers schedulers such as SCED and static priority some, albeit limited opportunities to leverage those residual delays to further reduce bandwidth.  The limited scope of those opportunities is also why SCED and static priority behave similarly.  As we shall see next, this does not necessarily hold when reprofiling decisions yield a richer set of service curves for SCED to leverage, \ie as is the case when \Greedy is used.
\begin{figure}[!h]
\centering
\begin{subfigure}{0.45\linewidth}
  \centering
  \includegraphics[width=\linewidth]{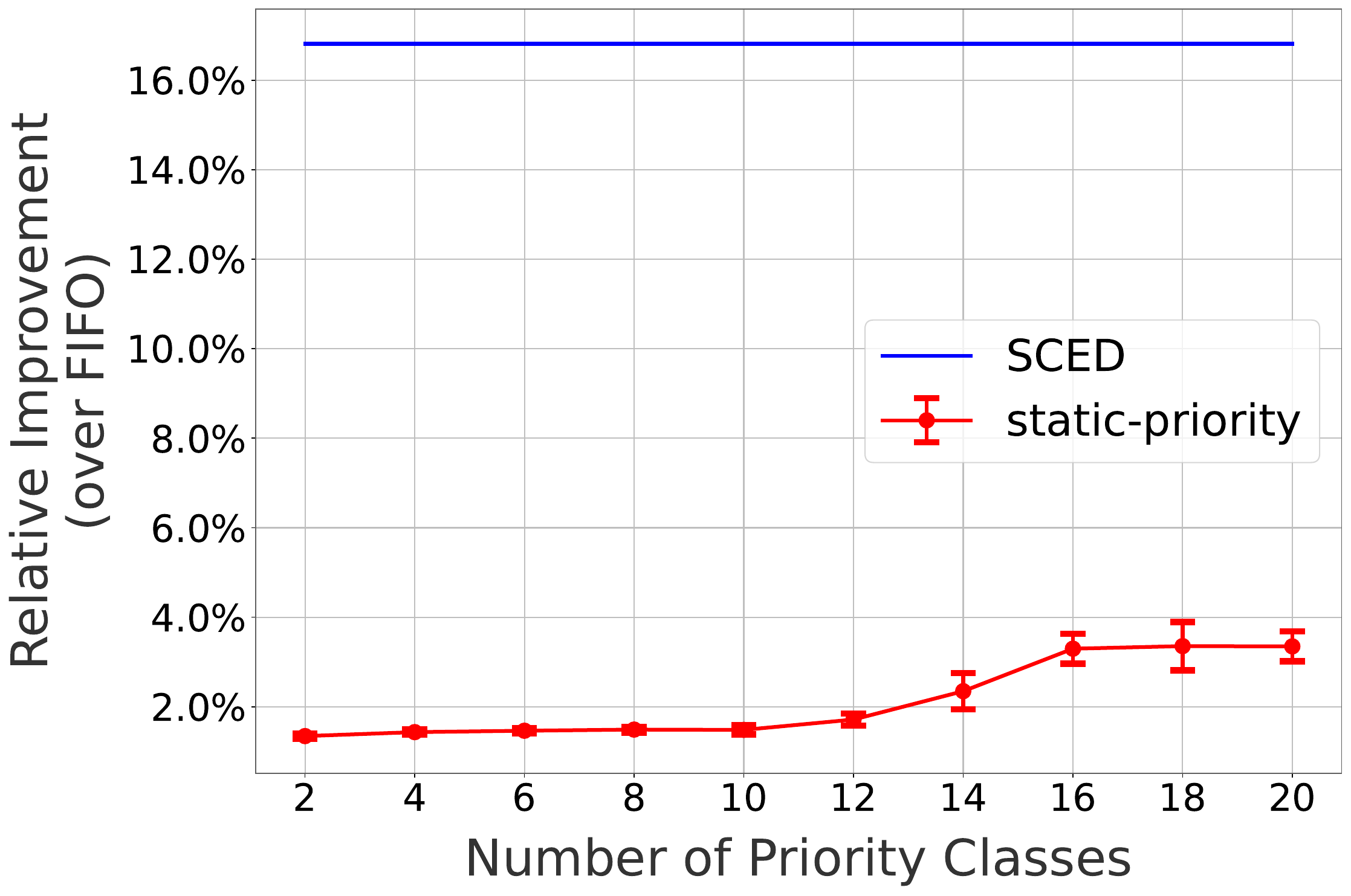}
  \caption{Orion CEV (3100 flows)}
  \label{fig:cev_num_class}
\end{subfigure}
\begin{subfigure}{0.45\linewidth}
  \centering
  \includegraphics[width=\linewidth]{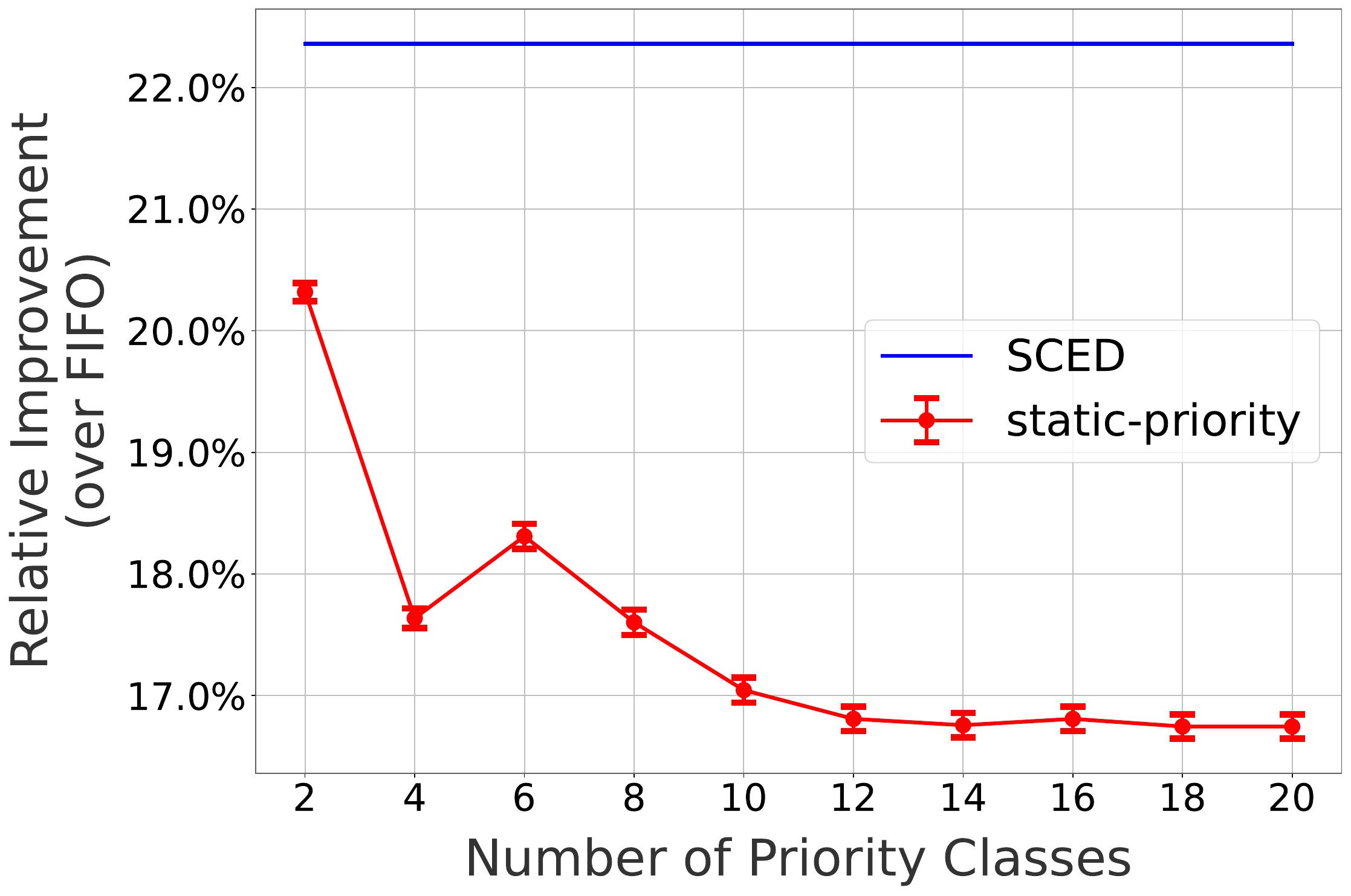}
  \caption{US-Topo (3000 flows)}
  \label{fig:ustopo_num_class}
\end{subfigure}
\caption{SCED and static priority schedulers' bandwidth improvement (with \Greedy) compared to FIFO (with FS) as a function of the number of priority classes.}
\label{fig:scheduling_gap_num_class}
\end{figure}

This is explored in~\fig{fig:scheduling_gap_num_class} that compares the three schedulers, but now using \Greedy for SCED and static-priority, and FS for FIFO (FS is ``optimal'' for FIFO).  Comparing \figs{fig:scheduling_gap_num_class}{fig:scheduling_gap_num_class_fr} illustrates the benefits of \Greedy over FS, primarily for SCED, as known from~\cite{qiu2024benefits}, and to a lesser extent for static priority, consistent with \fig{fig:ddl_scale}.

\fig{fig:scheduling_gap_num_class} reveals another interesting aspect, namely, under \Greedy, unlike FS, a performance gap persists\footnote{Additional comparisons investigating the impact of network scale using the parking lot topology are available in Appendix~\ref{app:performance_gap}.} between SCED and static priority, irrespective of how many priority classes the latter is allowed.  This is in part because \Greedy allows richer flow profiles than FS (distinct peak and long-term rates). Under SCED, the joint optimization of making flows smoother while preserving some scheduling flexibility can leverage those richer profiles.

Finally, \fig{fig:ustopo_num_class} also illustrates that increasing the number of priority classes does not always benefit the performance of static-priority scheduling.  The figure shows a drop in performance as more priority classes are introduced.  This is counter-intuitive and an artifact of the $k$-means algorithm on which we rely.  It seeks to utilize all classes without considering whether merging classes might yield better performance. As a result, flows with similar local deadlines may be unnecessarily separated, leading to poorer performance\footnote{A simple albeit expensive solution involves running $k$ versions of the clustering algorithm, one for each number of priorities, and try all solutions.}.
%


\section{Related Work}
\label{sec:related}

\subsection{Recent Advances in TSN/DetNet Systems}

Recent work on TSN/DetNet has targeted enabling deterministic communication over common network technologies through advances in system design and standardization~\cite{tsn18, zhang2024time, tsn_survey22, walrand23, egger2025wireless}. Examples include architectures to integrate TSN with software-defined control and cross-domain networking~\cite{guo2024software, li2024development, ihle2025p4}, as well as prototype implementations that expose practical constraints such as synchronization inaccuracies and hardware-induced delays~\cite{egger2025wireless, eppler2025impact}. The focus is on system-level realizations and the challenges of preserving theoretical guarantees in deployed solutions. In contrast, we explore the fundamental relationship between shaping and scheduling, providing insights that can guide system designs.

\subsection{Traffic Shaping and Scheduling Optimization}

A large body of work studies network-wide scheduling and shaping configuration in TSN. Scheduling is often formulated as a global optimization problem, particularly for time-aware shaping (TAS), and solved using ILP, constraint programming, or heuristics~\cite{li2024development, stuber2023survey, chahed2023tsn, xue2025survey}. Complementary efforts focus on configuring traffic shaping mechanisms such as credit-based shaping (CBS) and asynchronous traffic shaping (ATS) using analytical models or simulation-based approaches~\cite{yan2025optimizing, hirofuchi2026implementation}.

These works typically focus on computing feasible or efficient configurations under a fixed scheduling model. In contrast, we consider multiple schedulers and investigate the role of proactively adjusting flow shaping profiles to minimize the bandwidth required to meet delay bounds. Our analysis also highlights how scheduler's flexibility influences the effectiveness of shaping in reducing bandwidth.

\subsection{Bandwidth Optimization under QoS Constraints}

Bandwidth optimization under QoS constraints has been studied in several contexts beyond TSN/DetNet. In deterministic settings, network calculus and effective bandwidth theory have been used to derive the minimum service rates required to meet delay guarantees for regulated traffic. Related formulations arise in admission control and resource allocation, where the goal is to determine whether a set of flows can be supported under given bandwidth constraints~\cite{wang2024bpnn}, or to allocate resources efficiently while satisfying QoS requirements~\cite{hernandez2003new, gong2011joint}. In parallel, traffic engineering and network utility optimization frameworks address bandwidth allocation across paths and flows, often optimizing utilization, congestion, or throughput under capacity constraints~\cite{sahni2007bandwidth, noormohammadpour2018minimizing, bethanabhotla2013utility}.

While these works provide important insights into resource-efficient network design, they typically assume fixed traffic profiles and scheduling models, and focus on determining the minimum resources needed to ensure feasibility. In contrast, our work treats traffic profiles as decision variables and jointly considers how to adjust them under different scheduling disciplines. This allows us to study the trade-off between reducing traffic burstiness while tightening network delays across schedulers with varying levels of flexibility and, therefore, ability to leverage differences in-network delays.

\subsection{Priority Assignment and Queue Management}

Priority assignment has been extensively studied in real-time systems as a means to ensure schedulability under fixed-priority scheduling. Classical approaches include rule-based schemes such as rate-monotonic and deadline-monotonic policies, which assign priorities according to task periods or deadlines and are optimal under specific assumptions~\cite{davis2016review}. For instance, works targeting hard delay bounds often apply a deadline-monotonic policy and assign a flow with a tighter deadline to a higher priority class~\cite{song2024benefits, zhu2017workloadcompactor}. More general settings are addressed by optimal priority assignment algorithms, such as Audsley’s algorithm~\cite{audsley2001priority}, which iteratively constructs a feasible priority ordering using schedulability tests. When analytical methods become intractable, heuristic and optimization-based approaches, including genetic algorithms and other search techniques, have been proposed to explore the large space of possible priority assignments~\cite{lee2022optimal, kumar2024optimal}. Other works perform admission control by assigning incoming flows to pre-configured priority queues with specified service guarantees, with the objective of maximizing link utilization or admission rate~\cite{grosvenor2015queues, van2020chameleon}.

These works focus on identifying priority assignments that ensure feasibility or improve schedulability under a fixed system model. In contrast, we consider priority assignment as part of a broader optimization problem, where priorities are determined jointly with traffic shaping to minimize bandwidth. Rather than relying on feasibility-driven or rule-based policies, our approach, while similar to deadline-monotonic policies, adjusts local deadlines and priority assignments towards reducing resource requirements.

\subsection{End-to-end Deadline Allocations}

Prior work on delay-constrained networking typically decomposes end-to-end delay into per-hop contributions using analytical models~\cite{guck2016function, gore2022network}, or implicitly determines delay allocation through resource optimization~\cite{kumar2017joint, petreska2019bound}. While some studies reveal structural properties such as delay balancing across hops~\cite{du2011statistical}, they do not explicitly treat deadline allocation as a decision variable. In contrast, we seek to optimize the allocation of delay budgets across hops and study its interaction with traffic profiles and scheduling.


\section{Conclusion}
\label{sec:conclusion}

This paper investigates bandwidth minimization in networks with hard end-to-end delay guarantees, focusing on the role of traffic profiles under FIFO and static priority schedulers. The benefits of reprofiling, \ie adjusting a flow's traffic profile before it enters the network, had been previously established~\cite{qiu2024benefits} for service curve schedulers (SCED).  This work extends the results to two FIFO and static priority schedulers.  As with SCED, we formulate a joint optimization accounting for reprofiling, scheduling, and bandwidth provisioning, and develop efficient solutions for both schedulers.

The results offer insights beyond confirming the benefits of reprofiling. Under FIFO, full shaping, \ie allocating as much of a flow's delay budget to making it smoother, realizes a near-optimal. This is intuitive as full shaping equalizes residual flows' delays, minimizing delay differences that FIFO cannot exploit. Full shaping is also effective under static priority, though an efficient greedy heuristic can outperform it, especially in networks with short path lengths.  Finally, although reprofiling helps FIFO and static priority narrow their performance gap with SCED, SCED's greater scheduling flexibility can leverage more sophisticated reprofiling solutions that allow it to continue outperforming the two simpler schedulers.

\bibliographystyle{IEEEtran}
\bibliography{MultiHop_StaticPriority+FIFO}
\vfill

\clearpage

\appendices


\section{Summary of Notation and Acronyms}
\label{app:glossary}

\begin{center}
\begin{supertabular}{ c | c }
\hline
{\textbf{Acronyms}} & {\textbf{Definition}} \\ \hline
2SRC & Two-Slope Reprofiling Curve \\ \hline
PBOO & Pay Burst Only Once \\ \hline
FIFO & First In First Out \\ \hline
SP & Static Priority \\ \hline
SCED & Service Curve Earliest Deadline first \\ \hline\hline
{\textbf{Notation}} & {\textbf{Definition}} \\ \hline
$\alpha$ & (token bucket) arrival curve  \\ \hline
$\beta$ & service curve \\ \hline
$\sigma$ & two-slope reprofiling curve (2SRC) \\ \hline
\multirow{2}{*}{$\Delta(\alpha,\beta)$} & delay upper bound for \\ & arrival curve $\alpha$ under service curve $\beta$ \\ \hline
$\sigma_i$ & 2SRC of flow $i$ \\ \hline
$S$ & minimal service function \\ \hline
$m$ & number of flows in the network \\ \hline
$n$ & number of links in the network \\ \hline
$k$ & number of priority classes in the network \\ \hline
$i$ & flow index \\ \hline
$j$ & link (hop) index \\ \hline
$h$ & priority class index \\ \hline
$r$ & flow long-term rate \\ \hline
$b$ & flow burst size \\ \hline
$d$ & flow end-to-end latency target \\ \hline
\multirow{2}{*}{$\widehat{d}$} & flow maximum reprofiling delay \\ & $\min(d, b/r)$ \\ \hline
$\mathcal{P}_i$ & set of links on flow $i$'s path (route) \\ \hline
$\pmb{\mathcal{P}}$ & flow path matrix \\ \hline
$\mathcal{F}_j$ & set of flows on link~$j$ \\ \hline
$(r_i, b_i, d_i)$ & profile of flow~$i$ \\ \hline
$(\mathbf{r}, \mathbf{b}, \mathbf{d})$ & vector of flow profiles \\ \hline
$\Gamma_j$ & priority assignment at link~$j$ \\ \hline
$\mathbf{\Gamma}$ & priority assignment for the network \\ \hline
$t$ & time \\ \hline
$R_i$ & 2SRC short-term rate of flow $i$ \\ \hline
$B_i$ & 2SRC burst size of flow $i$ \\ \hline
$D_i$ & shaping delay (same with $b/R$) of flow $i$ \\ \hline
$T_{hj}$ & scheduling deadline of priority $h$ at hop $j$ \\ \hline
$\widetilde{T}_{ij}$ & local deadline of flow $i$ at hop $j$ \\ \hline
\multirow{3}{*}{$\widetilde{T}'_{ihj}$} & $T_{hj}$ when $i=0$ \\ & $T_{hj} + D_i$ when $i \in G_h(\Gamma_j)$ \\ & $D_i$ when $i \in G_{h'}(\Gamma_j), h' < h$ \\ \hline
$T^*_{hj}$ & critical value of $T_{hj}$ for adjustment \\ \hline
\multirow{2}{*}{$T^+_{hj}$} & intersection between $C_jt$ and \\ & the aggregate higher-priority traffic \\ \hline
$\check{T}_{hj}$ & deadline boundary for priority assignment \\ \hline
$C_j$ & transmission link bandwidth capacity \\ \hline
$C^{*}_j$ & minimum required bandwidth capacity \\ \hline
$\mathbf{C}$ & total network bandwidth capacity \\ \hline
$\widetilde{s}_{ihj}$ & flow $i$ slack for priority $h$ \\ \hline
\end{supertabular}
\end{center}


\section{Proofs}
\label{app:proofs}

The propositions introduced to the paper focus on a single hop inside the network. To simplify the notations, we revise the propositions and present the proofs with the link-wise subscript $j$ removed.

\subsection{Proof of Proposition 1}
\label{app:proof1}

\textit{Proposition~\ref{prop:min_bw}:} Consider a hop equipped with a static-priority scheduler with $k$ priority classes indexed in decreasing order of priority from $1$ to $k$ (priority~$1$ being the highest), serving the set of flows $\mathcal{F}$. Given a priority assignment $\Gamma$ and the corresponding minimal service function $S_{h}$ for each priority class, the hop must provision a bandwidth of at least $C^*$ in order to satisfy the deadlines of all classes, where
\begin{equation}
\label{eq:min_bw1}
C^* =
\max_{1 \le h \le k}
\left(
\sum_{i \in \mathcal{F}} r_i,\;
C^*_{h}
\right),
\end{equation}
where
\begin{equation*}
C^*_{h} = \sup_{t > T_{h}} \frac{S_{h}(t)}{t},
\end{equation*}
denotes the bandwidth required to serve traffic from priority class $h$ by its deadline $T_{h}$.

To ensure finite worst-case delay for all packets, the available bandwidth $C^*$ must be no smaller than the aggregate long-term token rates, i.e., $C^* \geq \sum_{i \in \mathcal{F}} r_i$, which establishes the first part of~\Eqref{eq:min_bw1}.
    
We start from the canonical delay bound guarantee from Network Calculus that relates worst-case delay to the link bandwidth at one hop, and then establish its equivalence to~\Eqref{eq:min_bw1}. According to Network Calculus~\cite[Theorem 1.4.2]{le2001network}, satisfying the deadline $T_h$ of priority class $h$ requires that
\begin{align}
& \sup_{t \geq 0}\big\{\inf_{\tau \geq 0}\{\alpha_h(t) \le \beta_h(t+\tau)\}\big\} \leq T_h\\
\iff & \sup_{t \geq 0}\big\{\inf_{\tau \geq 0}\{\alpha_h(t - T_h) \le \beta_h(t+\tau)\}\big\} \leq 0\\
\label{eq:priority_condition}
\iff & \beta_h(t) \geq \alpha_h(t - T_h), \forall t \geq 0.
\end{align}
where $\alpha_h$ and $\beta_h$ denote the arrival and service curves of priority class $h$, respectively. 
    
For 2SRC-shaped flows, we have
\begin{equation}
\label{eq:alpha_h}
\alpha_h(t) = H_h(t) = \sum_{i \in G_h(\Gamma)} \sigma_i(t)
\end{equation} 
where $\sigma_i(t)$ is the 2SRC of flow $i$.
    
Following Network Calculus~\cite[Proposition 1.3.4]{le2001network}, the service curve of a non-preemptive static-priority scheduler is given by
\begin{equation}
\label{eq:beta_h}
\beta_h(t) = [Ct - \sum_{1 \le h' < h}H_{h'}(t)]^+,
\end{equation}
representing the residual bandwidth after serving all higher-priority traffic.
    
We next show that Inequality~\eqref{eq:priority_condition} is equivalent to
    \begin{equation}
    \label{eq:bandwidth_condition}
        Ct \geq S_h(t), \forall t > T_h,
    \end{equation}
with $S_h(t)$ being the minimal service function of priority class $h$, implying that $C \geq \sup_{t > T_h}\{S_h(t)/t\}$ is sufficient to meet the deadline of class $h$, thereby proving~\Eqref{eq:min_bw1}.

Towards proving
\begin{align*}
& \beta_h(t) \geq \alpha_h(t - T_h), \forall t \geq 0\\
\iff & Ct \geq S_h(t), \forall t > T_h,
\end{align*}
we start with
\begin{align*}
& \beta_h(t) \geq \alpha_h(t - T_h), \forall t \geq 0\\
\Rightarrow & Ct \geq S_h(t), \forall t > T_h,
\end{align*}
Substituting $\alpha_h$ and $\beta_h$ from~\Eqref{eq:alpha_h} and~\eqref{eq:beta_h} into $\beta_h(t) \geq \alpha_h(t - T_h), \forall t \geq 0$,
we obtain
\begin{equation}
\label{eq:priority_condition1}
[Ct - \sum_{1 \le h' < h}H_{h'}(t)]^+ \geq \sum_{i \in G_{h}(\Gamma)}\sigma_i(t - T_h) = H_h(t - T_h).
\end{equation}
Let $\beta'_h(t) = Ct - \sum_{1 \le h' < h}H_{h'}(t)$. We first note that 
\begin{align*}
& \beta_h(t) \geq \alpha_h(t - T_h), \forall t \geq 0\\
\Rightarrow & \beta'_h(t) \geq \alpha_h(t - T_h), \forall t \ge T_h
\end{align*}
since $\alpha_h(t)\geq 0, \forall t\geq 0$.
    
According to the definition of 2SRC, $H_h(t - T_h) > 0$ for $t > T_h$. Because Inequality~\eqref{eq:priority_condition1} holds from our assumption that $\beta_h(t) \geq \alpha_h(t - T_h), \forall t \geq 0$, this implies that $\beta'_h(t) > 0$ for $t > T_h$. Hence, Inequality~\eqref{eq:priority_condition1} directly implies
\begin{align}
\label{eq:priority_condition2}
& Ct - \sum_{1 \le h' < h}H_{h'}(t) \geq H_h(t - T_h), \forall t > T_h\\
\iff & Ct \geq S_h(t), \forall t > T_h
\end{align}
according to the definition of $S_h(t)$ given in~\Eqref{eq:minimal_service_function}.

Next, we show that
\begin{align*}
& Ct \geq S_h(t), \forall t > T_h\\
\Rightarrow & \beta_h(t) \geq \alpha_h(t - T_h), \forall t \geq 0,
\end{align*}
In other words, we assume that Inequality~\ref{eq:bandwidth_condition} (and therefore Inequality~\ref{eq:priority_condition2}) holds, which implies
\begin{equation*}
\beta'_h(t) = Ct - \sum_{1 \le h' < h}H_{h'}(t) \geq H_h(t - T_h), \forall t > T_h
\end{equation*}
Since $H_h(t - T_h) = 0$ when $t \le T_h$, and $[\beta'_h(t)]^+ \ge 0$ for all $t$~\footnote{By definition, $[\beta'_h(t)]^+ = \max(0, \beta'_h(t)), \forall t$.}, this implies
\begin{equation*}
[Ct - \sum_{1 \le h' < h}H_{h'}(t)]^+ \geq H_h(t - T_h),
\end{equation*}
or $Ct \geq S_h(t), \forall t > T_h
\Rightarrow
\beta_h(t) \geq \alpha_h(t - T_h), \forall t \geq 0$.

This establishes the equivalence between Inequalities~\ref{eq:priority_condition} and~\ref{eq:bandwidth_condition}, and therefore proves Proposition \ref{prop:min_bw}.

\subsection{Proof of Proposition 2}
\label{app:proof2}

\textit{Proposition~\ref{prop:opt_assign}:} Consider a hop employing a static-priority scheduler with $k$ priority classes indexed in decreasing order of priority from $1$ to $k$ (with priority $1$ being the highest), serving a set of $m$ flows $\mathcal{F}$. Suppose the flows are 2SRC-shaped and indexed in non-decreasing order of their local deadlines $\widetilde{T}_{i}$, i.e., $\widetilde{T}_{i} \le \widetilde{T}_{i'}$ for $i \le i'$. Then there exists a priority assignment $\Gamma^*$ that minimizes the required link bandwidth $C$ while satisfying all local deadlines, such that a flow $i$ is assigned a strictly higher priority than flow $i'$ only if $\widetilde{T}_{i} < \widetilde{T}_{i'}$.

We next prove the proposition based on a more generalized packet model that accounts for the packet size of each flow instead of a fluid model. Under a packet model, the minimal service function becomes:
\begin{align*}
& S_{h}(t) =\\
& \begin{cases}
0, & t \le T_{h}, \\
l_h^{max}(\Gamma) + \sum_{1 \le h' < h} H_{G_{h'}(\Gamma)}(t) + H_{G_{h}(\Gamma)}(t - T_{h}), & t > T_{h},
\end{cases}
\end{align*}
where $l^{(max)}_h(\Gamma)$ represents the maximum packet size among all flows belonging to priority classes strictly lower than $h$ due to non-preemptive scheduling, and
\begin{equation*}
H_{G_h(\Gamma)}(t) = \sum_{i \in G_h(\Gamma)}\sigma_i(t).
\end{equation*}
denotes the aggregate arrival curves from all the higher priority flows, with $G_h(\Gamma)$ denoting the subset of flows assigned priority level $h$ under $\Gamma$.

For a given priority assignment $\Gamma$, we denote $\widetilde{T}^{(max)}_h(\Gamma) = \max_{i \in G_h(\Gamma)} \widetilde{T}_i$ and $\widetilde{T}^{(min)}_h(\Gamma) = \min_{i \in G_h(\Gamma)} \widetilde{T}_i$. To meet the deadline of all flows in priority class $h$, we need the scheduling deadline $T_h = \widetilde{T}^{(min)}_h(\Gamma)$. We prove the proposition by induction on the number $k$ of priority classes. Let the induction hypothesis $I(k)$ be formulated as follows:

\begin{flushleft}
$I(k)$: For any set of flows sharing a link, there exists an optimal $k$-priority assignment $\Gamma_k$ such that for all $1 \le h < h' \le k$, $\widetilde{T}^{(max)}_h(\Gamma_k) < \widetilde{T}^{(min)}_{h'}(\Gamma_k)$.
\end{flushleft}
\begin{itemize}[wide=0pt]
\item \textbf{Base case ($k=2$):} For $k=2$, we show that for any assignment $\Gamma_2$, if $\widetilde{T}^{(max)}_1(\Gamma_2) \ge \widetilde{T}^{(min)}_2(\Gamma_2)$, then there exists another assignment $\Gamma'_2$ satisfying $\widetilde{T}^{(max)}_1(\Gamma'_2) < \widetilde{T}^{(min)}_2(\Gamma'_2)$ and $C^*(\Gamma'_2) \le C^*(\Gamma_2)$.

According to Proposition~\ref{prop:min_bw}, the minimum bandwidth required for assignment $\Gamma_2$ is
\begin{align*}
& C^*(\Gamma_2)=\\
& \max
\begin{cases}
    \sum_{i = 1}^n r_i,\\
    \sup_{t > \widetilde{T}^{(min)}_1(\Gamma_2)}\left\{\frac{l_1^{max}(\Gamma_2) + H_{G_1(\Gamma_2)}(t - \widetilde{T}^{(min)}_1(\Gamma_2))}{t}\right\},\\
    \sup_{t > \widetilde{T}^{(min)}_2(\Gamma_2)}\left\{\frac{H_{G_1(\Gamma_2)}(t) + H_{G_2(\Gamma_2)}(t - \widetilde{T}^{(min)}_2(\Gamma_2))}{t}\right\}
\end{cases}.
\end{align*}

Define $G'_2(\Gamma_2) = \{ i \in G_1(\Gamma_2) \mid \widetilde{T}_i \ge \widetilde{T}^{(min)}_2(\Gamma_2)\}$ (the set of flows in $G_1(\Gamma_2)$ with deadlines larger than or equal to the smallest deadline of flows in $G_2(\Gamma_2)$),  and $G'_1(\Gamma_2) = G_1(\Gamma_2) \setminus G'_2(\Gamma_2)$ (the set $G_1(\Gamma_2)$ from which flows with deadlines larger than or equal to the smallest deadline of flows in $G_2(\Gamma_2)$ have been removed).
Next, construct $\Gamma'_2$ by setting $G_1(\Gamma'_2) = G'_1(\Gamma_2)$ and $G_2(\Gamma'_2) = G_2(\Gamma_2) \cup G'_2(\Gamma_2)$. This then yields $\widetilde{T}^{(max)}_1(\Gamma'_2) < \widetilde{T}^{(min)}_2(\Gamma'_2)$ and $\widetilde{T}^{(min)}_i(\Gamma'_2) = \widetilde{T}^{(min)}_i(\Gamma_2)$ for $i=1,2$, so that the required bandwidth under $\Gamma'_2$ is given by
\begin{align*}
& C^*(\Gamma'_2)=\\
& \max
\begin{cases}
    \sum_{i = 1}^n r_i,\\
    \sup_{t > \widetilde{T}^{(min)}_1(\Gamma_2)}\left\{\frac{l_1^{max}(\Gamma'_2) + H_{G_1(\Gamma'_2)}(t - \widetilde{T}^{(min)}_1(\Gamma_2))}{t}\right\},\\
    \sup_{t > \widetilde{T}^{(min)}_2(\Gamma_2)}\left\{\frac{H_{G_1(\Gamma'_2)}(t) + H_{G_2(\Gamma'_2)}(t - \widetilde{T}^{(min)}_2(\Gamma_2))}{t}\right\}
\end{cases}.
\end{align*}
We next show $C^*(\Gamma'_2) \le C^*(\Gamma_2)$.
Because $G_2(\Gamma_2) \subset G_2(\Gamma'_2)$, we have $l_1^{max}(\Gamma'_2) \ge l_1^{max}(\Gamma_2)$.
Two cases arise:

\begin{itemize}
\item If $l_1^{max}(\Gamma'_2) = l_1^{max}(\Gamma_2)$, then since $G_1(\Gamma'_2) \subset G_1(\Gamma_2)$, we have
\begin{align*}
& l_1^{max}(\Gamma_2) + H_{G_1(\Gamma_2)}(t - \widetilde{T}^{(min)}_1(\Gamma_2))\\
\ge & l_1^{max}(\Gamma'_2) + H_{G_1(\Gamma'_2)}(t - \widetilde{T}^{(min)}_1(\Gamma_2)).
\end{align*}

\item If $l_1^{max}(\Gamma'_2) > l_1^{max}(\Gamma_2)$, there exists $\hat{i} \in G'_2(\Gamma_2)$ such that $l_{\hat{i}} > l_1^{max}(\Gamma_2)$.
Because $\sigma_{\hat{i}}(t) \ge l_{\hat{i}}$ for all $t > 0$ and $ G_1(\Gamma'_2) \subseteq G_1(\Gamma_2) - \hat{i}$,
\begin{equation*}
\begin{aligned}
& l_1^{max}(\Gamma_2) + H_{G_1(\Gamma_2)}(t - \widetilde{T}^{(min)}_1(\Gamma_2))\\
= & l_1^{max}(\Gamma_2) + H_{G_1(\Gamma_2) - \hat{i}}(t - \widetilde{T}^{(min)}_1(\Gamma_2)) \\
& + \sigma_{\hat{i}}(t - \widetilde{T}^{(min)}_1(\Gamma_2))\\
\ge & l_1^{max}(\Gamma_2) + H_{G_1(\Gamma'_2)}(t - \widetilde{T}^{(min)}_1(\Gamma_2)) + l_{\hat{i}}\\
\ge & l_1^{max}(\Gamma'_2) + H_{G_1(\Gamma'_2)}(t - \widetilde{T}^{(min)}_1(\Gamma_2)).
\end{aligned}
\end{equation*}
\end{itemize}

Finally, since $H_{G'_2(\Gamma_2}(t) \ge H_{G'_2(\Gamma_2)}(t - \widetilde{T}^{(min)}_2(\Gamma_2))$, it follows that
\begin{equation*}
\begin{aligned}
& H_{G_1(\Gamma_2)}(t) + H_{G_2(\Gamma_2)}(t - \widetilde{T}^{(min)}_2(\Gamma_2))\\
= & H_{G_1(\Gamma'_2)}(t) + H_{G'_2(\Gamma_2)}(t) + H_{G_2(\Gamma_2)}(t - \widetilde{T}^{(min)}_2(\Gamma_2))\\
\ge & H_{G_1(\Gamma'_2)}(t) + H_{G'_2(\Gamma_2)}(t - \widetilde{T}^{(min)}_2(\Gamma_2)) \\
& + H_{G_2(\Gamma_2)}(t - \widetilde{T}^{(min)}_2(\Gamma_2))\\
= & H_{G_1(\Gamma'_2)}(t) + H_{G_2(\Gamma'_2)}(t - \widetilde{T}^{(min)}_2(\Gamma_2)).
\end{aligned}
\end{equation*}
Combining the two ensures $C^*(\Gamma'_2) \le C^*(\Gamma_2)$.

\item \textbf{Induction Step:} Assume $I(k)$ holds for $k \ge 2$. We show that it also holds for $k+1$.

For any $(k+1)$-priority assignment $\Gamma_{k+1}$, the required bandwidth is
\begin{equation}
C^*(\Gamma_{k + 1}) = 
\max_{1 \le h \le k}\begin{cases}
\sum_{i = 1}^n r_i,\\
\sup_{t > \widetilde{T}^{(min)}_h(\Gamma_{k+1})}\{S_h(t)/t\}
\end{cases},
\end{equation}
where $S_h(t) = \sum_{1 \le h' < h}H_{G_{h'}(\Gamma_{k+1})}(t) + l_h^{max}(\Gamma_{k+1}) + H_{G_{h}(\Gamma_{k+1})}(t - \widetilde{T}^{(min)}_h(\Gamma_{k+1}))$.

We first show that there exists a $(k+1)$-priority assignment $\Gamma'_{k+1}$ satisfying
\smallskip
\begin{itemize}
    \item \textbf{Condition~$1$}: $C^*(\Gamma'_{k+1}) \le C^*(\Gamma_{k+1}) \,\forall\,\Gamma_{k+1}$, and $\exists\, \hat{m}$ such that $G_1(\Gamma'_{k+1}) = \{i \mid \widetilde{T}_1 \le \widetilde{T}_i < \widetilde{T}_{\hat{m}}\}$.
\end{itemize}
\smallskip
Under a slight abuse of notation, Condition~$2$ then states that for any assignment $\Gamma_{k+1}$ satisfying Condition~$1$, there exists a $(k+1)$-priority assignment $\Gamma'_{k+1}$ satisfying
\smallskip
\begin{itemize}
    \item \textbf{Condition~$2$}: $C^*(\Gamma'_{k+1}) \le C^*(\Gamma_{k+1})$ and $\widetilde{T}^{(max)}_h(\Gamma'_{k+1}) < \widetilde{T}^{(min)}_{h'}(\Gamma'_{k+1}),\ \forall 1 \le h < h' \le k+1$.
\end{itemize}    
\smallskip
Once established, combining these two conditions completes the induction and proves $I(k+1)$.
\begin{enumerate}[wide=0pt]
\item We first show the existence of an assignment $\Gamma'_{k+1}$ satisfying Condition~$1$. If $\Gamma_{k+1}$ satisfies Condition~$1$, then $\Gamma_{k+1} = \Gamma'_{k+1}$. Otherwise, for all $1 \le \hat{m} \le m$, $G_1(\Gamma_{k+1}) \neq \{ i \ | \ \widetilde{T}_1 \leq \widetilde{T}_i < \widetilde{T}_{\hat{m}}\}$. Define $\hat{i} = \min\{ 1\leq i \leq n \ | \ i \notin G_1(\Gamma_{k+1})\}$ and suppose $\hat{i} \in G_{\hat{h}}(\Gamma_{k+1})$. Further define $G'_{\hat{h}}(\Gamma_{k+1}) = \{ i \in G_1(\Gamma_{k+1})\ | \ \widetilde{T}_i \ge \widetilde{T}^{(min)}_{\hat{h}}(\Gamma_{k+1})= \widetilde{T}_{\hat{i}} \}$ and $G'_1(\Gamma_{k+1}) = G_1(\Gamma_{k+1}) - G'_{\hat{h}}(\Gamma_{k+1})$.

Consider the assignment $\Gamma'_{k+1}$ such that 1) $G_1(\Gamma'_{k+1}) = G'_1(\Gamma_{k+1})$, 2) $G_{\hat{h}}(\Gamma'_{k+1}) = G'_{\hat{h}}(\Gamma_{k+1}) \cup G_{\hat{h}}(\Gamma_{k+1})$, and 3) $G_i(\Gamma'_{k+1}) = G_i(\Gamma_{k+1})$, when $2 \le i \le k + 1$ and $i \neq \hat{h}$. Note that 
\begin{itemize}[wide=0pt]
\item when $h > \hat{h}$, neither the higher-priority traffic $\sum_{1 \le h' < h}H_{G_{h'}(\Gamma_{k+1})}(t)$, nor the lower-priority maximum packet size $l_h^{(max)}(\Gamma_{k+1})$, nor the traffic from priority class $h$ itself $H_{G_{h}(\Gamma_{k+1})}(t - \widetilde{T}^{(min)}_h(\Gamma_{k+1}))$ change. The minimal service function of priority class $h$ is, therefore, unchanged or $S_h(\Gamma_{k+1}) = S_h(\Gamma'_{k+1})$;
\item when $h = \hat{h}$, the lower-priority maximum packet size $l_{\hat{h}}^{(max)}(\Gamma_{k+1})$ remains unchanged. For the higher priority traffic, we have
\begin{equation*}
\begin{aligned}
    & \sum_{1 \le h' < \hat{h}}H_{G_{h'}(\Gamma_{k+1})}(t) - \sum_{1 \le h' < \hat{h}}H_{G_{h'}(\Gamma'_{k+1})}(t)\\
    = & H_{G_1(\Gamma_{k+1})}(t) - H_{G'_1(\Gamma_{k+1})}(t)\\
    = & H_{G'_{\hat{h}}(\Gamma_{k+1})}(t),
\end{aligned}
\end{equation*}
and for traffic from priority class $\hat{h}$
\begin{equation*}
\begin{aligned}
    & H_{G_{\hat{h}}(\Gamma_{k+1})}(t - \widetilde{T}^{(min)}_{\hat{h}}(\Gamma_{k+1}))\\
    & - H_{G_{\hat{h}}(\Gamma'_{k+1})}(t - \widetilde{T}^{(min)}_{\hat{h}}(\Gamma_{k+1}))\\
    = & - H_{G'_{\hat{h}}(\Gamma_{k+1})}(t - \widetilde{T}^{(min)}_{\hat{h}}(\Gamma_{k+1})).
\end{aligned}
\end{equation*}
Since $H_{G'_{\hat{h}}(\Gamma_{k+1})}(t) \ge H_{G'_{\hat{h}}(\Gamma_{k+1})}(t - \widetilde{T}^{(min)}_{\hat{h}}(\Gamma_{k+1}))$, we have
\begin{equation*}
\begin{aligned}
    & S_{\hat{h}}(\Gamma_{k+1}) - S_{\hat{h}}(\Gamma'_{k+1})\\
    = & H_{G'_{\hat{h}}(\Gamma_{k+1})}(t) - H_{G'_{\hat{h}}(\Gamma_{k+1})}(t - \widetilde{T}^{(min)}_{\hat{h}}(\Gamma_{k+1}))\\
    \ge & 0.
\end{aligned}
\end{equation*}
\item when $h < \hat{h}$, $l_h^{(max)}(\Gamma'_{k+1}) = \max\left\{l_h^{(max)}(\Gamma_{k+1}),\  \max_{i \in G'_{\hat{h}}} l_i \right\}$.
\begin{itemize}[wide=0pt]
    \item when $h > 1$, the traffic from class $h$ itself does not change. Let
    \begin{equation*}
    \begin{aligned}
        \Delta S_h
        &= \sum_{1 \le h' < h}H_{G_{h'}(\Gamma_{k+1})}(t) - \sum_{1 \le h' < h}H_{G_{h'}(\Gamma'_{k+1})}(t)\\
        &= H_{G_1(\Gamma_{k+1})}(t) - H_{G'_1(\Gamma_{k+1})}(t)\\
        &= H_{G'_{\hat{h}}(\Gamma_{k+1})}(t)
    \end{aligned}
    \end{equation*}
    denote the reduction on higher-priority traffic by replacing $\Gamma_{k+1}$ with $\Gamma'_{k+1}$.
    \item when $h = 1$, class $h$ is the highest priority class. Let
    \begin{equation*}
    \begin{aligned}
        & \Delta S_h\\
        = & H_{G_1(\Gamma_{k+1})}(t - \widetilde{T}^{(min)}_1(\Gamma_{k+1})) - H_{G_1(\Gamma_{k+1})}(t - \widetilde{T}^{(min)}_1(\Gamma'_{k+1}))\\
        = & H_{G'_{\hat{h}}(\Gamma_{k+1})}(t - \widetilde{T}^{(min)}_1(\Gamma'_{k+1}))
    \end{aligned}
    \end{equation*}
    denote the traffic reduction from class~$1$
    itself by replacing $\Gamma_{k+1}$ with $\Gamma'_{k+1}$.
\end{itemize}
In both cases, when $l_h^{(max)}(\Gamma'_{k+1}) = l_h^{(max)}(\Gamma_{k+1})$, we have $S_h(\Gamma'_{k+1}) \le S_h(\Gamma_{k+1})$ since $\Delta S_h \ge 0$. When there exists a flow 
$\tilde{i} \in G'_{\hat{h}}(\Gamma_{k+1})$ such that $l_h^{(max)}(\Gamma'_{k+1}) = 
l_{\tilde{i}}>  l_h^{(max)}(\Gamma_{k+1})$. Since both 
$\sigma_{\tilde{i}}(t)\ge l_{\tilde{i}}$ and 
$\sigma_{\tilde{i}}(t - \widetilde{T}^{(min)}_h(\Gamma'_{k+1})) \ge l_{\tilde{i}}$ when $t > \widetilde{T}^{(min)}_h(\Gamma'_{k+1})$, we have $\Delta S_h \ge 
l_{\tilde{i}}$, and therefore
    \begin{equation*}
    \begin{aligned}
        & S_h(\Gamma_{k+1}) - S_h(\Gamma'_{k+1})\\
        = & l_h^{max}(\Gamma_{k+1}) - l_h^{(max)}(\Gamma'_{k+1}) + \Delta S_h\\
        = & l_h^{max}(\Gamma_{k+1}) + \Delta S_h - l_{\tilde{i}}\\
        \ge & 0,
    \end{aligned}
    \end{equation*}
    which also implies $S_h(\Gamma'_{k+1}) \le S_h(\Gamma_{k+1})$.
\end{itemize}

We have, therefore, shown that for all priority classes $S_h(\Gamma'_{k+1}) \le S_h(\Gamma_{k+1})$, which establishes that $C^*(\Gamma'_{k+1}) \le C^*(\Gamma_{k+1})$ and proves the existence of an assignment $\Gamma'_{k+1}$ satisfying Condition~$1$.

\item Next we show that for any $(k+1)$-priority assignment $\Gamma_{k+1}$ satisfying Condition 1, there exists a $(k+1)$-priority assignment $\Gamma'_{k+1}$ satisfying Condition 2.

For $\Gamma_{k+1}$, there exists $1 \leq \hat{m} \leq m$ such that $G_1(\Gamma_{k+1}) = \{ i\ | \ \widetilde{T}_1 \leq \widetilde{T}_i < \widetilde{T}_{\hat{m}}\ \}$. If $G_1(\Gamma_{k+1}) = \emptyset$, by induction of hypothesis $I(k)$ we have $I(k+1)$. Hence, we focus on the case where $G_1(\Gamma_{k+1}) \neq \emptyset$.

Consider the subset of flows $\tilde{\mathcal{F}} = \{i | \hat{m} \le i \le m \}$, i.e., $\tilde{\mathcal{F}} = \mathcal{F} - G_1(\Gamma_{k+1})$. According to $I(k)$, there exists a $k$-priority assignment $\Gamma'_k$ for $\tilde{\mathcal{F}}$ such that $\forall 1 \le h < h' \le k$, $\widetilde{T}^{(max)}_h (\Gamma'_k) < \widetilde{T}^{(min)}_{h'}(\Gamma'_k)$. $\Gamma'_k$ gives a minimum required bandwidth of 
\begin{equation*}
    C^*(\Gamma'_k) = \max_{1 \le h \le k}\left\{\sum_{i = \hat{m}}^{k+1} r_i, \sup_{t > \widetilde{T}^{(min)}_h(\Gamma'_k)}\{S_h(t|\Gamma'_k)/t\}\right\}.
\end{equation*}

Consider the $k$-priority assignment $\Gamma_k$, where $G_h(\Gamma_k) = G_{h+1}(\Gamma_{k+1})$ for all $1 \leq h \leq k$. Applying $\Gamma_k$ to $\tilde{\mathcal{F}}$ and using the fact that $C^*(\Gamma_k) \geq C^*(\Gamma'_k)$, we then have 
\begin{equation*}
\begin{aligned}
& C^*(\Gamma_k)\\
= & \max_{1 \leq h \leq k}\left\{\sum_{i = \hat{m}}^{k+1} r_i, \sup_{t > \widetilde{T}^{(min)}_h(\Gamma_k)}\{S_h(t|\Gamma_k)/t\}\right\}\\
\ge & \max_{1 \leq h \leq k}\left\{\sum_{i = \hat{m}}^{k+1} r_i, \sup_{t > \widetilde{T}^{(min)}_h(\Gamma'_k)}\{S_h(t|\Gamma'_k)/t\}\right\}\\
\iff & \max_{1 \leq h \leq k}\left\{\sup_{t > \widetilde{T}^{(min)}_h(\Gamma_k)}\{S_h(t|\Gamma_k)/t\}\right\}\\
\ge & \max_{1 \leq h \leq k}\left\{\sup_{t > \widetilde{T}^{(min)}_h(\Gamma'_k)}\{S_h(t|\Gamma'_k)/t\}\right\},
\end{aligned}
\end{equation*}
which further gives
\begin{equation}
\label{order:relation}
\begin{aligned}
& \max_{1 \leq h \leq k}\left\{\sup_{t > \widetilde{T}^{(min)}_h(\Gamma_k)}\Big\{\big(S_h(t|\Gamma_k) + H_{G_1(\Gamma_{k+1})}(t)\big)/t\Big\}\right\}\\
\ge & \max_{1 \leq h \leq k}\left\{\sup_{t > \widetilde{T}^{(min)}_h(\Gamma'_k)}\Big\{\big(S_h(t|\Gamma'_k) + H_{G_1(\Gamma_{k+1})}(t)\big)/t\Big\}\right\}.
\end{aligned}
\end{equation}

Now consider the $(k+1)$-priority assignment $\Gamma'_{k+1}$, where $G_{h+1}(\Gamma'_{k+1}) = G_h(\Gamma'_k)$ for all $1 \leq h \leq k$, and $G_1(\Gamma'_{k+1}) = G_1(\Gamma_{k+1})$. By the definition of $G_1(\Gamma_{k+1})$ and $\Gamma'_k$, we know that $\widetilde{T}^{(max)}_h(\Gamma'_{k+1}) < \widetilde{T}^{(min)}_{h'}(\Gamma'_{k+1}), \forall 1 \le h < h' \le k+1$. Next we show that $C^*(\Gamma'_{k+1}) \leq C^*(\Gamma_{k+1})$.

Since $S_{h+1}(\Gamma'_{k+1}) = S_h(\Gamma'_k) + H_{G_1(\Gamma_{k+1})}(t), \forall 1 \le h \le k$, combined with Inequality~\ref{order:relation}, we have
\begin{equation*}
\begin{aligned}
& C^*(\Gamma'_{k+1})\\
= & \max_{1 \leq h \leq k + 1}\left\{\sum_{i = 1}^n r_i, \sup_{t > \widetilde{T}^{(min)}_h(\Gamma'_{k+1})}\{S_h(t|\Gamma'_{k+1})/t\}\right\}\\
= & \max_{2 \leq h \leq k + 1}
\begin{cases}
    \sum_{i = 1}^n r_i,\\
    \sup_{t > \widetilde{T}^{(min)}_h(\Gamma'_k)}\Big\{\big(S_h(t|\Gamma'_k) + H_{G_1(\Gamma_{k+1})}(t)\big)/t\Big\}\\
    \sup_{t > \widetilde{T}^{(min)}_1(\Gamma'_{k+1})}\{S_1(t|\Gamma'_{k+1})/t\}
\end{cases}\\
\le & \max_{2 \leq h \leq k + 1}
\begin{cases}
    \sum_{i = 1}^n r_i,\\
    \sup_{t > \widetilde{T}^{(min)}_h(\Gamma_k)}\Big\{\big(S_h(t|\Gamma_k) + H_{G_1(\Gamma_{k+1})}(t)\big)/t\Big\},\\
    \sup_{t > \widetilde{T}^{(min)}_1(\Gamma_{k+1})}\{S_1(t|\Gamma_{k+1})/t\}
\end{cases}\\
= & \max_{1 \leq h \leq k + 1}\left\{\sum_{i = 1}^n r_i, \sup_{t > \widetilde{T}^{(min)}_h(\Gamma_{k+1})}\{S_h(t|\Gamma_{k+1})/t\}\right\}\\
= & C^*(\Gamma_{k+1}).
\end{aligned}
\end{equation*}
Hence we show the existence of a assignment $\Gamma'_{k+1}$ satisfying Condition 2.

\end{enumerate}
\end{itemize}

\section{Technical Details}
\label{app:technical_details}

\subsection{NLP Formulation for the FIFO Case}
\label{app:fifo_nlp}

The formulation of Non-Linear Programs (NLPs) for solving $\textbf{MIN}_{FIFO}$ closely follows the methodology in Section IV.B and Appendix B.E of~\cite{qiu2024benefits}. We first note that, according to~\Eqref{eq:min_bw}, the link bandwidths $C_j$ ($1 \leq j \leq n$) can be derived from the scheduling delay bounds $T_j$ and the flow shaping delays $D_i$ ($1 \leq i \leq m$). According to~\Eqref{eq:min_bw} and Lemma 7 of~\cite{qiu2024benefits}, the required bandwidth $C^*_j$ is given by
\beq
\label{eq:min_bw_fifo}
C^*_j = \max_{\hat{i} \in \mathcal{F}_j}\left\{\sum_{i \in \mathcal{F}_j} r_i, \frac{\sum_{i \in \mathcal{F}_j} \sigma_{i}(D_{\hat{i}})}{\widetilde{T}'_{\hat{i}j}} \right\},
\eeq
where $\widetilde{T}'_{ij} = T_j + D_i$ is the inflection point of the minimal service function $S_{j}$.

Solving $\textbf{MIN}_{FIFO}$ entails minimizing $\sum_{j=1}^{n} C_j$ using~\Eqref{eq:min_bw_fifo}, by exploring all feasible $T_j$ and $D_i$ combinations that satisfy~\Eqref{eq:opt_fifo}. To formulate $\textbf{MIN}_{FIFO}$ as an NLP, we first require a closed-form expression for~\Eqref{eq:min_bw_fifo}. This is achievable when the relative order of the inflection points $\widetilde{T}'_{ij}$ is fixed—denoted as condition \textbf{ORD} in Section IV.B of~\cite{qiu2024benefits}—which depends solely on $D_i$ because $T_j$ is identical for all flows sharing link $j$ under FIFO scheduling. Consequently, once an ordering of $D_i$ ($1 \leq i \leq m$) is specified, the ordering of all inflection points across all links is determined, enabling a closed-form NLP formulation for $\textbf{MIN}_{FIFO}$.

We now consider a concrete example of a link $j$ with two flows. Suppose the shaping delays satisfy:
\bequn
D_1 \leq D_2.
\eequn
The required bandwidth $C_j$fig:dd can be represented as a set of nonlinear constraints, beginning with the stability constraint:
\bequn
C_j \geq \sum_{i \in \mathcal{F}_j} r_i = r_1 + r_2,
\eequn
Additional constraints arise from the inflection points on $S_j$:
\bequn
\begin{aligned}
    C_j &\geq \frac{b_1 + R_2D_1}{\widetilde{T}'_{1j}}\\
    &= \frac{b_1D_2 + b_2D_1}{(T_j + D_1)D_2}\\
    C_j &\geq \frac{b_1 + r_1(D_2-D_1) + b_2}{\widetilde{T}'_{2j}}\\
    &= \frac{b_1 + r_1(D_2 - D_1) + b_2}{T_j + D_2}.
\end{aligned}
\eequn
Applying the same procedure to all network links yields the complete set of nonlinear constraints. Together with the global shaping delay ordering constraint, these form an NLP instance for solving the problem.

Finalizing a solution to $\textbf{MIN}_{FIFO}$ entails enumerating all permutations of $D_i$ and solving the corresponding NLPs, a process that is inherently combinatorial. To balance solution quality with computational efficiency, we adopt the randomized combinatorial search strategy proposed in~\cite{qiu2024benefits}, which evaluates only a logarithmic subset of feasible orderings.

\section{Supplementary Results}

\subsection{Performance of $k$-means Clustering on Priority Assignment}
\label{app:k_means_performance}

\begin{figure*}[!h]
\centering
\begin{subfigure}{0.3\linewidth}
  \centering
  \includegraphics[width=\linewidth]{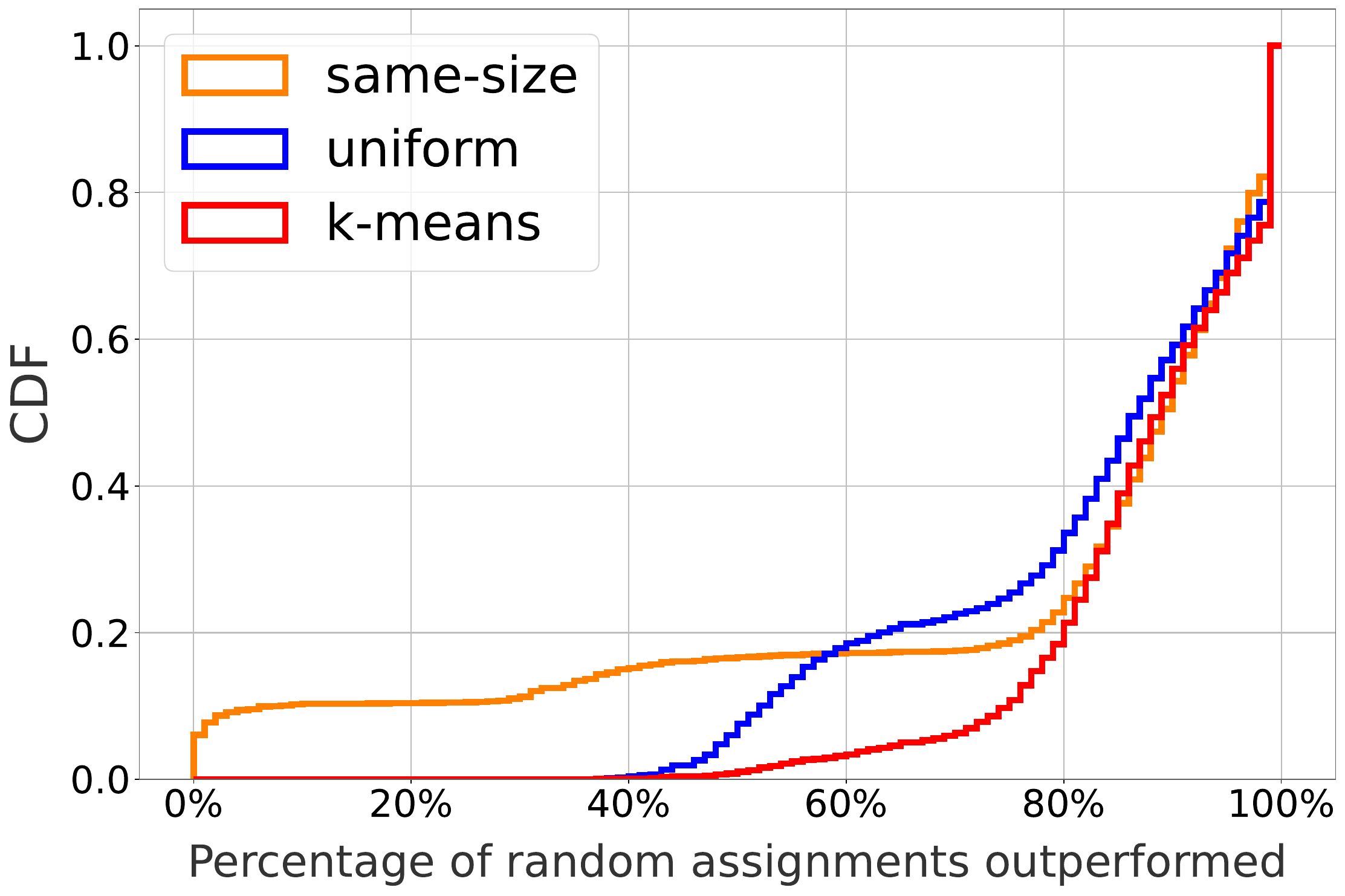}
  \caption{2 priority classes}
  \label{fig:priority_assignment_2}
\end{subfigure}
\begin{subfigure}{0.3\linewidth}
  \centering
  \includegraphics[width=\linewidth]{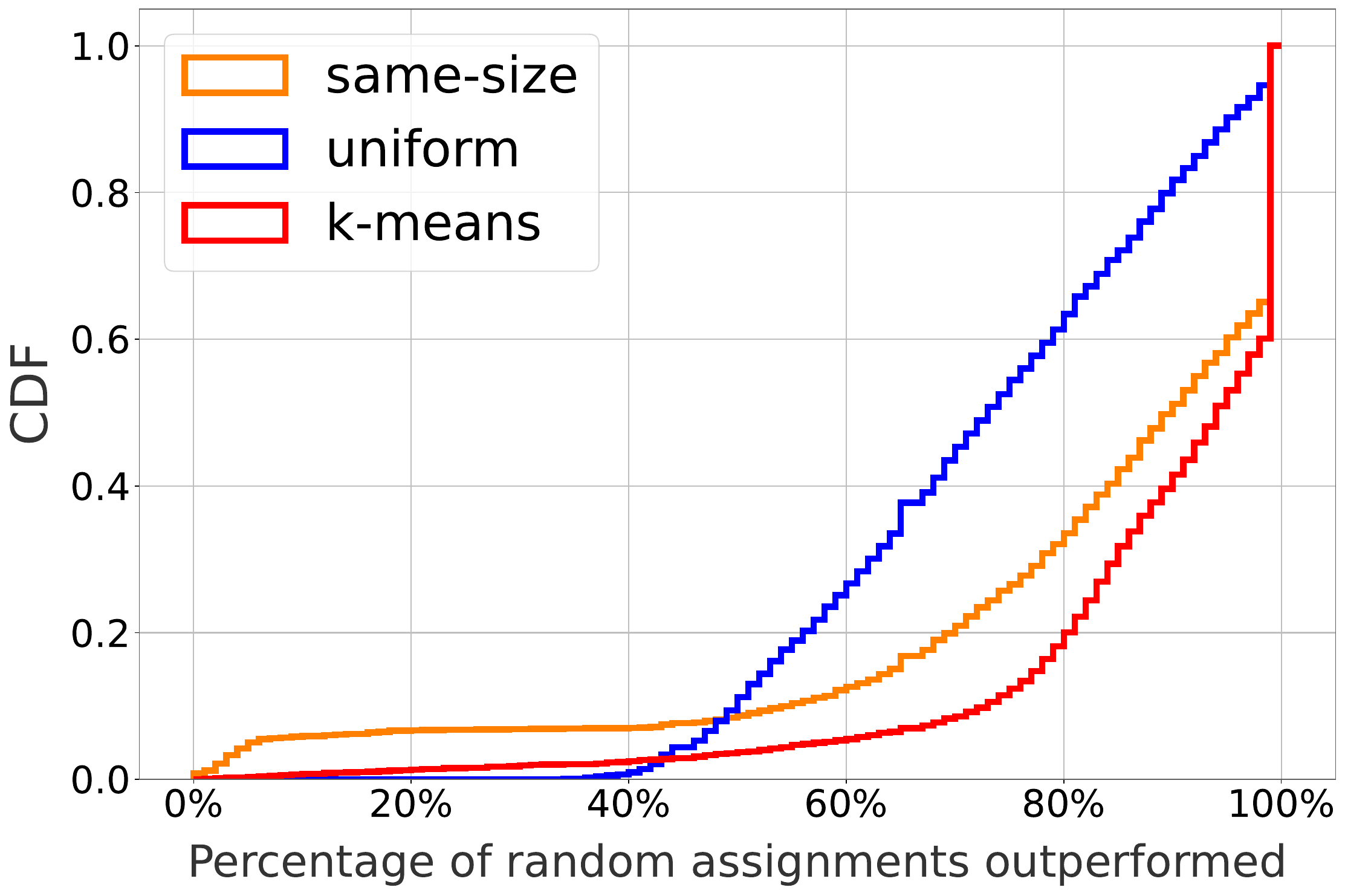}
  \caption{4 priority classes}
  \label{fig:priority_assignment_4}
\end{subfigure}
\begin{subfigure}{0.3\linewidth}
  \centering
  \includegraphics[width=\linewidth]{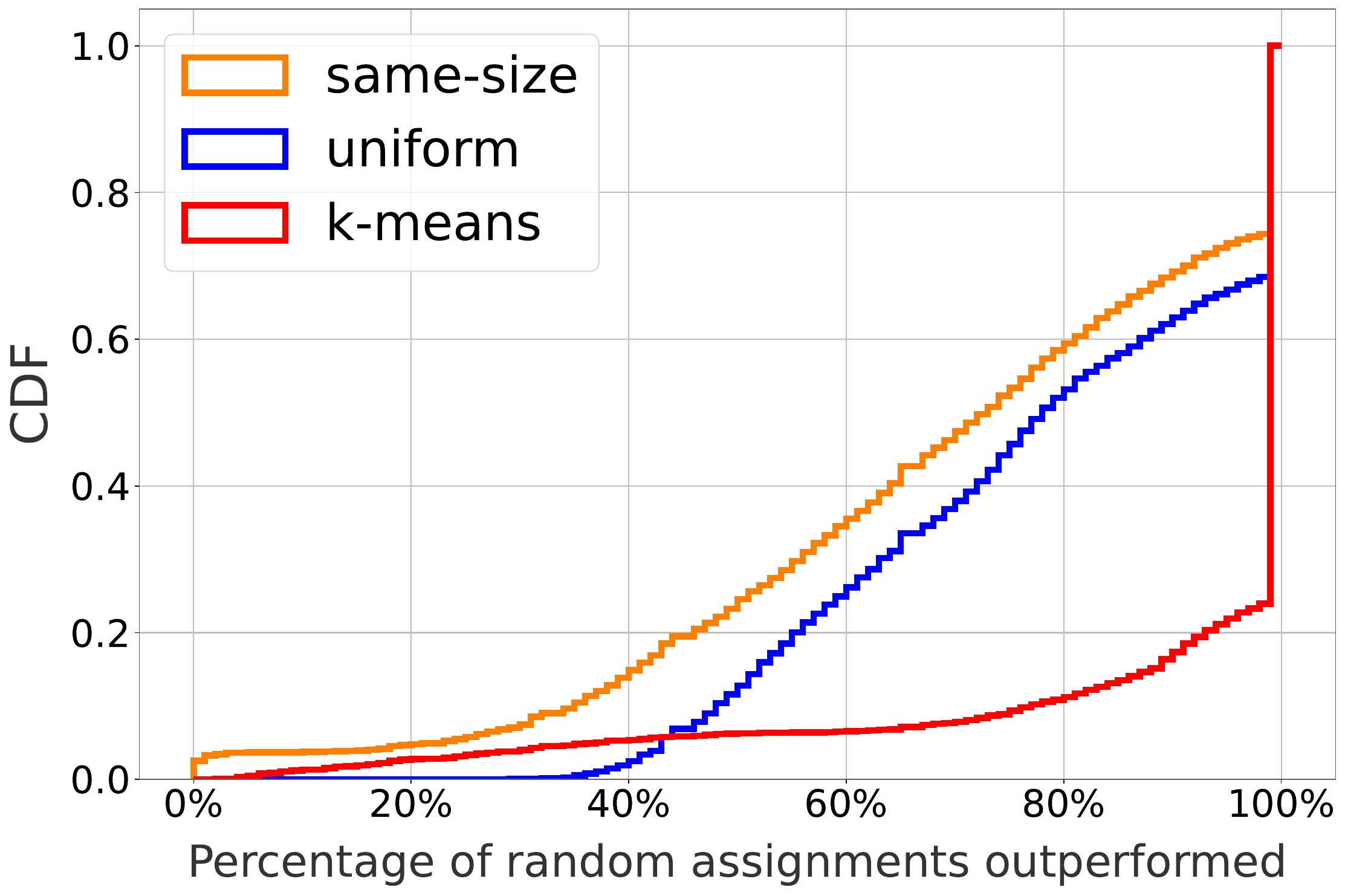}
  \caption{8 priority classes}
  \label{fig:priority_assignment_8}
\end{subfigure}
\caption{Distributions on the percentage of random assignments outperformed across different priority assignment strategies.}
\label{fig:priority_assignment}
\end{figure*}

Recall from Lemma~\ref{lemma:assign_ddl_boundary} that assigning flows to priority classes at each hop amounts to selecting $k+1$ boundary values $\{\check{T}_{hj}\}_{h=1}^{k+1}$ to partition flows' deadlines.  To assess the performance of $k$-means in selecting these boundaries, we compare it against three alternative strategies: \emph{same-size} that selects boundaries to ensure that each priority class is assigned the same number of flows; \emph{uniform} that selects equidistant boundaries in the range of local deadlines, and \emph{random} that randomly places boundaries in the range of local deadlines.  We use the \emph{random} strategy as a baseline and report the performance of the other three strategies relative to it.  Specifically, we report the percentage of random assignments they outperform (\ie assignments requiring as much or more bandwidth).

We fix the number of flows to $3000$, and sample $10$ random source–destination pairs (s-d pairs) from the US-Topo topology, each with flow profiles drawn from the corresponding application distributions. For each of these $10$ configurations, we generate $10$ random initial deadline allocations\footnote{As described in Section~\ref{sec:solution_sp}, initial allocations are obtained by randomly selecting shaping delays and evenly distributing the remaining delay budget across hops.}, resulting in a total of $100$ instances.

For each instance, we apply the three candidate strategies ($k$-means, same-size, and uniform) to determine priority assignments, and compare them against $100$ randomly generated assignments. The required bandwidth at each link is computed using~\Eqref{eq:min_bw}. For each strategy, we record the percentage of random assignments it outperforms at each link, and aggregate this metric across all links. The resulting cumulative distribution functions (CDFs) are plotted in~\fig{fig:priority_assignment} for $k=2$, $4$, and $8$ priority classes.

A lower CDF indicates a higher likelihood of outperforming random assignments, and thus better bandwidth minimization performance. Across all values of $k$, $k$-means consistently outperforms the other strategies, with its advantage increasing with the number of priority classes. For example, $k$-means outperforms $90\%$ of random assignments with probabilities approximately $50\%$, $60\%$, and $85\%$ for $k=2$, $4$, and $8$, respectively. This suggests that, as the number of priority classes grows and the boundary selection problem becomes more complex, simple rule-based or random strategies become less effective, while $k$-means better captures the impact of the underlying distribution of local deadlines.

Intuitively, $k$-means groups flows with similar local deadlines into the same priority class, creating a clearer separation of scheduling requirements across classes. This structure allows the scheduler to more effectively absorb burstiness from higher-priority traffic before serving lower-priority classes, thereby reducing the overall bandwidth requirement.

\subsection{Shaping Ratio across Priority Classes}
\label{app:shaping_ratio}

\begin{figure}[!h]
\centering
\includegraphics[width=0.6\linewidth]{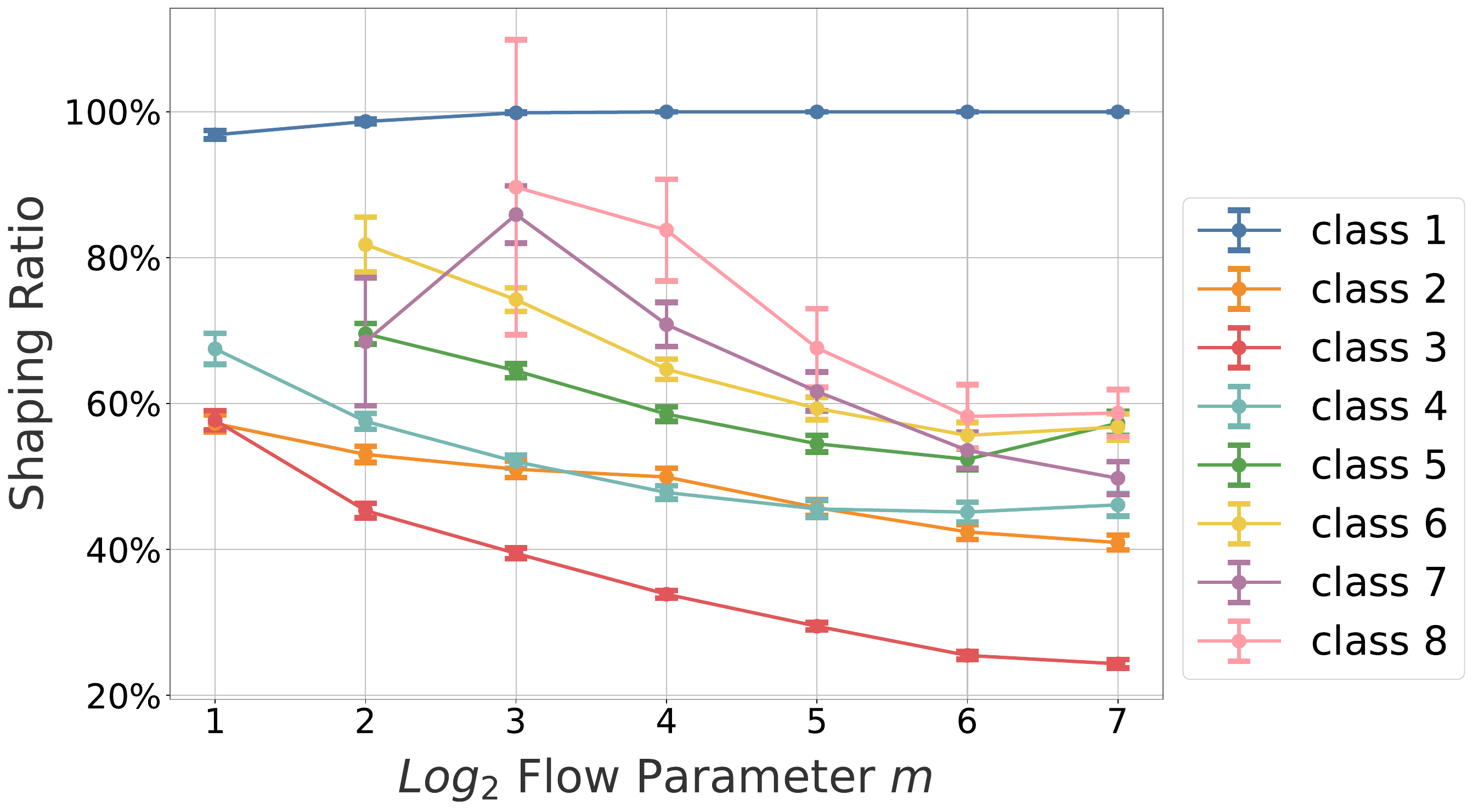}
\caption{Shaping ratio for each priority class from \Greedy's solution on parking lot topology (with the number of links $n=2$).}
\label{fig:parking_lot_shaping_ratio}
\end{figure}

Focusing on the case $n=2$ from~\fig{fig:parking_lot_fr}, when flows traverse short paths, we analyze the \emph{shaping ratio} $D_i/\widehat{d}_i$ of flows within each priority class. Since priority assignment is performed independently at each hop, a flow may belong to different priority classes along its path. We therefore compute the average shaping ratio for each class as a weighted average over flows, where the weight corresponds to the fraction of hops at which a flow is assigned to that class. 

The results are shown in~\fig{fig:parking_lot_shaping_ratio}. Some higher-index (lower-priority) classes, such as classes~6–8, may not appear for small numbers of flows, as they are only populated when sufficient flow diversity exists. We observe that flows in the highest priority class (class~1) are almost always fully shaped. This is intuitive: highest-priority flows experience no interference from lower-priority traffic and effectively see FIFO-like service\footnote{Under the fluid model, lower-priority traffic does not affect higher-priority classes.}. As a result, their optimal solution aligns closely with FS, consistent with the observations in Section~\ref{sec:evaluation_fifo}.

For lower-priority classes, the shaping ratio generally decreases as the number of flows increases. This trend mirrors the bandwidth improvements observed in~\fig{fig:parking_lot_fr}: with more flows, greater heterogeneity in deadline requirements allows the scheduler to exploit in-network delay allocation more effectively, reducing reliance on shaping.

Interestingly, higher-priority classes do not always exhibit higher shaping ratios. In fact, for classes~3 through~8, lower-priority classes often have higher shaping ratios. This is due to the discrete deadline classes used in the parking lot setup ($10$, $25$, $50$, and $100$\,ms). Flows with larger deadlines can afford more shaping while still retaining relatively large local deadlines, which results in their assignment to lower-priority classes despite having higher shaping ratios.

\subsection{Performance Gap between Static Priority and SCED}
\label{app:performance_gap}

\begin{figure}[!h]
\centering
\includegraphics[width=0.6\linewidth]{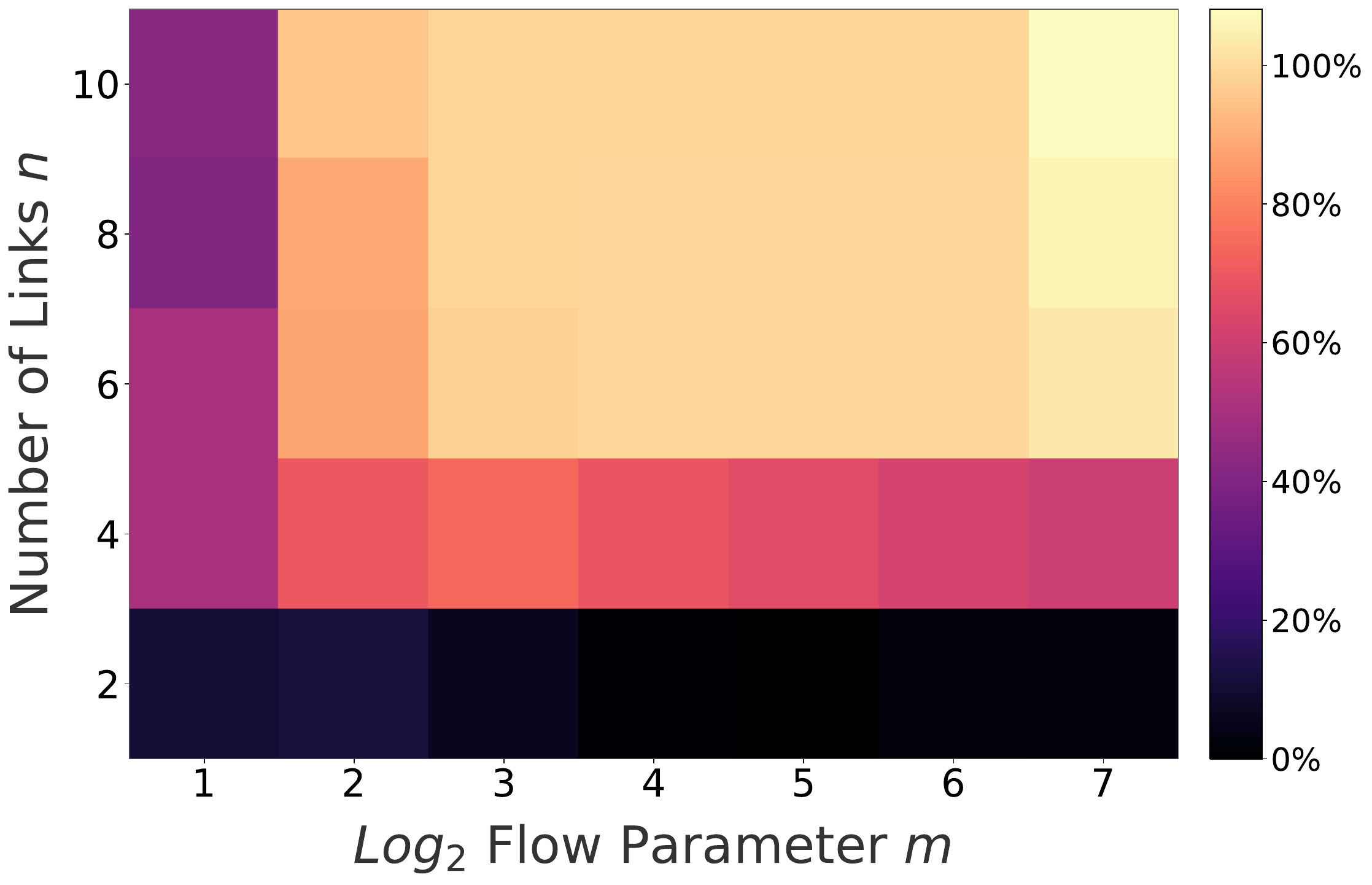}
\caption{Performance gap between static priority and SCED (with regard to relative bandwidth improvement over FIFO) on the parking lot topology.}
\label{fig:scheduler_gap}
\end{figure}

As \fig{fig:parking_lot_fifo_nr} and \fig{fig:parking_lot_baseline}, we seek to more systematically explore the impact of scale, both hop count $(n)$ and number of flows $(m)$ on the performance among schedulers.  As before, we rely on the parking lot topology of \fig{fig:parking_lot}, and report on the performance gap between SCED and static priority as $m$ and $n$ vary, relative to the improvement SCED yields over FIFO\footnote{Let the required bandwidths under FIFO, static-priority, and SCED be $x$, $y$, and $z$, respectively. The performance gap between SCED and static priority relative to SCED's improvement over FIFO is, therefore, of the form $\frac{y-z}{x-z}$.}. In other words, a gap of $100\%$ means that static priority performs no better than FIFO, while a gap of $0\%$ means that static priority achieves the same improvement over FIFO as SCED.

The results are reported in \fig{fig:scheduler_gap}, which, consistent with earlier observations, shows that increasing the number of flows introduces greater heterogeneity in delay requirements, which SCED is better able to leverage.  Additionally, as the number of hops increases, static-priority solutions tend to converge towards FS, while SCED remains able to exploit scheduling flexibility at individual hops, thereby, increasing its ability to outperform static priority. 

\end{document}